# UNIVERSIDAD NACIONAL DEL CALLAO

FACULTAD DE CIENCIAS NATURALES Y MATEMÁTICA

ESCUELA PROFESIONAL DE FÍSICA

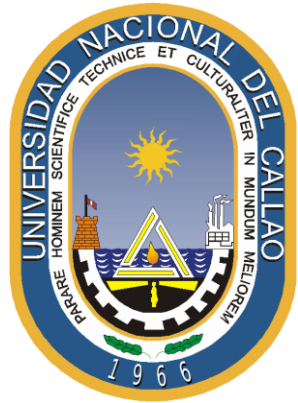

"UNA INTERPRETACIÓN DE COLAPSO OBJETIVO PARA LOS PROBLEMAS DE LA MEDIDA Y LA CLASICALIZACIÓN DONDE NO SE CONSERVA LA INFORMACIÓN"

TESIS PARA OPTAR EL TÍTULO PROFESIONAL DE LICENCIADO EN FÍSICA

SOTELO BAZÁN EDUARDO FRANCO

Callao, Diciembre, 2018

PERU

DEDICATORIA

Al estado peruano, para que fortalezca la investigación científica.

# AGRADECIMIENTO

# ÍNDICE

# RELACIÓN DE FIGURAS

# RELACIÓN DE TABLAS

# RESUMEN

En este trabajo se abordó los problemas de la clasicalización y la medida unificándolos en un solo problema: *El problema del colapso*, para esto se desarrolló un programa de trabajo – *el programa de la clasicalización* – basado en una interpretación de colapso objetivo que se propuso para este fin, y que superó las limitaciones de la interpretación oficial.

Este programa explica que la clasicalización surge por alternar la evolución unitaria y los colapsos, debido a la redistribución de la energía, así la mecánica clásica emerge como una buena aproximación; no obstante, se ha definido el *caos profundo* como el régimen donde los sistemas macroscópicos son tan sensibles a las condiciones iniciales que son afectados por los colapsos y el indeterminismo de la mecánica cuántica, permitiendo así la emergencia de la termodinámica con su irreversibilidad; a este nivel la observación ocurre porque el aparato de medida (clásico) extiende sus colapsos al microsistema a través del entrelazamiento cuántico.

En este trabajo se hizo una extensión del programa de la clasicalización a los agujeros negros, y se explicó – a un nivel aun en desarrollo – sobre el rol de los colapsos en su termodinámica; además, se hace una crítica al programa de la decoherencia, y se presentó una *teoría de la información física* que se desarrolló.

PALABRAS CLAVES: Proceso U, proceso R, clasicalización, medida cuántica, indeterminismo, entropía, flecha del tiempo, entrelazamiento.
6

# ABSTRACT

In this work, was researched the problems of clasicalization and measurement unifying them in a single problem: *The problem of collapse*, for this purpose was developed a working program – *The clasicalization's program* – in an interpretation of objective collapse that was proposed for this purpose, and which overcomes the limitations of the official interpretation.

This program explain that the classicization arises by alternating the unitary evolution and the collapses, due to the redistribution of energy, so the classical mechanics emerges as a good approximation; however, was defined *deep chaos* as the regime where macrosystems are so sensitive to the initial conditions that are affected by the collapses and indeterminism of quantum mechanics, allowing the emergence of thermodynamics with its irreversibility; at this level the observation occurs because the measurement device (classic) extends their collapses at microsystem through quantum entanglement.

In this work, was did an extension of the clasicalization's program at the black holes, and was explained – at a level still in development – about the role of the collapses in its thermodynamic; in addition, the program of quantum decoherence was criticized, and was presented a *theory of physical information* that was developing.

# CAPÍTULO I

# PLANTEAMIENTO DE LA INVESTIGACIÓN

## 1.1. Identificación del problema

La interpretación de Copenhague (I.C.) es incompleta porque no explica dos problemas: la clasicalización y el problema de la medida[1], en el experimento mental del gato de Schrödinger[2] exhiben la raíz del problema: la superposición de estados es incompatible con los resultados de una medida y los estados del mundo clásico.

La I.C. falla al abordar el experimento mental comentado líneas arriba, evidenciando que no tiene una explicación del acto de observación (deja en total misterio el mecanismo de la observación, también si todo el sistema en conjunto microsistema-observador puede evolucionar con una única función de onda o si el colapso lo irrumpe), además extiende la superposición y el enredo a sistemas clásicos sin alguna restricción explícita que evite la contradicción con la física clásica, y sobre todo, no explica la transición al mundo clásico: si todos los sistemas clásicos-macroscópicos están constituidos por partículas y sistemas microscópicos ¿cómo emerge la evolución del mundo clásico, la flecha termodinámica del tiempo, a partir de la evolución reversible de la función de onda?

El principio de correspondencia de Bohr (sección [2.2.1]) es insatisfactorio para resolver el problema de la clasicalización porque sólo aborda el problema de la discontinuidad del salto cuántico en la transición al mundo clásico "suave y continuo", sin abordar cómo la superposición de estados (incluso sin coherencia) se pierde en el mundo clásico.

---

[1] No obstante, los investigadores asumen ciertas posturas específicas para tratar estos problemas, sin embargo, la interpretación oficial es un consenso (de las interpretaciones de los fundadores) que no tiene una postura clara y determinada sobre estos problemas.
[2] En el anexo A.1 se explica esta paradoja.

En la I.C. no queda claro si la función de onda tiene una correspondencia con el mundo físico en el sentido ontológico[3], por el contrario, una interpretación objetiva (sección [2.2.3] y [2.2.4]) es consistente con que (el cuadrado de su módulo) da la distribución espacial de la "partícula" en el espacio, así al medir y encontrarlo en un autoestado del observable (en un estado clásicamente definido, al menos para la resolución del aparato de medida) ocurre el colapso como un cambio instantáneo del estado inicial previo a la medición $|\psi\rangle$, al autoestado $|a\rangle$, como se muestra:

$$|\psi\rangle = \sum_i \lambda_i |i\rangle \;\rightarrow\; |a\rangle \qquad [1.1]$$

Equivalentemente, es un cambio abrupto en la distribución del microsistema:

$$|\langle x|\psi\rangle|^2 \;\rightarrow\; |\langle x|a\rangle|^2 \qquad [1.2]$$

Inicialmente Schrödinger interpretó el cuadrado del módulo de la función de onda, solución de su ecuación, con la partícula misma (el electrón) pero luego abandonó la idea por consideraciones espaciales, pues el electrón podría expandirse hasta ser más grande que todo el átomo (y eso le pareció absurdo), sin embargo un electrón o un fotón se puede expandir a escalas macroscópicas para interferir consigo mismo en el experimento de la doble rendija, y reducirse a un punto localizado en la pantalla al colapsar su función de onda, de manera que se puede interpretar el colapso como un proceso físico real en vez de aparente o relativo al observador.

Este tipo de interpretación es de *Colapso Objetivo*[4], como en la teoría GRW (Ghigrardi, Rimini, Weber), ver sección [2.2.3], que surgieron de manera opuesta a preservar la evolución unitaria cuando se realiza una medición o se aplica a sistemas clásicos (como lo defiende la decoherencia), existiendo varios de estos programas en pleno desarrollo; no obstante, en este trabajo se ha propuesto una nueva interpretación de colapso objetivo con un enfoque propio, y con especial atención en la información física y la reversibilidad del tiempo.

---

[3] Esto es probablemente el aspecto más filosófico: ¿la función de onda es sólo una herramienta matemática que contiene la información del sistema, o existe físicamente antes de la observación?

[4] Se prefiere denominar *objetivo* al colapso en vez de la función de onda, ya que esta es una magnitud compleja al que no se le asocia una cantidad física, a menos que se insista en su interpretación objetiva y se asuma la existencia de dos amplitudes físicas desfasadas $\pi/2$ entre si (la parte real e imaginaria) en el microsistema.

Un problema cardinal, aunque sutil, es de la asimetría del tiempo en el mundo clásico-termodinámico que emerge a partir del mundo cuántico[5], cuya evolución unitaria es reversible, por simetría la entropía debería aumentar en ambos sentidos del tiempo o simplemente conservarse (sección [2.5.2]); por otro lado, si se desea "naturalizar" el rol del observador, se debe desarrollar un marco donde los colapsos están ocurriendo siempre en el mundo físico independiente a un aparato de medida, pero deja de ocurrir al aislar un microsistema[6] (esto es justamente lo que las teorías de colapso objetivo proponen), las condiciones y el mecanismo de la activación de los colapsos es el problema central en toda investigación de colapso objetivo.

### 1.2. Formulación del problema

#### 1.2.1. Problema general

¿Cómo solucionar los problemas de la medida y la clasicalización en una interpretación de colapso objetivo donde no se conserva la información?

#### 1.2.2. Problemas específicos

1. ¿Se debe extender la mecánica cuántica al mundo clásico-macroscópico y asociar una función de onda a los sistemas clásicos, o de forma inversa, se debe extender la mecánica clásica al mundo cuántico-microscópico asociando trayectorias a las partículas microscópicas y por lo tanto considerar variables ocultas?
2. Si todo el universo es un sistema aislado que evoluciona aumentando su entropía ¿cómo podría tener una función de onda que evolucione unitaria y reversiblemente?
3. Todos los observadores son sistemas clásicos, pero ¿son todos los sistemas clásicos observadores, es el mundo clásico-macroscópico un "mundo de observadores", es por eso que los sistemas macroscópicos están siempre clásicamente definidos?

---

[5] En esta interpretación se asume que la mecánica cuántica es fundamental y la física clásica se deriva de ella, por consideraciones de emergencia de sistemas y sus leyes; no obstante, esta no es una postura de unanimidad, y hay defensas muy acérrimas de que la mecánica cuántica es derivable de una teoría determinista, realista e incluso local, como la mecánica clásica; en consecuencia, esta elección es una prerrogativa del investigador motivada y sustentada en una interpretación.

[6] Al menos para escalas de tiempos de laboratorio, la teoría GRW propone un colapso espontáneo en los microsistemas aislados en un tiempo del orden de cada 100 millones de años, sección [2.2.3].

4. ¿Es el acto de observación un caso particular de la clasicalización, del "mundo de los observadores"?

5. ¿Cómo se relaciona la irreversibilidad en la medición de un microsistema (de la mecánica cuántica) con la irreversibilidad termodinámica (de la física clásica), están conectados causalmente o es sólo una coincidencia?

6. ¿Puede el mecanismo de la decoherencia resolver el problema de la medida y la clasicalización, con su interpretación del colapso aparente?

7. El colapso de la función de onda es un proceso instantáneo, y si es objetivo se le asocia un acontecimiento físico ¿puede violar el colapso objetivo la causalidad relativista, como se interpreta su ocurrencia instantánea en toda la extensión de la función de onda cuando la simultaneidad es relativa?

8. ¿Cuál es el rol de los colapsos en la termodinámica de los agujeros negros, pueden los agujeros negros desarrollar fenómenos irreversibles sin colapsos en una eventual teoría de gravedad cuántica?

## 1.3. Objetivos de la investigación

### 1.3.1. Objetivo general

Resolver y explicar el problema de la medida y la clasicalización a través de una interpretación de colapso objetivo donde no se conserva la información.

### 1.3.2. Objetivos específicos

1. Reconocer que cada teoría (cuántica y clásica) funcionan correctamente en sus respectivos dominios sin necesidad de extender cada teoría al otro dominio.

2. Demostrar que el universo (como un sistema aislado) no evoluciona unitariamente según la ecuación de Schrödinger, al ser irreversible termodinámicamente.

3. Establecer que los sistemas clásicos actúan como observadores de la mecánica cuántica, y por eso no se encuentran en superposición de estados al nivel clásico-macroscópico.
4. Explicar el mecanismo de la observación, donde el sistema clásico extiende su clasicalización al microsistema, observándolo.
5. Demostrar que la irreversibilidad termodinámica del aumento de la entropía se origina, a un nivel fundamental, en los colapsos de las funciones de ondas de los microsistemas que componen el mundo físico.
6. Responder y criticar a la interpretación de usar la decoherencia para explicar la clasicalización y el problema de la medida.
7. Explicar el colapso de la función de onda en el marco relativista, de forma consistente con la localidad y causalidad relativista.
8. Explicar el rol del colapso objetivo en la termodinámica de los agujeros negros, donde no se conserva la información.

## 1.4. Justificación

Existe un permanente debate sobre la interpretación de la mecánica cuántica (cuyas interpretaciones luchan por establecerse como la oficial), en el plano teórico se pueden exhibir las diferencias de cada interpretación que podrían conducir a resultados divergentes, un buen ejemplo es sobre una teoría de gravedad cuántica: sus condiciones son tan extremas que es posible que algún elemento teórico introducido por una interpretación de la mecánica cuántica, pueda interferir en la teorización de la gravedad cuántica (que debe abarcar tanto las escalas de Planck como las de cosmología cuántica), el hecho de desarrollar una interpretación con los elementos teóricos adecuados es justificable porque permite avanzar en los dominios donde no se ha teorizado aun satisfactoriamente.

Una segunda justificación es la consistencia teórica en la transición de la teoría cuántica a la teoría clásica: cada teoría es válida en su propio dominio, no se debe extender una teoría a la otra (como la I.C.) sino que se debe construir un tránsito del mundo microscópico (con estados superpuestos, evolución unitaria-reversible, y el irreversible colapso) al mundo macroscópico con trayectoria y evolución irreversible (la flecha termodinámica del tiempo); la consistencia de la clasicalización se convierte en un problema de mucho interés teórico puesto que demanda una explicación (al igual que el problema de la medida) que permita el paso entre sistemas macroscópicos y microscópicos con total normalidad.

Como refuerzo, una tercera justificación, se sostiene que no se podrá llegar a una teoría completa de gravedad cuántica mientras que no se solucione el problema de la clasicalización (y la medida) porque unificar la relatividad general con la mecánica cuántica (en una teoría cuántica de campos) requiere conocer una correcta transición entre la física clásica (no relativista) y la teoría cuántica, que no se dispone en la actualidad[7]; de hecho, el problema de conciliar la teoría cuántica de campos con la relatividad general se alimenta del desconocimiento de la clasicalización.

### 1.5. Importancia

La importancia de este trabajo se encuentra por un lado en el nivel interpretativo, sobre los fundamentos y problemas de la mecánica cuántica, y por otro lado en tener un modelo versátil que pasa del dominio cuántico al clásico, además de explicar el mecanismo de la observación; no obstante, no son los únicos problemas que permiten ser abordados, así por ejemplo *la paradoja de la pérdida de información en los agujeros negros* que despierta un interés de alto nivel en la física teórica y que se han presentado muchas soluciones (principio holográfico, el muro de fuego, etc) es respondido por el *teorema de la clasicalización* incluso desde el nivel no-relativista (sección [5.4.2]), el cual resume el programa de la clasicalización.

Un segundo argumento de la importancia es que permite abordar desde la física lo que para muchos físicos era sólo un problema filosófico sobre el indeterminismo y la información del universo[8]: El problema de si el indeterminismo es relativo al observador (pero que el sistema aislado microsistema-observador siga siendo determinista) o si por el contrario es una característica intrínseca del universo, ha sido evadido como propio de lo filosófico y no de lo físico (incluso la complejidad de la naturaleza es tal que da lo mismo si a su nivel fundamental es determinista o indeterminista porque es caótico-impredecible a efectos prácticos), sin embargo al concluir que el indeterminismo es responsable tanto de la irreversibilidad termodinámica como de la creación de información al nivel cósmico[9], se pone de manifiesto que es una cuestión física muy seria.

---

[7] El problema de la clasicalización y de la medida unificados no es investigado exhaustivamente aun cuando es claramente un precursor para una teoría de gravedad cuántica.

[8] Filosóficamente se ha planteado: ¿por qué el universo es de esta manera y no de otra?, lo cual puede ser físicamente abordado como: ¿por qué el universo tiene esta información física y no otra?

Finalmente, hay una importancia especial al abordar un cambio de paradigma físico: mientras que en las posturas predominantes de la I.C. se sigue aceptando el determinismo en la evolución en cualquier sistema aislado con la consecuente conservación de la información[10], en esta interpretación de colapso objetivo la información no se conserva, además que se interpreta al horizonte de eventos como una singularidad física, siendo ambos probablemente un *"sacrilegio"* en el paradigma de la física, pero son consistentes e incluso deseables para explicar los atributos astronómicos (succionar la materia circundante) y termodinámicos (evaporación por radiación de Hawking) en los agujeros negros.

---

[9] Si el universo es determinista la información se conserva y toda la información del universo de hoy se encontraría (transformado) en el pasado hasta el *Big-Bang*, por el contrario, si es indeterminista podemos decir que casi no había información al inicio y que se fue creando aleatoriamente durante millones de años en cada colapso de cada función de onda.

[10] La mejor evidencia de esto se encuentra por un lado en el programa de la decoherencia (que defiende la evolución unitaria tanto en la medida como en la clasicalización, lo cual conserva el determinismo y la información física) que viene siendo ya aceptado como parte de la mecánica cuántica estándar, y por otro lado en la paradoja de la pérdida de información en los agujeros negros (resulta en paradoja porque se aferran en defender la evolución unitaria en una dinámica clásica: la termodinámica de los agujeros negros) que incluso hizo retroceder al mismo Stephen Hawking en su postura inicial donde se violaba la unitaridad en los agujeros negros, además en las principales soluciones a la paradoja siempre se conserva la unitaridad, en vez de simplemente resolver por no conservarla.

# CAPÍTULO II
# MARCO TEÓRICO

## 2.1. Evolución e interacción cuántica de un sistema microscópico

### 2.1.1. El proceso unitario (proceso U)

Casi toda la mecánica cuántica está basada en procesos unitarios[11], sea el ket $|\psi(t)\rangle$ el estado del microsistema a un tiempo dado $t$, la ley que gobierna su evolución está dada por la ecuación de Schrödinger, en notación de ket:

$$i\hbar \frac{d}{dt}|\psi(t)\rangle = \widehat{H}|\psi(t)\rangle \qquad [2.1]$$

Donde $\widehat{H}$ es el Hamiltoniano del sistema, esta evolución es determinista porque si se conoce el estado del microsistema a un tiempo dado $t_0$ se puede conocer su estado en cualquier otro tiempo $t$, mediante el uso del operador de evolución cuántico unitario $\widehat{U}(t,t_0)$ como se muestra:

$$|\psi(t)\rangle = \widehat{U}(t,t_0) \cdot |\psi(t_0)\rangle$$

$$\widehat{U}(t,t_0)^\dagger \widehat{U}(t,t_0) = \widehat{1} \qquad \widehat{U}(t,t_0)^\dagger = \widehat{U}(t,t_0)^{-1} \qquad [2.2]$$

Este operador va a cumplir con la ecuación de Schrödinger al reemplazar [2.2] en [2.1] (al tiempo inicial $t_0$, $\widehat{U}$ es operador de identidad):

$$i\hbar \frac{d}{dt}\widehat{U}(t,t_0) = \widehat{H} \cdot \widehat{U}(t,t_0) \qquad \widehat{U}(t_0,t_0) = \widehat{1} \qquad [2.3]$$

La ecuación [2.3] puede ser resuelta de forma general sólo si el Hamiltoniano es independiente del tiempo, entonces adquiere la forma siguiente:

---

[11] Son transformaciones que preservan el módulo del ket de estado; además, cualquier operación determinista sobre un microsistema se expresa con transformaciones unitarias.

$$\widehat{U}(t,t_0) = e^{-\frac{i}{\hbar}(t-t_0)\widehat{H}} \qquad [2.4]$$

El operador de evolución es reversible porque para cualquier estado posterior en el tiempo siempre existe otro operador de evolución que lo devuelva a su estado inicial, que es el operador de evolución inverso $\widehat{U}(t,t_0)^{-1} = \widehat{U}(t_0,t)$; [2.5] es la condición necesaria y suficiente para que un operador $\widehat{U}(t,t_0)$ sea unitario:

$$\widehat{U}(t,t_0)^\dagger \widehat{U}(t,t_0) = \widehat{1} \qquad \widehat{U}(t,t_0)^\dagger = \widehat{U}(t,t_0)^{-1} \qquad [2.5]$$

La forma explícita de la ecuación de Schrödinger se obtiene de multiplicar el bra $\langle \vec{r}|$ con la ecuación [2.1] en notación ket, como se muestra:

$$i\hbar \frac{d}{dt}\psi(\vec{r},t) = \left[-\frac{\hbar^2}{2m}\nabla^2 + V(\vec{r},t)\right]\psi(\vec{r},t) \qquad [2.6]$$

### 2.1.2. El colapso o proceso de reducción (proceso R)

La antítesis de la evolución unitaria es el colapso – o proceso R– que es una discontinuidad en la evolución unitaria, un salto cuántico instantáneo, irreversible e indeterminista de un estado del microsistema $|\psi\rangle$ a cualquier otro estado $|\varphi\rangle$, lo único que lo caracteriza es la probabilidad de ocurrencia[12]:

$$P[\psi \to \varphi] = |\langle\varphi|\psi\rangle|^2 \qquad [2.7]$$

Sólo la probabilidad es invariante ante la inversión del proceso R:

$$P[\varphi \to \psi] = |\langle\psi|\varphi\rangle|^2 = |\langle\varphi|\psi\rangle|^2 = P[\psi \to \varphi] \qquad [2.8]$$

No obstante, el proceso R es irreversible, si se pretende usar transformaciones unitarias para regresar al estado inicial:

$$\nexists\ \widehat{U}\ /\ \widehat{U}\cdot|\varphi\rangle = |\psi\rangle \qquad [2.9]$$

---

[12] Esta probabilidad está dada por la regla de Born, y debe ser no nula para acontecer.

El indeterminismo del proceso R no permite conocer o predecir el estado al que colapsará (sólo se puede conocer las probabilidades), y tampoco el resultado de una medida desde que el acto de medición es un proceso R y no un proceso U; el proceso U es la evolución de la probabilidad del proceso R en el tiempo.

**2.1.3. Estados enredados y separables debidos a procesos U y R**

El proceso R (a diferencia del proceso U) puede crear y destruir estados, cuando se trata de dos o más microsistemas se pueden obtener estados entrelazados o separables; dos microsistemas 1 y 2 tienen estados separables si el estado general $|\psi(1,2)\rangle$ se puede expresar como el producto tensorial de dos estados $|\psi(1)\rangle$ y $|\psi(2)\rangle$ que sólo dependen de los microsistemas 1 y 2 respectivamente; sea la base común a ambos $\{|u_k\rangle\}, k = 1, 2, ...$ autoestados del observable $\hat{A}$, de manera que:

$$|\psi(1)\rangle = \sum_k \lambda_k^1 |u_k\rangle \qquad |\psi(2)\rangle = \sum_k \lambda_k^2 |u_k\rangle$$

Entonces los estados separables cumplen las ecuaciones [2.10]:

$$|\psi(1,2)\rangle = |\psi(1)\rangle \otimes |\psi(2)\rangle \qquad \begin{array}{c} |\psi(1,2)\rangle = \sum_{k,k'} C_{k,k'} |u_k\rangle |u_{k'}\rangle \\ \\ C_{k,k'} = \lambda_k^1 \lambda_{k'}^2 \end{array} \qquad [2.10]$$

El estado general $|\psi(1,2)\rangle$ es entrelazado si los estados para 1 y 2 no pueden ser factorizados como en [2.10]; es decir, si no se cumple $C_{k,k'} = \lambda_k^1 \lambda_{k'}^2$; en [2.10] la evolución unitaria va a preservar la factorización del estado general, sin embargo un colapso puede crear estados entrelazados a partir de estados separables si se mide en otro observable $\hat{B}$ para todo el sistema general: sean dos electrones con estados separables: $|1\rangle = (\uparrow + \downarrow)/\sqrt{2}$ y $|2\rangle = (\uparrow + \downarrow)/\sqrt{2}$ donde $\uparrow$ y $\downarrow$ representan los estados de espín "arriba" y "abajo", el estado general viene dado por $|\psi\rangle = |1\rangle|2\rangle = (\uparrow\uparrow + \uparrow\downarrow + \downarrow\uparrow + \downarrow\downarrow)/2$ donde $\hat{A} = \hat{S}_z$, sea el operador $\hat{B} = \hat{J}_z$ que actúa sobre el sistema general $|\psi\rangle$, cuyos autoestados para este caso son los tripletes:

$$|\psi\rangle = (|1,1\rangle + \sqrt{2}|1,0\rangle + |1,-1\rangle)/2 \qquad \begin{aligned} |1,1\rangle &= \uparrow\uparrow \\ |1,0\rangle &= (\uparrow\downarrow + \downarrow\uparrow)/\sqrt{2} \\ |1,-1\rangle &= \downarrow\downarrow \end{aligned} \qquad [2.11]$$

Si al medir $\hat{J}_z$ (la componente vertical del momento angular total) en $|\psi\rangle$ (factorizable) colapsa al estado $|1,0\rangle$ (no factorizable), entonces la medida ha creado un estado entrelazado (la probabilidad de crear este estado enredado es 0.5, de lo contrario se crean estados separables con ambos espines para arriba $|1,1\rangle$ o para abajo $|1,-1\rangle$), pero no se puede obtener un estado singlete: $\langle 0,0|\psi\rangle = 0$; de igual manera se puede romper el entrelazamiento mediante la observación, si se mide $\hat{S}_z$ en $|1,0\rangle$; en estos casos, la evolución unitaria sólo preserva la factorización o el entrelazamiento, es posible que estados separables evolucionen unitariamente a estados entrelazados, como cuando incide un fotón en un átomo para inducir la absorción o emisión (sección [2.1.7]), mediante la teoría de perturbaciones dependientes del tiempo (sección [2.1.6]), pero debido a la reversibilidad del proceso U deben volver al estado separable como al inicio (una vez que el fotón atraviesa al átomo y se aleja de él, desaparece el entrelazamiento).

### 2.1.4. El operador de densidad

Al nivel estadístico no se trabaja con el estado cuántico de todo el ensemble[13], se emplea en cambio la distribución de probabilidad de que un microsistema se encuentre en un estado cuántico determinado $|\varphi_i\rangle$, $i = 1, 2, 3, \dots$, pero no en una superposición cuántica de ellos, representado por el operador de densidad $\hat{\rho}$:

$$\hat{\rho} = \sum_i p_i |\varphi_i\rangle\langle\varphi_i| \qquad [2.12]$$

Donde $p_i$ es la probabilidad de cualquier microsistema elegido al azar de encontrarse en el estado $|\varphi_i\rangle$ (antes de hacer una medida), cada producto $|\varphi_i\rangle\langle\varphi_i|$ representa un estado puro; el operador $\hat{\rho}$ no es una superposición de estados sino una mezcla estadística, y en consecuencia, aplicado sólo a ensembles de muchas partículas; una población de microsistemas, cada uno en el estado $|\psi\rangle = \sum_i \lambda_i |\varphi_i\rangle$ se encuentra en el estado puro siguiente:

$$|\psi\rangle\langle\psi| = \sum_{i,j} \lambda_i \lambda_j^* |\varphi_i\rangle\langle\varphi_i| \qquad [2.13]$$

---

[13] Debido a la gran cantidad de información cuántica que esto supone, de igual manera que en el caso clásico donde la información es trabajada sobre promedios y variables estadísticas.

Donde los términos $\lambda_i \lambda_j^*$ son reales si $i = j$ y complejos si $i \neq j$, estos últimos constituyen los *términos de interferencia* y exhiben la coherencia cuántica del estado puro, cuando estos términos desaparecen se obtiene una forma equivalente a [2.12], y se considera un estado mezcla (en un ensemble) o una superposición incoherente (en el marco de la decoherencia); la traza de $\hat{\rho}$ es la suma de las probabilidades e igual a la unidad:

$$Tr(\hat{\rho}) = \sum_i p_i = 1 \qquad [2.14]$$

El valor medio de un operador $\hat{A}$ en este ensemble (el valor esperado de una cantidad física) es la traza del producto del operador $\hat{A}$ y el operador de densidad:

$$\langle\hat{A}\rangle = Tr(\hat{A} \cdot \hat{\rho}) = Tr(\hat{\rho} \cdot \hat{A}) = \sum_i p_i A_{ii} \qquad [2.15]$$

Donde $A_{ij} = \langle\varphi_i|\hat{A}|\varphi_j\rangle$, la ecuación [2.15] interpreta el valor esperado $\langle\hat{A}\rangle$ como el promedio ponderado del observable $\hat{A}$ en cada uno de los estados puros $|\varphi_i\rangle\langle\varphi_i|$, también permite obtener [2.14] haciendo $\hat{A} = \hat{1}$ (el operador de identidad); un operador de mucho interés es la entropía de Von Neumann (la entropía por microsistema), que resulta de calcular $\hat{A} = -\ln(\hat{\rho})$ usando la ecuación [2.15]:

$$S_N(\hat{\rho}) = -Tr[\hat{\rho} \ln(\hat{\rho})] \qquad [2.16]$$

Puesto que $S_N(\hat{\rho})$ no contiene la constante de Boltzman $k_B$ es más apropiado interpretarlo como la entropía de información de Shannon, la entropía estadística usual viene dada simplemente por: $S = N k_B S_N(\hat{\rho})$, si el operador de densidad es un estado puro la entropía se anula ($S_N(\hat{\rho}) = 0$).

La evolución del operador de densidad está dada por la ecuación cuántica de Liouville o la ecuación de Von Neumann, en la imagen de Schrödinger:

$$i\hbar \frac{d}{dt}\hat{\rho} = [\hat{H}, \hat{\rho}] \qquad [2.17]$$

Considerando las ecuaciones [2.2] y [2.3], se tiene la solución general:

$$\hat{\rho}(t) = \hat{U}(t, t_0)\hat{\rho}(t_0)\hat{U}(t, t_0)^{-1} \qquad [2.18]$$

Si el Hamiltoniano no depende del tiempo, $\hat{U}(t, t_0)$ cumple [2.4] y dejan los autovalores $e^{-\frac{i}{\hbar}(t-t_0)E_k}$ para cada $|\varphi_k\rangle\langle\varphi_k|$ de $\hat{\rho}(t_0)$ en [2.18], luego los autovalores de $\hat{U}$ y $\hat{U}^{-1}$ se cancelan y se conservan $\hat{\rho}$ y su entropía:

$$\hat{\rho}(t) = \hat{\rho}(t_0) \qquad [\hat{H}, \hat{\rho}] = 0 \qquad \frac{d}{dt}S_N(\hat{\rho}) = 0 \qquad [2.19]$$

### 2.1.5. Perturbación dependiente del tiempo y la regla de oro de Fermi

Es posible tratar un Hamiltoniano dependiente del tiempo usando "pequeñas perturbaciones" al microsistema; sea el Hamiltoniano que no depende del tiempo $\hat{H}$ con los autoestados de energía $|n\rangle$, se cumple la ecuación de autovalores:

$$\hat{H}|n\rangle = E_n|n\rangle \qquad [2.20]$$

Antes de la perturbación (inicia al tiempo $t_0$), el microsistema se encuentra en un estado inicial de energía $|i\rangle$, su estado en el tiempo viene dado por el ket:

$$|\psi_i^0(t)\rangle = e^{-iE_i t/\hbar}|i\rangle, \qquad t \leq t_0 \qquad [2.21]$$

La perturbación introduce el Hamiltoniano de interacción dependiente del tiempo $\hat{H}^I(t)$, y el nuevo estado desconocido del microsistema $|\psi(t)\rangle$ se proyecta en la base del Hamiltoniano anterior (sin perturbar) ya conocido, con la introducción de unas amplitudes normalizadas dependientes del tiempo $a_n(t)$:

$$|\psi(t)\rangle = \sum_n a_n(t)\, e^{-iE_n t/\hbar}|n\rangle \qquad \sum_n |a_n(t)|^2 = 1 \qquad [2.22]$$

La ecuación [2.22] establece que una perturbación externa (como la interacción con una partícula portando energía) lleva a una superposición de distintos autoestados de energía; la ecuación de Schrödinger para el nuevo estado $|\psi(t)\rangle$ es:

$$i\hbar \frac{d}{dt}|\psi(t)\rangle = \left(\widehat{H} + \widehat{H}^I(t)\right)|\psi(t)\rangle \qquad [2.23]$$

al reemplazar la ecuación [2.22] en [2.23], considerando [2.1.5.1], y proyectando en el bra $\langle f|$, se tiene la derivada temporal de $a_f(t)$, ($j^2 = -1$):

$$\frac{d}{dt}a_f(t) = \frac{1}{j\hbar}\sum_n a_n(t)\langle f|\widehat{H}^I(t)|n\rangle e^{j\omega_{fn}t} \qquad \begin{array}{c} t \in [t_0, t'] \\ \\ a_n(t_0) = \delta_{in} \end{array} \qquad [2.24]$$

Donde $\omega_{fn} = (E_f - E_n)/\hbar$, el sistema de ecuaciones [2.24] es conocida como ***representación de interacción*** (es exacta a la ecuación de Schrödinger, sin aproximación alguna); las condiciones iniciales $a_n(t_0) = \delta_{in}$ son debido a la continuidad con la ecuación [2.21] al tiempo $t = t_0$; si el término de acoplamiento $\langle f|\widehat{H}^I(t)|n\rangle \sim 0$ entonces el coeficiente $a_f(t)$ tiende a ser constante; la interacción con un Hamiltoniano $\widehat{H}^I$ del que $|n\rangle$ no sea vector propio permite un acoplo no nulo[14]; en la teoría de perturbaciones se considera un pequeño acoplo comparado con los autovalores $E_n$:

$$\left|\langle f|\widehat{H}^I(t)|n\rangle\right| \ll E_n, E_f \qquad [2.25]$$

La condición [2.25] permite, en una *primera aproximación*, anular la derivada temporal de las amplitudes $\dot{a}_f(t) \sim 0$, luego [2.24] viene a ser[15]:

$$a_f(t') \sim \delta_{if} + \frac{1}{j\hbar}\int_{t_0}^{t'} dt \,\langle f|\widehat{H}^I(t)|i\rangle e^{j\omega_{fi}t} \qquad [2.26]$$

---

[14] Porque la acción del Hamiltoniano $\widehat{H}$ en el ket propio $|n\rangle$ devuelve un ket $E_n|n\rangle$ que es paralelo a $|n\rangle$, en cambio la acción de un Hamiltoniano distinto $\widehat{H}^I$ en el ket $|n\rangle$ dará un vector $\widehat{H}^I|n\rangle = |\widehat{H}^I n\rangle$ que en general no sea paralelo a $|n\rangle$, y tampoco ortogonal a $\langle f|$.

[15] Primero se integra ambos lados de [2.24], luego se considera $a_n(t) \sim a_n(t_0) = \delta_{in}$ lo que anula la sumatoria, y finalmente se considera $\int \dot{a}_f dt = a_f(t') - a_f(t_0) = a_f(t') - \delta_{if}$

Las amplitudes $a_f(t')$ permite el cálculo de la probabilidad de transición[16]; para obtener la resonancia la interacción perturbativa debe ser armónica:

$$\widehat{H}^I(t) = 2\widehat{W}\cos(\omega t) = \widehat{W}\left(e^{i\omega t} + e^{-i\omega t}\right) \quad [2.27]$$

Donde $\widehat{W}$ no depende del tiempo, [2.27] en [2.26] se tiene:

$$a_f(t') \sim \delta_{if} + \frac{\langle f|\widehat{W}|i\rangle}{j\hbar}\left[\frac{e^{j(\omega_{fi}+\omega)\Delta t} - 1}{j(\omega_{fi}+\omega)} + \frac{e^{j(\omega_{fi}-\omega)\Delta t} - 1}{j(\omega_{fi}-\omega)}\right] \quad [2.28]$$

Donde $\Delta t = t' - t_0$, esta ecuación muestra un comportamiento resonante cuando la frecuencia de la perturbación armónica $\omega$ se aproxima a $\omega_{fi}$ ó $-\omega_{fi}$ y se tiende anular conforme se aleja de estos valores[17], como se muestra en la Figura N° 2.1:

FIGURA N° 2.1

RESONANCIA DE LA AMPLITUD DE TRANSICIÓN

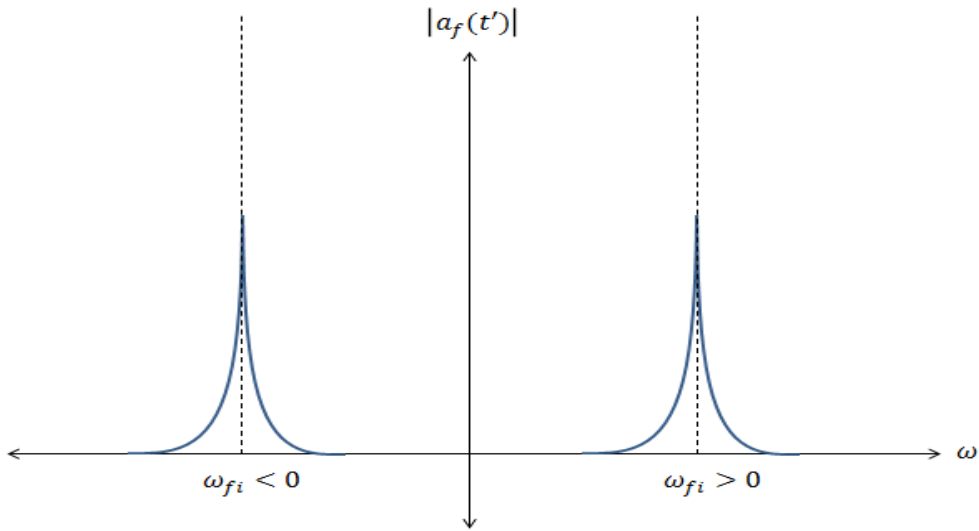

Fuente: Elaboración propia.

---

[16] En este caso el espectro de energías del Hamiltoniano no perturbado es discreto, si el espectro de energía es continuo entonces se debe tratar de una densidad de probabilidad de transición.

[17] Debido a las altas frecuencias $\omega \sim 10^6$ o $10^{15} Hz$ cada término $\omega_{fi} \pm \omega$ se hace muy grande comparado con la unidad y así los denominadores rebasan a los numeradores en [2.28], anulándose; por otro lado al aproximarse entre sí las frecuencias angulares entonces uno de los términos $\omega_{fi} \pm \omega = 0$ se anula y se tendría una indeterminación de tipo $0/0$ cuyo límite converge a $\Delta t$ que puede ser del orden del periodo de la perturbación armónica.

En la Figura N° 2.1 están representados la absorción ($\omega_{fi} > 0$) y la emisión estimulada[18] ($\omega_{fi} < 0$), la ecuación [2.28] es invariante a $\omega \to -\omega$, así que la absorción y emisión tienen la misma probabilidad de ocurrencia; la ecuación [2.28] puede ser aproximada de la siguiente manera[19]:

$$a_f(t') \sim \delta_{if} + \Delta t \frac{\langle f|\widehat{W}|i\rangle}{j\hbar} e^{j\theta} sinc(\theta) \qquad \theta = \frac{(\omega - |\omega_{fi}|)\Delta t}{2} \qquad [2.29]$$

Sea un microsistema perturbado por una radiación con una densidad de probabilidad de energía $\rho(E) = dP/dE$, la probabilidad de transición por unidad de tiempo del estado inicial a otro final distinto ($\delta_{if} = 0$) viene dado por:

$$w_{i \to f} = \frac{1}{\Delta t} \int_0^\infty |a_f(t')|^2 \rho(E) dE \qquad [2.30]$$

[2.29] en [2.30], y considerando que $dE = \hbar d\omega$ y $d\theta = d\omega \Delta t/2$:

$$w_{i \to f} = \frac{2}{\hbar} \int_{-\frac{|\omega_{fi}|\Delta t}{2}}^\infty |\langle f|\widehat{W}|i\rangle|^2 sinc^2(\theta) \rho(E) d\theta \qquad [2.31]$$

Tomando en cuenta unas consideraciones[20], la ecuación [2.31] toma la forma de la **regla de oro de Fermi**:

$$w_{i \to f} = \frac{2\pi}{\hbar} |\langle f|\widehat{W}|i\rangle|^2 \rho(E) \qquad [2.32]$$

---

[18] Claramente la emisión asociada a $\omega_{fi} < 0$ es estimulada por la existencia de una frecuencia $\omega$ externa al microsistema (inicialmente sin perturbar), esta frecuencia correspondiente a la perturbación puede ser interpretado como la de un fotón incidente con la energía igual a la diferencia entre los dos niveles.

[19] Si se emplea el artificio: $\frac{e^{j(\omega_{fi} \pm \omega)\Delta t} - 1}{j(\omega_{fi} \pm \omega)} = \Delta t e^{\frac{j(\omega_{fi} \pm \omega)\Delta t}{2}} sinc\left(\frac{(\omega_{fi} \pm \omega)\Delta t}{2}\right)$

[20] Por un lado desde que $\theta = -|\omega_{fi}|\Delta t/2$ supone un valor donde la función $sinc(\theta)$ se anula, se puede aproximar al límite $\theta \to -\infty$, y por otro lado que $sinc(\theta)$ se le considera una *función delta naciente* y mantiene aproximadamente constante el término de acoplamiento y la densidad de energía (en la proximidad en que ocurre la resonancia) de manera que ambos términos pueden salir de la integración, con todo esto sólo quedaría una integral tipo *Dirichlet*.

**2.1.6. Absorción y emisión estimulada y espontánea**

La emisión (estimulada o espontanea) de una partícula ocurre portando energía como un fotón (luminoso, si es emisión radiativa) o un fonón (térmico, si es emisión no radiativa), el proceso inverso es la absorción, la diferencia entre la emisión estimulada y espontánea se encuentra en el origen de la radiación estimulante: en ambos casos es necesario de una perturbación, en la emisión estimulada un fotón (del baño térmico o de una fuente radiativa) perturba al microsistema, en la emisión espontanea el fotón procede de las fluctuaciones cuánticas del campo electromagnético en el vacío[21] con la energía $\hbar(\omega_2 - \omega_1)$, con dirección y fase aleatoria, e induce su decaimiento para después desaparecer.

De acuerdo a la teoría de las perturbaciones dependientes del tiempo (sección [2.1.5]), un microsistema en un estado excitado $|2\rangle$ no puede decaer al estado fundamental $|1\rangle$ puesto que $\langle 1|2\rangle = 0$, sólo al interactuar con un fotón con la energía apropiada puede entrar en resonancia[22] y evolucionar unitariamente al estado $|\psi\rangle = \lambda_1(t)|1\rangle + \lambda_2(t)|2\rangle$ y tras el proceso R decaer a $|1\rangle$ y emita un fotón[23] con probabilidad $|\lambda_1(t)|^2$; la absorción es el proceso inverso: un microsistema en el estado de energía menor $|1\rangle$ al ser perturbado por un fotón pasa al estado $|\psi'\rangle = \lambda_2(t)|1\rangle + \lambda_1(t)|2\rangle$ puesto que las probabilidades de emisión y absorción $|\langle 1|\psi\rangle|^2$ y $|\langle 2|\psi'\rangle|^2$ son las mismas; así por ejemplo, un ensemble compuesto de $N$ microsistemas de este tipo forman un cuerpo negro de dos niveles a temperatura $T$, con el operador de densidad dado por:

$$\hat{\rho} = \left(e^{-\hbar\omega_1/kT}|1\rangle\langle 1| + e^{-\hbar\omega_2/kT}|2\rangle\langle 2|\right)/Z \qquad [2.33]$$

Donde $Z = e^{-\hbar\omega_1/kT} + e^{-\hbar\omega_2/kT}$ es la función de partición; los $N_1$ microsistemas, en el estado $|1\rangle$, al interactuar con los fotones del baño termal pasan al estado $|\psi'\rangle$ y de ahí una fracción $B_{12}N_1\rho(\Delta E)$ colapsan al estado $|2\rangle$ ocurriendo la absorción, $\rho(\Delta E)$ es la densidad de fotones con energía $\Delta E = \hbar(\omega_2 - \omega_1)$, de igual

---

[21] En la electrodinámica cuántica los fotones son creados y destruidos en el vacío cuántico, con un tiempo de vida inversamente proporcional a la indeterminación en la energía.
[22] La interacción de un fotón con un electrón es posible mediante el campo eléctrico (oscilante).
[23] Esta emisión es estimulada, y el fotón emitido tiene la misma fase y dirección del fotón incidente.

manera $B_{21}N_2\rho(\Delta E)$ lo es para los microsistemas que realizan la emisión estimulada, y $AN_2$ los que hacen emisión espontánea, los coeficientes $B_{12}$, $B_{21}$ y $A$ son llamados *coeficientes de Einstein*, tal como se muestra en la Figura N° 2.2:

FIGURA N° 2.2

ABSORCIÓN Y EMISIÓN

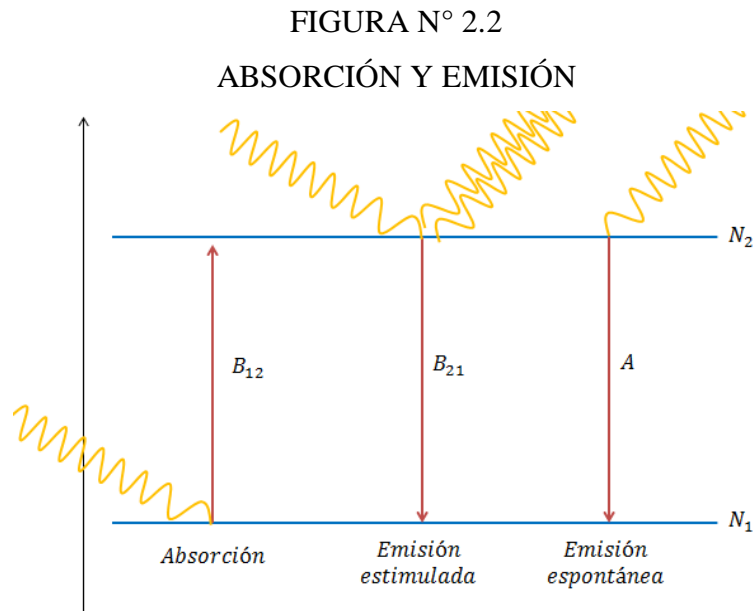

Fuente: Elaboración propia.

$B_{12}$ y $B_{21}$ son iguales, pues la probabilidad es la misma para ambos procesos como se deriva de la ecuación [2.28]; este ejemplo muestra cómo la combinación de los procesos U y R al nivel microscópico son necesarios para sostener la dinámica térmica del mundo macroscópico, pues el intercambio de energía es fundamental para todos los procesos del mundo clásico.

**2.2. Análisis sobre los argumentos del *límite clásico* y modelos de colapso objetivo precedentes**

**2.2.1. El principio de correspondencia de Bohr**

El primer intento por reconocer y abordar el problema de la clasicalización se encuentra en el principio de correspondencia, los *saltos cuánticos* en el modelo atómico de Bohr introducen discretización del espectro de energía que requería una transición hacia el mundo clásico "suave y continuo"; así, cuando un microsistema adquiere altos valores en sus números cuánticos pasa a comportar y reproducir los resultados de la física clásica, el límite entre ambas teorías se da de forma transitoria:

*"Este principio enunciado en 1923 por Bohr consiste de dos partes: 1. Las predicciones de la teoría cuántica para el comportamiento de cualquier sistema físico deberán corresponder a las predicciones de la física clásica en el límite en el cual los números cuánticos que especifican el estado del sistema se hacen muy grandes"* (Eisberg & Resnick, 2000, pág. 149).

En la actualidad es insatisfactorio o incompleto este principio de la mecánica cuántica, puesto que la diferencia entre ambas teorías no es sólo la cuantización del espectro de los observables, sino que la mecánica cuántica tiene evolución unitaria-reversible y se encuentran en superposición de estados, contrario al mundo clásico, lo cual no puede conciliarse haciendo sólo muy grandes a los números cuánticos.

### 2.2.2. El teorema de Ehrenfest

La ecuación dada por Paul Ehrenfest describe la evolución temporal del valor esperado de un observable $\hat{A}$, el cual para los observables de posición $\vec{r}$ y momento $\vec{p}$ exhiben un comportamiento consistente con la mecánica clásica:

$$\frac{d}{dt}\langle \hat{A}\rangle = \frac{1}{i\hbar}\langle [\hat{A},\hat{H}]\rangle + \langle \frac{\partial \hat{A}}{\partial t}\rangle \qquad [2.34]$$

$$m\frac{d}{dt}\langle \vec{r}\rangle = \langle \vec{p}\rangle \qquad \frac{d}{dt}\langle \vec{p}\rangle = -\langle \vec{\nabla} V(\vec{r})\rangle \qquad [2.35]$$

Estas ecuaciones son obtenidas de la ecuación de Schrödinger [2.1], de modo que la dinámica que describe es una consecuencia natural de la unitaridad.

*"Estas dos ecuaciones expresan el teorema de Ehrenfest. Su forma recuerda las clásicas ecuaciones de Hamilton-Jacobi para una partícula"* (Cohen-Tannoudji, Diu, & Laloe, 1997, pág. 242).

Si en la ecuación [2.35] al pasar a la escala macroscópica las indeterminaciones en los observables de posición y momento $\sigma(\hat{r})$ y $\sigma(\hat{p})$ se hacen despreciables entonces la partícula es descrita como localizada y con una trayectoria representada por $\langle \hat{r}\rangle$ y $\langle \hat{p}\rangle$ (los valores medios de sus observables de posición y momento), su dinámica seguiría las leyes de newton siempre que $\langle \vec{\nabla} V(\hat{r})\rangle$ (el valor esperado del gradiente del potencial) se aproxime al gradiente del potencial evaluado en $\langle \hat{r} \rangle$[24] (el valor esperado de la posición), pero esta es una situación muy limitada pues las indeterminaciones (como en la posición) se pueden expandir a niveles macroscópicos: por ejemplo, si la partícula impacta en diagonal a una barrera escalón (con su energía cinética siempre positiva) entonces se tendría una superposición coherente de "dos partículas", una atravesando con refracción y otra reflejándose, ambas funciones de onda alejándose una de otra, con los valores medios $\langle \hat{r} \rangle$ y $\langle \hat{p} \rangle$ en una "virtual trayectoria media" entre ambos paquetes de ondas (la amplitud evaluada en $\langle \hat{r} \rangle$ sería muy despreciable y no se encontraría a la partícula al medir en $\langle \hat{r} \rangle$), perdiendo así su significado físico.

No siempre las indeterminaciones serán despreciables en la escala macroscópica, en el experimento de la doble rendija un fotón puede deslocalizarse a escala macroscópica o una partícula libre como un paquete de ondas (incluso en reposo) se expandirá y podrá adoptar deslocalización a escala clásica aunque los valores medios de sus observables sigan cumpliendo la ecuación de Ehrenfest; nuevamente, se encuentra con el problema de la superposición de estados para los objetos del mundo clásico, que no se puede solucionar con la complementariedad.

### 2.2.3. La primera teoría de colapso objetivo GRW y su variante CSL

La teoría GRW (Ghigrardi, Rimini, Weber) postula al colapso como un evento espontaneo en todos los microsistemas, cada partícula elemental sufre un colapso aleatorio en el tiempo que sigue una distribución de Poisson, con una frecuencia $\lambda$ extremadamente escasa, en los objetos del mundo macroscópicos los colapsos conducen a un comportamiento estadísticamente regular que suprime las superposiciones en el mundo clásico. Para conseguir consistencia con la mecánica cuántica (donde el colapso sólo ocurre al realizar una medida), los autores sugieren que la tasa de colapso espontáneo para una partícula individual es del orden de una vez cada cien millones de años:

---

[24] Esto es: $\langle \vec{\nabla} V(\hat{r}) \rangle \sim \vec{\nabla} V(\langle \hat{r} \rangle)$ lo cual se puede obtener al nivel clásico si la función de onda luce lo suficientemente localizada (con incertidumbres en posición y momentos despreciables) como una delta de Dirac, de manera que el potencial prácticamente es constante a dimensiones microscópicas donde la función de onda es no-nula.

> *"Para la frecuencia de localización de sistemas microscópicos nosotros elegimos:*
>
> $$\lambda_{micro} \simeq 10^{-16} sec^{-1}$$
>
> *Esto significa que un sistema es localizado una vez cada $10^8 - 10^9$ años. (…) En lo que concierne a objetos macroscópicos (conteniendo un número de constituyentes del orden del número de Avogadro) (…) obtenemos como frecuencia de prueba característica: $\lambda_{macro} \simeq 10^7 sec^{-1}$"* (Ghirardi, Rimini, & Weber, 1986, pág. 480).

Una variante de esta teoría es el modelo de *localización espontanea continua* CSL, con la diferencia que CLS reemplaza el colapso abrupto de GRW, por una proceso suave y continuo. La evolución estocástica de CLS viene dado por:

$$\frac{d}{dt}|\psi_\omega(t)\rangle = \left[-\frac{i}{\hbar}\hat{H} + \hat{A}\omega(t) - \lambda\hat{A}^2\right]|\psi_\omega(t)\rangle \qquad [2.36]$$

Con $\lambda$ constante, $\hat{A}$ un operador hermitiano cuyos autoestados son los estados localizados para microsistema (a los que "colapsará") y $\omega(t)$ un proceso estocástico complejo (Okon & Sudarsky, 2014, pág. 4).

Es evidente en la ecuación [2.36] que el término $-\lambda\hat{A}^2$ le dará un carácter irreversible y continuo en el tiempo, siguiendo una distribución de Poisson; en cuanto a principios e interpretaciones, luce un programa muy prometedor:

> ***"Continuando citando a Bell: si la teoría se va a aplicar a cualquier cosa que no sean sólo operaciones de laboratorio altamente idealizadas, ¿no estamos obligados a admitir que procesos más o menos "como de medida" ocurren más o menos todo el tiempo, más o menos en todas partes? ¿No hemos saltado todo el tiempo?*** *La idea básica detrás del programa de reducción dinámica es precisamente esto: colapsos espontáneos y aleatorios de la función de onda ocurren todo el tiempo, a todas las partículas, ya sea aisladas o interactuando, ya sea que formen solo un pequeño átomo o un gran dispositivo de medición. Por supuesto, tales colapsos deben ser raros y suaves para los sistemas microscópicos, a fin de no alterar su comportamiento cuántico como lo predice la ecuación de Schrödinger. Al mismo tiempo, su efecto debe* *sumarse de tal manera que, cuando miles de millones de partículas se pegan entre sí para formar un sistema macroscópico, un solo colapso que se produce en una de las partículas afecte al sistema global. Entonces tenemos que miles de millones de tales colapsos actúan con mucha frecuencia en el macro-sistema, que juntos fuerzan a su función de onda a estar muy bien localizada en el espacio"* (Bassi, 2007, pág. 2).

**2.2.4. Interpretación de Roger Penrose**

Sir. Roger Penrose propone, en su artículo *Sobre el rol de la gravedad en la reducción del estado cuántico (1996)*, que al considerar la gravitación en la superposición de estados, conduce al colapso o reducción del estado cuántico de manera espontánea, una partícula deslocalizada genera un campo gravitacional (aunque sea muy débil) al que se le asocia una auto-energía potencial gravitatoria mayor que si estuviese localizada[25], en consecuencia los estados localizados son menos energéticos y más estables, el colapso ocurre porque el sistema tiende a ir de un estado deslocalizado-inestable a un estado localizado-estable.

Si el microsistema se encuentra en superposición de dos estados con posiciones más o menos localizadas $|1\rangle$ y $|2\rangle$, va decaer espontáneamente en un tiempo de vida característico $\tau$ en uno u otro de los dos estados, de manera inversamente proporcional a la *auto-energía gravitacional* $E_\Delta$ (que va a depender de la diferencia de masas asociado a los dos estados tomando una como positiva y la otra como negativa):

$$\tau \approx \frac{\hbar}{E_\Delta} \qquad [2.37]$$

No obstante, la propuesta de Penrose es considerada "minimalista" porque si bien ataca el problema crucial sobre la causa del colapso en la función de onda, no hace un desarrollo completo de su teoría, como el mismo lo señala:

---

[25] La energía gravitacional por ser atractiva es un potencial tipo "pozo", y en consecuencia las partículas que estén más dispersas tienen más energía gravitatoria que aquellas que están más juntas o unidas.

*"Sin embargo, debe quedar claro que esta propuesta no proporciona una teoría de la reducción del estado cuántico. Simplemente indica el nivel en el que se esperan las desviaciones de la estándar evolución lineal de Schrödinger (unitaria) debido a los efectos gravitacionales (…) la teoría correcta que une la relatividad general con la mecánica cuántica implicará un cambio importante en nuestra visión física del mundo de una magnitud al menos comparable con la implicada en el cambio de la física gravitacional de Newton a la de Einstein"* (Penrose, 1995, pág. 4).

Una crítica que se hace a esta propuesta es que sería una pista falsa sobre el origen del colapso: si el colapso ocurre por la transición a un estado gravitacional más estable, entonces podría no ser muy distinto a la emisión espontánea (sección [2.1.6]) que requiere del colapso, pero no lo explica realmente.

### 2.3. Mecánica clásica no relativista

#### 2.3.1. Simetría temporal de la mecánica clásica

La simetría temporal en la mecánica clásica se deriva de las ecuaciones canónicas de Hamilton; sea un sistema físico descrito por $2N$ variables de posición $q_i$ y momento lineal $p_i$ con el Hamiltoniano $H(q_i, p_i, t)$, la evolución temporal está dada por las ecuaciones canónicas de Hamilton:

$$\dot{q}_i = \frac{\partial H}{\partial p_i} \qquad \dot{p}_i = -\frac{\partial H}{\partial q_i} \qquad [2.38]$$

Las ecuaciones [2.38] son deterministas porque permiten conocer cada punto $(q_i, p_i)_{i=1}^{N}$ en cualquier tiempo tras conocer el estado inicial a un tiempo dado, estas ecuaciones [2.38] son simétricas en el tiempo, sus soluciones son reversibles, y no se puede distinguir pasado de futuro en ellas; al hacer una inversión temporal, de $t$ por $-t$ se obtienen:

$$q_i(-t) = q_i(t) \qquad p_i(-t) = -p_i(t) \qquad H(-t) = H(t) \qquad [2.39]$$

$$\dot{q}_i(-t) = \frac{dq_i(-t)}{d(-t)} = -\dot{q}_i(t) \qquad \dot{p}_i(-t) = \frac{dp_i(-t)}{d(-t)} = \dot{p}_i(t) \qquad [2.40]$$

Las ecuaciones [2.39] y [2.40] son intuitivas: la coordenada de posición $q_i$ no cambia por una inversión temporal como si lo hace la velocidad $\dot{q}_i$ o el momento lineal $p_i$, de igual manera el Hamiltoniano es invariante a la inversión del tiempo (basta con calcular la inversión de la energía cinética y potencial), las ecuaciones [2.39] y [2.40] reemplazadas en [2.38] devuelven las mismas ecuaciones para el tiempo invertido:

$$\dot{q}_i(-t) = \frac{\partial H(-t)}{\partial p_i(-t)} \qquad \dot{p}_i(-t) = -\frac{\partial H(-t)}{\partial q_i(-t)} \qquad [2.41]$$

Las ecuaciones [2.38] y [2.41] dicen que las leyes de la mecánica clásica funcionan igual si el tiempo va de pasado a futuro que de futuro a pasado, no contiene dentro de sí semilla alguna para una "flecha del tiempo" ya sea al nivel estadístico o caótico.

### 2.3.2. Teorema de Liouville

El Teorema de Liouville establece que una región $R$ dentro del espacio de fases $X$, como un conjunto de condiciones iniciales, evoluciona en el tiempo (moviéndose dentro del espacio de fases) conservando siempre su hipervolumen; esto es, evoluciona como un flujo incompresible.

FIGURA N° 2.3

EVOLUCIÓN DEL HIPERVOLUMEN EN EL ESPACIO DE FASES

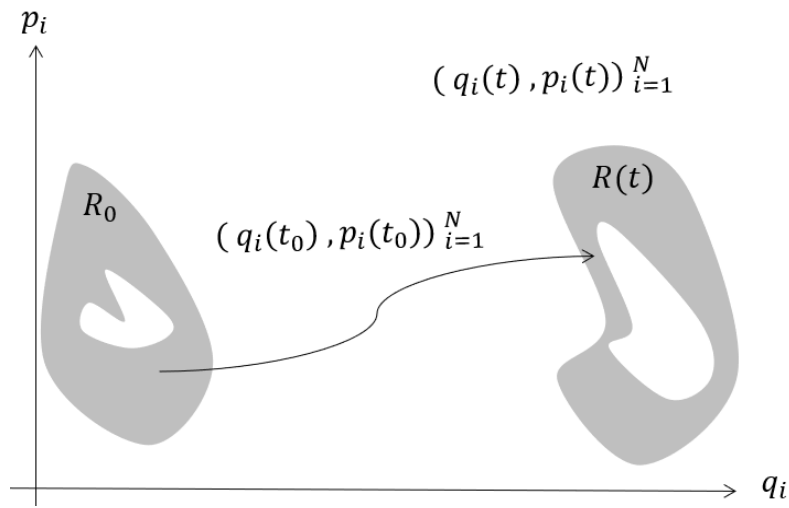

Fuente: Elaboración Propia.

En la Figura N° 2.3 la región $R_0$ a un tiempo dado $t_0$ se transforma en $R(t)$ al tiempo $t$ de acuerdo a las ecuaciones de Hamilton [2.38], entonces la medida del hipervolumen se conserva $|R_0| = |R(t)|$.

El teorema hace uso de la densidad de estados $\rho = \rho(q_i, p_j, t)$, definido como $\rho = dN/dV$, donde $dN$ es el número de estados[26] en el elemento de volumen $dV$ del espacio de fases, así el teorema de Liouville es presentado como (Goldstein, Poole, & Safko, 2014, pág. 421):

$$\frac{d\rho}{dt} = 0 \qquad\qquad \frac{\partial \rho}{\partial t} = -\{\rho, H\} \qquad\qquad [2.42]$$

Ambas ecuaciones de [2.42] son equivalentes; la función $\rho$ permite hacer una conexión con la mecánica estadística: por un lado la ecuación [2.42] describe también la evolución temporal de una densidad de probabilidad (normalizando $\rho$) y con ello tener una descripción estricta de la mecánica estadística basada directamente en las ecuaciones de la mecánica clásica y no en aproximaciones o tratamientos probabilísticos desconectado del determinismo de la mecánica clásica, y por otro lado, cuando el corchete de Poisson en [2.42] se anula se tiene el caso estacionario, lo cual corresponde con la distribución de muchas partículas en equilibrio termodinámico.

### 2.3.3. El exponente de Lyapunov

El teorema de Liouville (sección [2.3.2]) impide la aparición de atractores o repulsores en el espacio de fases para sistemas gobernados por las ecuaciones de Hamilton [2.38], pues los atractores son regiones al que las trayectorias de su vecindad convergen asintóticamente (el repulsor es un atractor en el sentido inverso del tiempo), y en consecuencia no conservan el hipervolumen de una región próxima a ellas, así que están prohibidas en la mecánica clásica y, por el contrario, ellos aparecen en sistemas disipativos (como un péndulo forzado oscilando en un medio con fricción) introduciendo procesos irreversibles en las dinámicas de los sistemas complejos; no obstante, es posible describir un comportamiento caótico en los sistemas hamiltonianos, sin atractores ni repulsores, donde las trayectorias descritas por las ecuaciones de Hamilton forman un "fibrado" que se enreda y retuerce (sin intersecarse nunca) de manera que dos puntos muy próximos en el espacio de fases pueden divergir rápidamente (por simetría en el tiempo, también habrán puntos lejanos que se aproximen mucho entre sí), el parámetro que caracteriza y cuantifica este comportamiento es el exponente de Lyapunov $\lambda$: sean $X_1(t_0)$ y $X_2(t_0)$ dos puntos muy próximos del espacio de fases al tiempo inicial $t_0$ y $\varepsilon_0 = |X_2(t_0) - X_1(t_0)|$ la separación entre ellos, la separación que tendrán a un instante posterior del tiempo $\varepsilon(t) = |X_2(t) - X_1(t)|$ entonces:

$$\varepsilon(t) = \varepsilon_0 \cdot e^{\lambda t} \qquad [2.43]$$

---

[26] Son concebidos en una interpretación probabilística o estadística, donde se asigna un "peso" o probabilidad al estado $(q_i, p_j)$ del espacio de fases de encontrar al estado del sistema en cuestión.

Donde $\lambda$ tiene dimensión $T^{-1}$, si $\lambda > 0$ entonces la dinámica es caótica, el caos se inicia para escalas de tiempo $\Delta t_{caos} \gg \tau = 1/\lambda$ (Goldstein, Poole, & Safko, 2014, págs. 492,493).

Así, $\lambda < 0$ corresponde al caso inverso y puede ser interpretado como el caos para un observador que mira de futuro a pasado, de manera que un $\lambda$ instantáneo puede variar (positiva y negativamente) en todo el espacio de fases.

### 2.4. Termodinámica y mecánica estadística

La termodinámica estudia la energía en los cuerpos macroscópicos, la mecánica estadística explica e interpreta a la termodinámica en términos de sistemas de muchas partículas; es conveniente mencionar que tanto la termodinámica como la física estadística están fundamentadas en el equilibrio termodinámico, constituyendo un problema de activa investigación los sistemas fuera del equilibrio.

> *"Las leyes de la termodinámica pueden ser fácilmente obtenidas a partir de los principios de la mecánica estadística, ya que son la expresión incompleta de estos principios"* – J.W. Gibbs.

### 2.4.1. El principio del incremento de la entropía

En los procesos termodinámicos se tiene el cambio de la entropía expresado en términos del calor y la temperatura, como se indica:

$$dS \geq \frac{\delta Q}{T} \qquad\qquad S_2 - S_1 \geq \int_1^2 \frac{\delta Q}{T} \qquad [2.44]$$

En [2.44] la condición de igualdad corresponde a un proceso reversible, y la desigualdad a un proceso irreversible. (Sonntag, Borgnakke, & Van Wylen, 2003, pág. 265); a la diferencia entre $dS$ y $\delta Q/T$ se le llama *generación de entropía* $S_{gen}$:

$$dS = \frac{\delta Q}{T} + \delta S_{gen} \qquad [2.45]$$

El cual siempre positivo (deducido de [2.44]) *"... esta generación interna puede ser causado por procesos tales como la fricción, expansión irrestricta, transferencias internas de energía (redistribución) o diferencias finitas de temperatura"* (Sonntag, Borgnakke, & Van Wylen, 2003, pág. 266); un sistema (masa de control) y su entorno (una temperatura circundante) al interactuar termodinámicamente tienen cambios de entropía $dS_{c.m}$ y $dS_{surr}$ respectivamente, de manera que la entropía neta o total del sistema general aislado (masa de control y fuente térmica) viene a ser la suma de ambos, el cual se reduce a las contribuciones de todas las generaciones de entropía que ocurren en todo el sistema general aislado:

$$dS_{net} = dS_{c.m} + dS_{surr} = \sum \delta S_{gen} \geq 0 \qquad [2.46]$$

La condición de igualdad corresponde a procesos reversibles, y la desigualdad a procesos irreversibles. (Sonntag, Borgnakke, & Van Wylen, 2003, pág. 269).

*"Esta es una ecuación muy importante, no sólo para la termodinámica sino también para el pensamiento filosófico. Esta ecuación [2.46] es referida como el principio del incremento de la entropía, El gran significado es que sólo ocurren los procesos en los que el cambio neto de entropía de la masa de control y su entorno circundante aumenta (o en el caso límite, permanece constante). El proceso inverso, en el cual ambos la masa de control y su entorno*

*circundante retornan a su estado original, puede nunca ocurrir, en otras palabras, la ecuación [2.46] dicta la única dirección en que cualquier proceso puede ocurrir. Así, el principio del incremento de la entropía puede ser considerado un enunciado general cuantitativo de la segunda ley del punto de vista macroscópico"* (Sonntag, Borgnakke, & Van Wylen, 2003, pág. 269).

La interpretación de la entropía en la termodinámica es desde luego relacionado a la flecha del tiempo: el aumento de la entropía determina el sentido en que pueden ocurrir los eventos, sólo los fenómenos donde la entropía del sistema aislado se conserva pueden ocurrir en ambos sentidos del tiempo; también se interpreta que una máquina es más eficiente si realiza el trabajo generando la mínima cantidad posible de entropía (en un proceso reversible, como el ciclo de Carnot, se tiene la máxima eficiencia posible), además conforme se llega al equilibrio térmico la energía libre de Gibbs decrece hasta un mínimo del que ya no se puede extraer más energía, y donde los procesos se detienen (la entropía alcanza un máximo y se conserva, a partir de ahí la descripción física es reversible, y también "aburrida" pues no acontece ningún proceso macroscópico).

### 2.4.2. La densidad de estados y densidad de probabilidad

Sea el sistema macroscópico descrito por $2N$ variables de coordenadas y momentos $q_j$ y $p_j$ en una variable generalizada $x_k$ como se muestra ($T$: transpuesta):

$$X = (x_k)_{k=1}^{2N} = (q_1, q_2, \ldots, q_N, p_1, \ldots, p_{N-1}, p_N)^T$$

$$\nabla = \left(\frac{\partial}{\partial x_k}\right)_{k=1}^{2N} = \left(\frac{\partial}{\partial q_1}, \frac{\partial}{\partial q_2}, \ldots, \frac{\partial}{\partial q_N}, \frac{\partial}{\partial p_1}, \ldots, \frac{\partial}{\partial p_N}\right)^T \quad [2.47]$$

Además, el elemento de hipervolumen $d\Gamma = \prod_{k=1}^{2N} dx_k$; en esta notación las ecuaciones canónicas de Hamilton [2.38] se escribe en una sola ecuación lineal:

$$\begin{bmatrix} \hat{0} & -\hat{1} \\ \hat{1} & \hat{0} \end{bmatrix} \dot{X} = \nabla H \quad [2.48]$$

En la ecuación [2.48] la matriz cuadrada de la izquierda es una matriz de rotación[27] de $\pi/2$ en el espacio de fases, su función es intercambiar las coordenadas de posición y momentos. Cada punto $X$ del espacio de fases $\xi(X)$ determina un microestado del sistema, y el Hamiltoniano es una función $H(X)$ sobre el espacio de fases; sea la región $\Pi \subset \xi(X)$, de hipervolumen $\Gamma$, donde la energía del sistema es $E$:

$$X \in \Pi(E) \ \leftrightarrow\ H(X) = E \qquad X \in \Pi_0(E) \ \leftrightarrow\ H(X) \leq E \qquad [2.49]$$

Se define el número de microestados $\Omega$ como el número de puntos $X$ que pertenecen al conjunto $\Pi$, $\Gamma_0(E)$ es la medida de la región $\Pi_0(E)$; la densidad de estados $\omega$ se define como el cambio de $\Gamma_0(E)$ al aumentar la energía E:

$$d\Gamma_0(E) = \omega(E) \cdot dE \qquad [2.50]$$

Puesto que el número de microestados $\Omega(E)$ puede ser teóricamente infinito, la ecuación [2.50] permite calcular la proporción entre $\Omega(E)$ y $\Omega(E')$:

$$\frac{\omega(E')}{\omega(E)} = \frac{d\Gamma_0(E')}{d\Gamma_0(E)} = \frac{\Omega(E')}{\Omega(E)} \qquad [2.51]$$

Ya que $d\Gamma_0(E) = \Gamma_0(E + \delta E) - \Gamma_0(E)$ es proporcional a $\Gamma(E)$, $\Pi(E)$ y $\Omega(E)$.

En la Figura N° 2.4 **a)** muestra al Hamiltoniano del sistema (eje vertical) como una función en el espacio de fases (plano horizontal), el plano $H(X) = E$ corta la función en los puntos $X \in \Pi(E)$, **b)** muestra el conjunto $\Pi(E)$ en el espacio de fases que contiene los $\Omega(E)$ microestados accesibles del sistema:

---

[27] De igual manera, en la ecuación de Schrödinger la unidad imaginaria al lado de la derivada temporal de la función de onda introduce una rotación de $\pi/2$ en el plano complejo, intercambiando términos real e imaginario.

FIGURA N° 2.4

FUNCIÓN HAMILTONIANO EN EL ESPACIO DE FASES

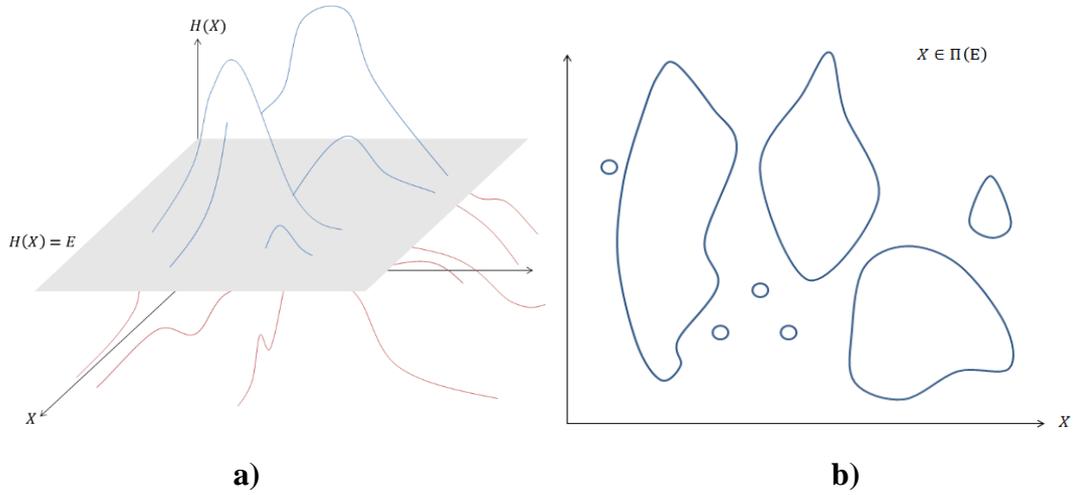

a)          b)

Fuente: Elaboración propia.

En general, se define la densidad de probabilidad $\rho(X,t)$ como el elemento de probabilidad por elemento de hipervolumen, para cualquier región de $\xi(X)$:

$$\rho(X,t) = \frac{dP}{d\Gamma} \qquad [2.52]$$

Una cantidad macroscópica $F(t)$ es el valor medio de una función $f(X,t)$ definida sobre cada microestado (la versión microscópica de esa cantidad):

$$F(t) = \langle f \rangle = \int_{\xi(X)} f(X,t) \cdot \rho(X,t) \cdot d\Gamma \qquad [2.53]$$

La entropía de información[28] se obtiene para $f(X,t) = -\ln \rho(X,t)$:

$$S(t) = -k_B \langle \ln \rho(X,t) \rangle \qquad [2.54]$$

El teorema de Liouville [2.42] establece que esta probabilidad evoluciona como un fluido incompresible, conservando siempre su hipervolumen, el caso del equilibrio térmico se formula cuando la densidad de probabilidad es estacionaria:

$$\frac{\partial \rho(X,t)}{\partial t} = 0 \qquad [2.55]$$

---

[28] Es equivalente a la entropía termodinámica por un factor $k_B$ (constante de Boltzmann).

## 2.5. Contradicciones teóricas

FIGURA N° 2.5

CONTRADICCIONES TEÓRICAS

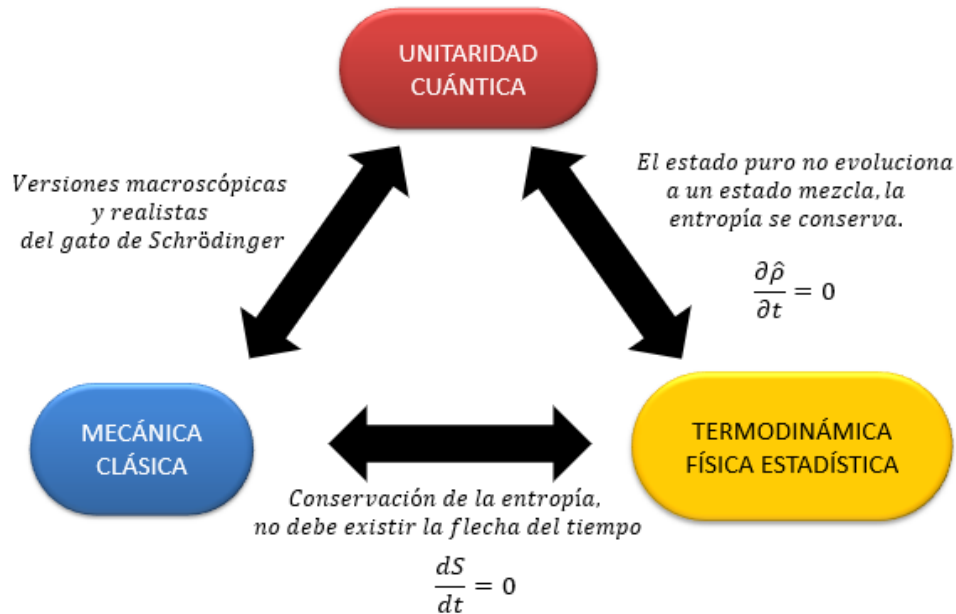

Fuente: Elaboración propia.

### 2.5.1. Contradicción: mecánica clásica vs. proceso U

Las contradicciones entre la mecánica clásica y la unitaridad de la mecánica cuántica se encuentran en la formulación de su estado y evolución, el estado cuántico es una superposición de autoestados clásicos-observables, y su evolución está dada por la ecuación de Schrödinger, si la unitaridad de la mecánica cuántica se siguiera aplicando a escala clásica se obtendrían versiones realistas del gato de Schrödinger: objetos moviéndose en muchas direcciones[29], trabajos desarrollándose en muchas formas superpuestas, objetos deslocalizados y entrelazados (…) estos fenómenos no necesariamente desaparecen al nivel macroscópico ($\hbar \to 0$), y no pueden ser resueltos por el principio de correspondencia de Bohr (sección [2.3.1]).

El rol del observador es igualmente contradictorio: en la mecánica clásica el observador es irrelevante para el mundo clásico, mientras que en el mundo cuántico es fundamental porque crea y destruye el estado y la medida, igualmente el colapso es un proceso discontinuo y contrario a la mecánica clásica (suave, continua y sin saltos) aunque esto si puede ser resuelto por la correspondencia.

---

[29] Si se deja un objeto macroscópico en reposo bastaría poco tiempo para que el paquete de ondas asociado al objeto se expanda (en todas direcciones) alcanzando dimensiones macroscópicas.

### 2.5.2. Contradicción: mecánica clásica vs. termodinámica-mecánica estadística

Al inicio la termodinámica y la mecánica clásica eran teorías incompatibles[30]; la mecánica estadística al explicar la termodinámica en base a la mecánica clásica introduce una contradicción de principio: leyes físicas reversibles al nivel fundamental conducen a fenómenos irreversibles al nivel estadístico.

> *"Boltzmann se propone derivar la segunda ley de la termodinámica usando las leyes de la mecánica a partir del enfoque estadístico de Maxwell. Boltzmann derivó una ecuación (...) que describe el proceso por medio del cual un gas tiende hacia su condición de equilibrio (...). Y aquí se encuentra la semilla de los problemas y de las disputas que tuvo que enfrentar Boltzmann, pues la pretensión de estar obteniendo resultados basados en principios mecánicos y los procesos irreversibles a los que se llega (el sistema siempre se mueve hacia el estado de equilibrio) conduce a una serie de paradojas (...). Para combatir a sus críticos, Boltzmann desarrolla aún más sus ideas estadísticas (...), dándole a su trabajo un carácter totalmente probabilístico. Estos esfuerzos lo conducirían a expresiones equivalentes a la famosa formula para la entropía S=k logW. (...) Boltzmann se defiende de sus oponentes bajo la idea de que la segunda ley de la termodinámica del aumento de la entropía es cierta en un sentido estadístico, no absoluto"* (Guzmán & Antonio, 2007, pág. 335).

El éxito experimental de la mecánica estadística la hizo ser aceptada y la crítica de la reversibilidad fue desestimada[31]; en la paradoja de la reversibilidad si se invierten las velocidades de todas las partículas se tendría la ocurrencia del proceso inverso y la disminución de la entropía (Guzmán & Antonio, 2007, pág. 338), para la mecánica estadística esto es cierto pero improbable[32], y que realmente existen muchas más condiciones iniciales que evolucionan maximizando la entropía que condiciones iniciales que evolucionan disminuyendo la entropía, así es matemáticamente esperado que la entropía aumente; no obstante, si al tiempo $t_0$ existieran $N$ condiciones iniciales $\{(q_{i_0}, p_{i_0})\}, i = 1:3n$ ($n$: número de partículas) que aumenten la entropía, entonces existirían también $N$ condiciones iniciales inversas $\{(q_{i_0}, -p_{i_0})\}, i = 1:3n$, las cuales hacen evolucionar "hacia atrás en el tiempo" disminuyendo la entropía con la misma ratio[33], luego al nivel estadístico se promedian estos fenómenos obteniendo la conservación de la entropía (para un sistema aislado), con pequeñas fluctuaciones en cualquier sentido del tiempo, en contradicción con el principio del incremento de la entropía (sección [2.4.1]); en general, como la mecánica clásica no privilegia ningún sentido del tiempo, no puede devenir en una flecha termodinámica del tiempo, ya que los mismos argumentos que un observador postula para explicar el aumento de la entropía de pasado a futuro pueden ser formulados por un observador que mira el tiempo de futuro a pasado, pues las ecuaciones canónicas de Hamilton se deben de satisfacer para ambos observadores.

---

[30] La mecánica clásica es reversible, mientras que la termodinámica es irreversible.
[31] En gran medida esto fue posible porque los detractores de la mecánica estadística eran positivistas y veían a los átomos como propios de la metafísica; al conseguir el éxito experimental, los positivistas cedieron a la nueva teoría por principio epistemológico: alinearse con los experimentos; y ahora, tales cuestionamientos de la reversibilidad parecen ser banales o propio de lo metafísico.

### 2.5.3. Contradicción: proceso U vs. termodinámica-mecánica estadística

Un universo regido sólo por la evolución unitaria[34] es incompatible con la termodinámica, se pueden usar los mismos argumentos que en la sección [2.5.2] sobre la reversibilidad del tiempo; además, para que la termodinámica o la mecánica estadística funcionen se necesita que las transformaciones unitarias aumenten la entropía del operador de densidad $\hat{\rho}$, así para un sistema aislado fuera del equilibrio el Hamiltoniano no depende del tiempo, de acuerdo a la ecuación [2.19], no existe ninguna transformación unitaria que aumente o cambie la entropía de $\hat{\rho}$, mientras que según la termodinámica la entropía debería aumentar en ese sistema aislado.

---

[32] A diferencia de la termodinámica, en la mecánica estadística la entropía puede disminuir con una probabilidad que cae exponencialmente con la cantidad de entropía disminuida, siendo sólo apreciable para disminuciones de entropía del orden de la constante de Boltzmann.

[33] Esto es, que si los $N$ primeros describen $S(t)$, entonces los $N$ segundos describen $S(t_0 - t)$.

[34] Donde el colapso no ocurre, o de acontecer no son objetivos sino como parte del proceso unitario de todo el sistema aislado.

## 2.6. Mecánica y termodinámica de los agujeros negros

### 2.6.1. La relatividad general es intrínsecamente reversible en el tiempo

En la relatividad general se tiene la ecuación del campo de Einstein:

$$G_{uv} = \frac{8\pi G}{c^4} T_{uv} \qquad G_{uv} = R_{uv} - \frac{1}{2} R g_{uv} \qquad [2.56]$$

Donde $G_{uv}$, $T_{uv}$, $R_{uv}$, $g_{uv}$ son el tensor de curvatura de Einstein, de energía-momento, de curvatura de Ricci y métrico respectivamente; al resolver [2.56] se encuentran los símbolos de Christoffel $\Gamma^{\sigma}_{uv}$ que establecen las ecuaciones geodésicas:

$$\frac{d^2}{d\lambda^2} x^{\sigma} + \Gamma^{\sigma}_{uv} \frac{dx^u}{d\lambda} \frac{dx^v}{d\lambda} = 0 \qquad [2.57]$$

Donde $\lambda$ es un parámetro afín que puede ser el tiempo propio, las soluciones de [2.57] son geodésicas y son las trayectorias inerciales de las "partículas libres", el concepto de fuerza puede ser entendido como pasando de una geodésica a otra[35], esta ecuación es determinista y conserva la información[36]; la ecuación [2.57] es invariante ante la inversión $\lambda \to -\lambda$, esto significa (al igual que en la sección [2.3.1]) que de los fenómenos que describe la ecuación [2.57] sus fenómenos inversos (hacia atrás en el sentido del parámetro afín $\lambda$) también son soluciones de la misma ecuación, así sus soluciones son simétricas y reversibles en el tiempo[37].

### 2.6.2. Los agujeros negros como singularidades físicas

Los agujeros negros son objetos astronómicos ampliamente estudiados por diversas disciplinas, al nivel físico-teórico son descritos por la relatividad general como singularidades[38] en las soluciones a las ecuaciones del campo de Einstein [2.56]; las singularidades son regiones del espacio-tiempo con comportamientos patológicos como que las geodésicas *timelike* ($ds > 0$) o nulas ($ds = 0$) tras un recorrido finito del parámetro afín o tiempo propio no pueden continuar, esto es incompletitud de geodésicas causales (Wald, 1984, pág. 212), o que la curvatura se haga infinita (singularidad de curvatura).

---

[35] Por analogía a la dinámica Newtoniana donde la Fuerza cambia el estado inercial del movimiento.
[36] De acuerdo a la sección [B.3.1].
[37] Si el parámetro afín $\lambda$ es el tiempo (ya sea para un observador externo o el propio).
[38] Las singularidades son aquellos puntos del espacio-tiempo donde no se puede hacer análisis matemático y, en consecuencia, no se puede calcular (derivadas, límites, etc.) ni definir una cantidad gravitacional adecuadamente.

La primera solución exacta en el vacío a la ecuación [2.56] (esto es, donde el tensor de energía-momento se anula), es la métrica de Schwarzschild para un agujero negro estático, sin carga eléctrica ni rotación, con simetría esférica, esta solución es estacionaria (el campo gravitatorio no cambia en el tiempo), y devuelve las coordenadas esféricas para $M = 0$, su elemento de línea $ds^2$ está dada por [2.58]:

$$ds^2 = \alpha(r)dt^2 - \frac{dr^2}{\alpha(r)} - r^2 d\Omega^2 \qquad \alpha(r) = 1 - \frac{2M}{r} \qquad [2.58]$$
$$d\Omega^2 = d\theta^2 + sen^2(\theta)d\phi^2$$

Donde $c = G = 1$; $\alpha(r)$ adquiere valores singulares para $r = 0$ y $r = 2M$ porque se anula o se hace infinito, lo cual impide que distintos diferenciales $dt$ y $dr$ se expresen analíticamente en $ds^2$; este agujero negro consiste en una singularidad central ($r = 0$) rodeada por un horizonte de eventos ($r = 2M$).

Para una dinámica radial ($d\Omega = 0$) se grafican las geodésicas nulas $ds^2 = 0$:

FIGURA N° 2.6

SOLUCIÓN DE SCHWARZSCHILD RADIAL

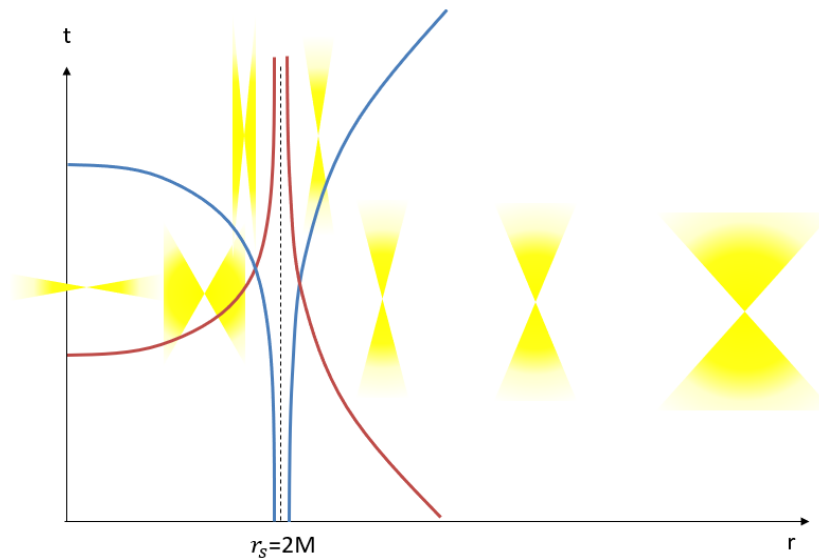

Fuente: Elaboración propia.

En la Figura N° 2.6 las líneas rojas representan haces de luz que se dirigen hacia la singularidad central, y las azules a las que se alejan[39], las geodésicas son reversibles y simétricas en el tiempo para una inversión del tiempo propio; los conos de luz (amarillo) representan el sentido de la causalidad, y se estiran verticalmente y comprimen horizontalmente en la proximidad exterior al horizonte de eventos; al interior del agujero negro $r < 2M$ la función $\alpha(r)$ en [2.58] se hace negativa, lo cual es interpretado como sigue: la coordenada $r$ se hace *timelike* (y establece el orden causal de los eventos) mientras que la coordenada $t$ hace *spacelike* (sin orden causal entre los eventos), así un observador se mueve inexorablemente hacía $r = 0$ de la misma manera en que un observador externo "se mueve" hacia el futuro, por simetría otro observador que en el exterior se mueve "hacia el pasado", dentro del agujero negro se está alejando inexorablemente de $r = 0$ y dirigiéndose al horizonte de eventos, este último es un agujero blanco; en ambos casos la ortodoxia interpreta que el espacio-tiempo (con los observadores en él) se mueve más rápido que la luz ya sea para acercarse o alejarse de la singularidad central.

En el diagrama de Penrose-Carter (o simplemente, diagrama de Penrose) para esta misma métrica se puede mostrar la información completa de forma explícita, la cual generaliza la reversibilidad expuesta anteriormente:

---

[39] En el interior del agujero negro, la causalidad sólo conduce a la singularidad central, de manera que ir hacia el centro (rojo) "retrocede en el tiempo" mientras que tratar de alejarse de la singularidad central (azul) sólo posterga su caída.

FIGURA N° 2.7

DIAGRAMA DE PENROSE PARA LA SOLUCIÓN DE SCHWARZSCHILD

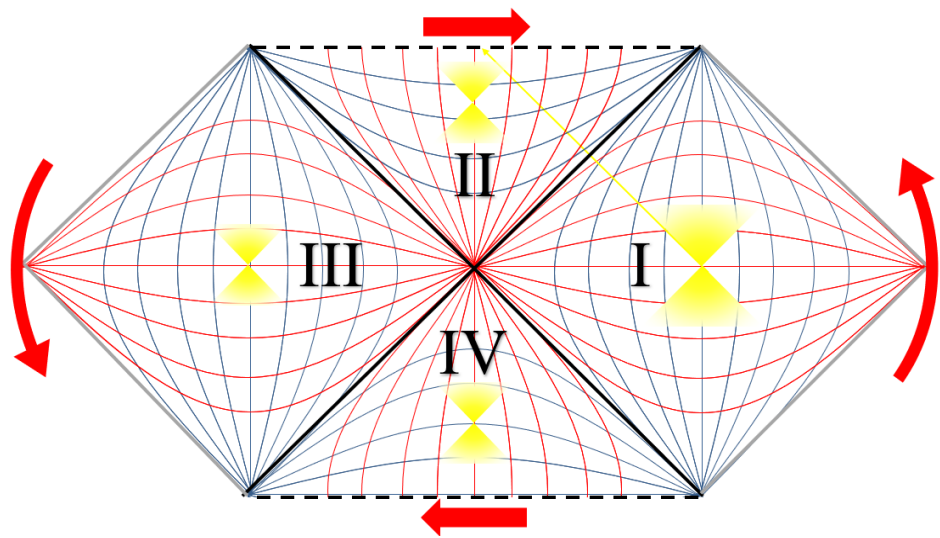

Fuente: Elaboración propia.

En la Figura N° 2.7 las curvas rojas son todos los eventos para $t$ constante, las curvas azules corresponden a un valor de $r$ constante, las líneas negras continuas corresponden a los horizontes de eventos ($r = 2M$), las líneas negras discontinuas representan las singularidades centrales ($r = 0$), y las flechas rojas (fuera del diagrama) representan el sentido del tiempo; en este diagrama los conos de luz no sufren deformación en cualquier parte del espacio-tiempo; la región I corresponde al espacio-tiempo de partículas que avanzan hacia adelante en el tiempo (y contiene partículas alejándose del agujero blanco y partículas cayendo al agujero negro), mientras que la región III corresponde a las antipartículas[40] (y contiene antipartículas alejándose del agujero blanco y antipartículas cayendo al agujero negro; o equivalentemente, partículas alejándose del agujero negro y partículas cayendo al agujero blanco), la región II corresponde al interior del agujero negro donde el cono de luz para cualquier punto a su interior conduce a la singularidad central, por el contrario la región IV es un agujero blanco en el que cualquier punto a su interior inevitablemente conduce al horizonte de eventos (en dirección al exterior); las partículas en la proximidad exterior al agujero negro (en la región I) pueden aproximarse asintóticamente al horizonte de eventos ($t \to \infty$, $r \to 2M$), el proceso inverso se da en la parte inferior de la región I en la proximidad al agujero blanco, donde una partícula se viene alejando del horizonte de eventos tomándose un tiempo infinito (desde el pasado infinito); este diagrama exhibe la reversibilidad y simetría en el tiempo, pues ante una inversión temporal $t \to -t$ se conservan todos los fenómenos; es importante resaltar que la descripción de una partícula atravesando el horizonte de eventos corresponde a una partícula que se toma un tiempo infinito en llegar al horizonte de eventos, y que luego retrocede un tiempo infinito en el interior del agujero negro[41] hasta $r = 0$, esto se nota también en la Figura N° 2.6 donde la caída al agujero negro de un fotón se representa con la línea roja (la línea azul también corresponde a un fotón saliendo del agujero blanco), sin embargo no es posible establecer la continuidad en el horizonte de eventos (aunque en el diagrama de Penrose-Carter luzca continuo), esto es mostrado en la sección [2.6.4] (que no existe flujo de energía-momento sobre el horizonte de eventos, y que la métrica sigue siendo singular en el horizonte de eventos) y abordado en la sección [5.7.4] (con colapso objetivo); así, una partícula que intenta entrar al agujero negro se queda en una eterna caída; en la Figura N° 2.6 ambas curvas rojas deben aproximarse entre sí asintóticamente para $t \to \infty$ pero sin tocarse nunca (no se pueda pasar de una geodésica a otra, exterior e interior desconectados).

---

[40] Esto es, partículas viajando hacia atrás en el tiempo; es normal que la relatividad incluya esta simetría del tiempo, así por ejemplo al considerar la relatividad especial en la mecánica cuántica se obtienen soluciones de las antipartículas como una consecuencia natural.

La interpretación ortodoxa de la relatividad en los agujeros negros ha dicho que la métrica de Schwarzschild no es buena para describir a un agujero negro (sin carga ni rotación) porque no describe el interior ni el ingreso de materia, además que presenta una singularidad en el horizonte de eventos, y que por eso se hace uso de las coordenadas de Eddington-Finkelstein que describe tanto el interior como el ingreso de materia, y que la singularidad en $r = 2M$ es sólo de coordenadas (removiéndose con estas nuevas); en esta investigación se ha usado la Figura N° 2.6 para representar el interior del agujero negro (se obtiene un agujero blanco al revertir el sentido del tiempo propio) de manera que la representación en estas coordenadas contiene toda la información requerida y el cambio a otras coordenadas sólo mejora su representación; las singularidades físicas y de coordenadas se encuentran combinadas en la métrica $g_{uv}$ y se distinguen porque las singularidades físicas dependen de la existencia de un campo gravitatorio (y varían con él), y desaparecen en un espacio-tiempo plano, mientras que las singularidades de coordenadas permanecen invariantes incluso en ausencia de gravedad, así en la métrica de Schwarzschild (ecuación [2.58]) las coordenadas esféricas introducen singularidades de coordenadas en el origen y en los polos[42]: $r = 0$ y $\theta = 0, \pi$, y el campo gravitatorio introduce las singularidades físicas en $r = 0$ y $r = 2M$ mediante el término $\alpha(r)$, de manera que el horizonte de eventos es en efecto una singularidad física, estas ideas han sido inspiradas en la sección [2.6.4]; para las coordenadas de Eddington-Finkelstein se emplea la coordenada tortuga $r^*$:

$$dr^* = \frac{dr}{1 - \frac{2M}{r}} \qquad r^* = r + 2M \ln\left|\frac{r - 2M}{2M}\right| \qquad [2.59]$$

---

[41] Al interior del agujero negro ($\alpha < 0$) el espacio y el tiempo intercambian roles, de manera que se puede avanzar o retroceder en la coordenada $t$, pero no se puede detener el "avance" en la coordenada $r$.

Donde $r^*$ no es analítica en $r = 2M$, e introduce (o remueve) singularidad en ese punto (sección [2.6.4]), así las coordenadas entrantes de Eddington-Finkelstein viene dado por $(v, r)$ en reemplazo de $(t, r)$ donde $v = t + r^*$, y para $(u, r)$ con $u = t - r^*$ se obtienen las coordenadas salientes de Eddington-Finkelstein que describe un agujero blanco, a continuación las coordenadas entrantes:

FIGURA N° 2.8

COORDENADAS ENTRANTES DE EDDINGTON-FINKELSTEIN

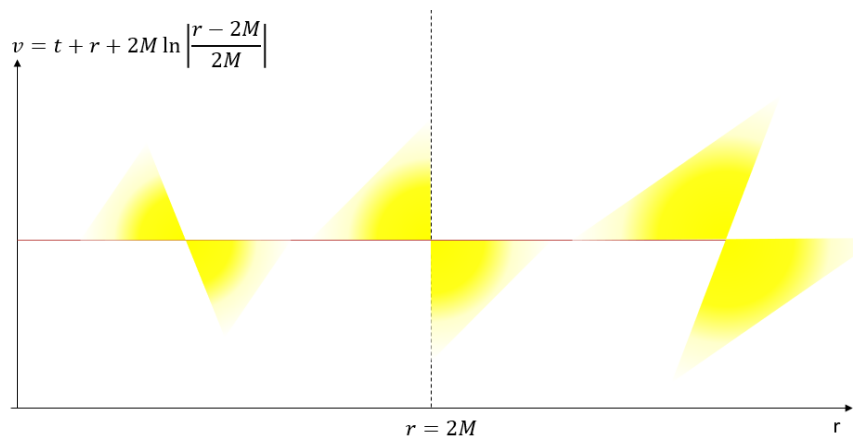

Fuente: Elaboración propia.

---

[42] Tiene la métrica en el vacío $g_{uv} = diagonal(1, 1, r^2, r^2 sen^2\theta)$, con determinante $g = -r^4 sen^2\theta$.

En la Figura N° 2.8 no se exhibe alguna singularidad en $r = 2M$ (porque fue removida por $r^*$) y se puede ingresar al interior, la comparación con la métrica de Schwarzschild revela una descripción equivalente: en ambos, el tiempo propio para la geodésica que va desde $r > 2M$ hasta $r \to 2M^+$ es finito mientras que para la coordenada $t$ el aumento es infinitamente positivo, luego el tiempo propio para una geodésica que va desde un $r \to 2M^-$ hasta $r = 0$ es finita, pero para la coordenada $t$ es un cambio infinitamente negativo, y el cambio total en la coordenada $t$ es finito; así, en la Figura N° 2.8 la coordenada $v = t + r^*$ lo que hace es compensar ese infinito aumento y decremento en la coordenada $t$ mediante la coordenada $r^*$ (que disminuye infinitamente desde $r > 2M$ hasta $r \to 2M^+$, y aumenta infinitamente desde $r \to 2M^-$ hasta $r = 0$) evitando el problema de la singularidad y ocultando la discontinuidad en $r = 2M$, pues el término logarítmico de $r^*$ es indeterminado en ese punto, y las coordenadas de Eddington-Finkelstein dejan de ser válida en el horizonte de eventos (son válidas sólo en su vecindad), luego "no tienen autoridad" para establecer continuidad ahí, ni "remueven la singularidad".

De acuerdo a lo establecido al inicio de esta sección sobre las singularidades, la discontinuidad de las geodésicas establece la *incompletitud de geodésicas causales* porque en la prolongación del tiempo propio las geodésicas se acaban abruptamente.

La supuesta continuidad de las geodésicas se usa para explicar el ingreso[43] a un agujero negro "forzando la relatividad general"; un cascarón simétrico de masa $m$ que cae esféricamente al agujero negro de masa $M$ se representa en la Figura N° 2.9: antes de $t_0$ se cae sobre el horizonte de eventos $r_1 = 2M$ (curva roja), para $t > t_0$ ya se encuentra al interior, y aumenta el radio de Schwarzschild a $r_2 = 2(M + m)$, pero el cascarón en el exterior debe caer sobre $r_2$ (curvas azules), entonces sigue la curva roja sólo hasta $t_0$ y continua con la curva azul, un observador externo puede interpretar que el cascarón ingresa al tiempo $t_0$ porque aumenta el radio de $r_1$ a $r_2$, aun así no se requiere de la continuidad de las geodésicas en el horizonte de eventos[44].

---

[43] En la sección [5.7.4] propongo un mecanismo de ingreso al interior del agujero negro que requiere del proceso R de la mecánica cuántica, ya que la relatividad por si sola no lo puede explicar.

[44] Ya que las geodésicas se prolongan hasta el infinito de la coordenada $t$, y requieren de un agujero negro eterno y prácticamente inmutable, así cuando el agujero negro desaparezca (por evaporación)

FIGURA N° 2.9

CAÍDA DE UN CASCARÓN SIMÉTRICO AL AGUJERO NEGRO

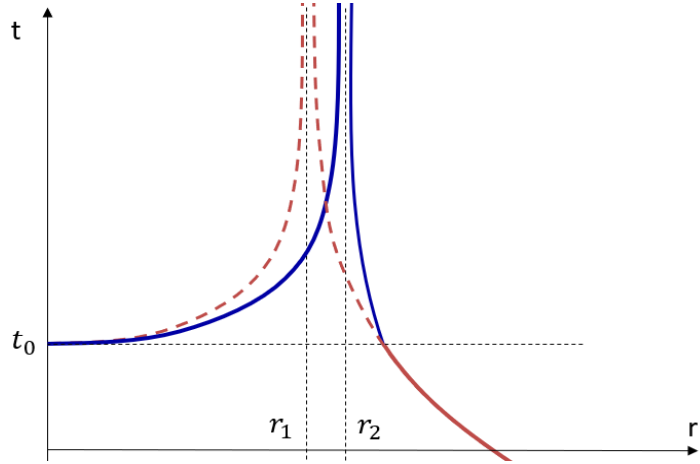

Fuente: Elaboración propia.

El caso más general para un agujero negro es tener masa $M$, carga eléctrica $Q$ y momento angular $L$, la rotación rompe la simetría esférica, pero preserva simetría axial, en este caso se usan las coordenadas de *Boyer-Lindquist*:

$$x = \sqrt{r^2 + a^2}sen(\theta)\cos(\phi)$$
$$y = \sqrt{r^2 + a^2}sen(\theta)sen(\phi) \qquad a = \frac{L}{M} \qquad [2.60]$$
$$z = r\cos(\theta)$$

Con las ecuaciones [2.60], y $4\pi\varepsilon = 1$ la solución general está dado por la métrica de Kerr-Newman[45], cuyo elemento de línea es:

$$ds^2 = \frac{\Delta}{\Sigma}(dt - asen^2(\theta)d\phi)^2 - \frac{sen^2(\theta)}{\Sigma}\big((r^2 + a^2)d\phi - adt\big)^2$$
$$- \frac{\Delta}{\Sigma}dr^2 + \Sigma d\theta^2 \qquad [2.6.2.4]$$

$$\Delta = r^2 - 2Mr + Q^2 + a^2 \qquad \Sigma = r^2 + a^2\cos^2(\theta)$$

---

se interrumpiría el ingreso (la eterna caída); además, el que aumente de tamaño es inconsistente con que las primeras partículas que lo formaron sigan cayendo en el exterior cuando el agujero negro era microscópico, y deban seguir cayendo en el exterior cuando el agujero negro sea macroscópico e incluso supermasivo; en todo caso, la idea de la continuidad de las geodésicas puede usarse como una buena aproximación mientras no se avanza mucho en la coordenada $t$.

[45] Para $Q = 0$ se obtiene la métrica de Kerr, para $a = 0$ se obtiene la métrica de Reissner-Nordström.

En esta métrica aparecen dos horizontes de eventos con sus respectivas ergoesferas que rodean a la singularidad central, la cual no es puntual sino circular (todos con simetría axial); en [2.6.2.4] los horizonte de eventos aparecen para $r = r_\pm = M \pm \sqrt{M^2 - Q^2 - a^2}$ y las ergoesferas aparecen para $r = r_\pm^\theta = M \pm \sqrt{M^2 - Q^2 - a^2 cos^2(\theta)}$; si $M^2 - Q^2 - a^2 < 0$ no existen horizontes de eventos y se tiene una singularidad desnuda[46], se le llama *agujero negro extremal*.

**2.6.3. Transformaciones reversibles e irreversibles en los agujeros negros**

La mínima masa que un agujero negro puede tener es la que queda si se le quita su carga y rotación, y se le llama *masa irreducible $M_{irr}$* (de Christodoulou-Ruffini):

$$M^2 = \left(M_{irr} + \frac{Q^2}{4M_{irr}}\right)^2 + \frac{M^2 a^2}{4M_{irr}^2}$$

$$M_{irr}^2 = \frac{\left(M^2 - \frac{Q^2}{2} + M\sqrt{M^2 - Q^2 - a^2}\right)}{2} = \frac{r_+^2 + a^2}{2}$$

[2.61]

La $M_{irr}$ se conserva frente a procesos reversibles en los agujeros negros (por ejemplo, la asistencia gravitatoria sin ingreso al agujero), y aumenta con procesos irreversibles[47] (como el ingreso de materia o energía a su interior), además que no hay proceso que disminuya $M_{irr}$ con el paso del tiempo (de pasado a futuro), entonces se puede definir una flecha del tiempo en base a los procesos que aumentan $M_{irr}$, y tiene un comportamiento análogo a la entropía termodinámica, sin embargo, $M_{irr}$ se refiere a los procesos relativos a un agujero negro en particular, mientras que la entropía se refiere a todos los procesos en conjunto; así la bifurcación de un agujero negro en dos (o más) podría conservar la suma de las $M_{irr}$ de todos los agujeros negros bifurcados, pero sería una disminución para el agujero negro original, y este es un proceso prohibido[48], así que no es lineal con la entropía.

> *"Esta nota reporta cinco conclusiones: (1) La masa energía de un agujero negro de momento angular L puede ser expresado en la forma:*

---

[46] Prohibida por la conjetura del censor cósmico, la cual es deseable, pero podría ser falsa.

[47] En principio no debería haber procesos irreversibles en los agujeros negros al ser obtenidos de la relatividad general, pero se admite su existencia por razones más fenomenológicas.

[48] Es prohibido por la segunda ley mecánica de los agujeros negros, sección [2.6.5].

$$m^2 = m_{ir}^2 + L^2/4m_{ir}^2$$

*Donde $m_{ir}$ es la masa irreducible del agujero negro. (…) (3) El rango alcanzable de transformaciones reversibles se extiende de $L = 0$, $m^2 = m_{ir}^2$ a $L = m^2$, $m^2 = 2m_{ir}^2$. (4) Una transformación irreversible es caracterizada por un incremento en la masa irreducible del agujero negro. (5) No existe proceso que disminuya la masa irreducible"* (Christodoulou, 1970, pág. 1).

### 2.6.4. El teorema del agujero negro

En las secciones precedentes se estableció la reversibilidad en las soluciones de la relatividad general (secciones [2.6.1] y [2.6.3]) y su efecto en la singularidad de los agujeros negros (sección [2.6.2]), en esta sección se establece la motivación de esta interpretación: que los agujeros negros son cerrados si están regidos únicamente por la relatividad general, de acuerdo al *Teorema del agujero negro*, que contradice a la ortodoxia sobre ellos:

*__Teorema 4.1__ (Teorema del agujero negro). Asumiendo la validez de la teoría de Einstein de la relatividad general, entonces las siguientes afirmaciones son ciertas:*

1) *Los agujeros negros son cerrados: la materia no puede entrar ni salir de sus interiores,*
2) *Los agujeros negros son innatos: no nacen de la explosión de objetos cósmicos, ni nacen de un colapso gravitacional,*
3) *Los agujeros negros son llenos e incompresibles, y si el campo de materia no está distribuido homogéneamente en un agujero negro, entonces debe haber agujeros negros secundarios en el interior del agujero negro.* (Ma & Wang, 2014, pág. 43)

Los autores afirman que la singularidad en el horizonte de eventos es física y real (de ahí que se impide el ingreso de materia al interior, y en consecuencia que sean innatos), ellos rechazan el uso de las coordenadas de Eddington-Finkelstein y kruskal en el horizonte de eventos por ser "*coordenadas singulares matemáticamente prohibidas*" y en consecuencia carece de validez su uso para remover la singularidad en el horizonte de eventos, pues el espacio-tiempo sigue siendo no-diferenciable en ellos[49]; como se vio en la sección [2.6.2] la métrica del espacio-tiempo plano en coordenadas esféricas $(t, r, \theta, \phi)$ ya contiene singularidades de coordenadas en el origen $r = 0$ y los polos $\theta = 0, \pi$ (que pueden ser removidas por rotaciones y traslaciones) y que son invariantes ante el campo gravitacional, en contraste la métrica de Schwarzschild, ecuación [2.58], introduce las singularidades en $r = 0; 2M$ (debido al factor $\alpha = 1 - 2M/r$) que son explícitamente dependientes del campo gravitacional, y que no debería ser considerado "singularidad de coordenadas" como la ortodoxia lo ha establecido, así este trabajo se ha sumado a la interpretación de que la singularidad en $r = 2M$ es física (porque desaparece en ausencia de gravedad).

Para los autores del teorema del agujero negro, los agujeros negros regidos solamente por la relatividad general no son creados (a partir del colapso gravitacional) y no se pueden aniquilar (ni bifurcar o fusionarse) conservándose eternamente tanto hacia el futuro infinito como hacia el pasado infinito; en esta investigación sólo se ha tomado interés por la primera y segunda afirmación del teorema (la tercera afirmación es independiente de los dos primeros y no es requerido para esta investigación).

Para sustentar que los agujeros negros son cerrados, basta con demostrar que el flujo de energía-momento en el horizonte de eventos se anula, esta demostración es realizada para el agujero negro de Schwarzschild, pero su validez es generalizable; a continuación, la demostración de los autores:

Por la conservación de la energía-momento se tiene:

$$\frac{\partial E}{\partial \tau} + \text{div}\boldsymbol{P} = 0 \qquad [2.62]$$

Donde $E$ y $\boldsymbol{P}$ son las densidades de energía y momento, tomando la integral de volumen en $B = \{x \in \mathbb{R}^3 | R_s < |x| < R_1\}$:

---

[49] En efecto, tales coordenadas son singulares en $r = 2M$ porque el uso de la coordenada tortuga (ecuación [2.59]) es singular: claramente $r^*$ es indeterminado en $r = 2M$, las coordenadas que incluyen $r^*$ en sus ecuaciones no quitan la indeterminación sino que la esconden para aparentar una continuidad en el espacio-tiempo entre el interior y el exterior de un agujero negro, se puede crear o remover singularidades usando transformaciones singulares.

$$\int_B \left[\frac{\partial E}{\partial \tau} + \text{div}\mathbf{P}\right] d\Omega = 0 \, , d\Omega = \sqrt{g} dr d\theta d\phi \qquad [2.63]$$

Donde $g = \det(g_{ij}) = \alpha r^4 \text{sen}^2(\theta), \alpha = (1 - 2MG/c^2 r)^{-1}$, y por la fórmula de gauss para la divergencia de **P**:

$$\int_B \text{div}\mathbf{P} d\Omega = \int_{S(R_1)} \sqrt{\alpha(R_1)} P_r dS(R_1) - \lim_{r \to R_s^+} \int_{S(r)} \sqrt{\alpha} P_r dS(r)$$

Aquí $S(r) = \{x \in \mathbb{R}^3 | |x| = r\}$ en vista de [2.63] deducimos que el cambio de la energía total

$$\int_B \frac{\partial E}{\partial \tau} d\Omega = \lim_{r \to R_s^+} \int_{S(r)} \sqrt{\alpha} P_r dS(r) - \sqrt{\alpha(R_1)} \int_{S(R_1)} P_r dS(R_1) \qquad [2.64]$$

Así [2.64] puede ser escrita como:

$$\lim_{r \to R_s^+} \int_{S(r)} P_r dS(r) = \lim_{r \to R_s^+} \frac{1}{\sqrt{\alpha(r)}} \left[\int_B \frac{\partial E}{\partial \tau} d\Omega + \sqrt{\alpha(R_1)} \int_{S(R_1)} P_r dS(R_1)\right] \qquad [2.65]$$

El lado derecho de [2.65] se anula porque al evaluar $\frac{1}{\sqrt{\alpha(r)}}$ en el horizonte de eventos se hace cero (los términos entre corchetes son claramente finitos), y en consecuencia el lado izquierdo de [2.65] se anula en el horizonte de eventos:

$$\lim_{r \to R_s^+} P_r = 0 \qquad [2.66]$$

En otras palabras, no hay flujo de energía-momento $P_r$ en la superficie del horizonte de eventos de Schwarzschild, y se ha mostrado que los agujeros negros son cerrados: la energía no puede penetrar el horizonte de eventos (Ma & Wang, 2014, págs. 43,44).

**2.6.5. Las cuatro leyes mecánicas de los agujeros negros**

El área del horizonte de eventos, para un agujero negro de Kerr-Newman, se calcula con la siguiente integral, usando además $M_{irr}^2$ en las ecuaciones [2.61]:

$$A = \int_0^\pi \int_0^{2\pi} \sqrt{g_{\theta\theta} g_{\phi\phi}}\Big|_{r=r_+} d\theta d\phi = 4\pi(r_+^2 + a^2) = 8\pi M_{irr}^2$$

$$A = 4\pi \left(2M^2 + 2M\sqrt{M^2 - Q^2 - a^2} - Q^2\right)$$

[2.67]

Larry Smarr propuso expresar la masa-energía como una función del área, la carga y el momento angular, primero se despeja $M$ de $A$ en [2.67]:

$$M = \left[\frac{A}{16\pi} + \frac{4\pi L^2}{A} + \frac{Q^2}{2} + \pi \frac{Q^4}{A}\right]^{1/2}$$

Diferenciando $dM$ se obtiene la fórmula de Smarr (Smarr, 1973, págs. 1,2):

$$dM = TdA + \Omega dL + \Phi dQ$$

$$T = \frac{\partial M}{\partial A} = \frac{1}{M}\left[\frac{1}{32\pi} - \frac{2\pi L^2}{A^2} - \frac{\pi Q^4}{2A^2}\right] \quad \Omega = \frac{\partial M}{\partial L} = \frac{4\pi L}{MA} \quad \Phi = \frac{\partial M}{\partial Q} = \frac{1}{M}\left[\frac{Q}{2} + \frac{2\pi Q^3}{A}\right]$$

[2.68]

En [2.68] el diferencial total describe una suma de 3 componentes de la energía de tensión superficial, rotacional y electrostática:

> *"Es visto también que si T es interpretado como la tensión superficial, entonces $M_{ir}$ sería interpretado como la energía superficial de un agujero negro. (…) La interpretación de T como una tensión superficial y de Ω como una velocidad angular sugiere una comparación de la relatividad general de los agujeros negros rotacionales con las Newtonianas gotas líquidas rotacionales"* (Smarr, 1973, pág. 2).

J. Bardeen, B. Carter y S. Hawking reinterpretaron las ecuaciones [2.68] y lo presentaron en el marco de las 4 leyes mecánicas de los agujeros negros:

> *"**La segunda ley.** El área A del horizonte de eventos de cada agujero negro nunca disminuye con el tiempo: $\delta A \geq 0$ ; si dos agujeros negros colapsan, el área del horizonte de eventos final es más grande que la suma de las áreas de los horizontes iniciales (…) no se puede transferir área de un agujero negro a* *otro desde que los agujeros negros no se pueden bifurcar. (…)* **La primera ley**. $\delta M = \frac{\kappa}{8\pi} \delta A + \Omega_H \delta J_H$ *puede ser visto que $\frac{\kappa}{8\pi}$ es análogo a la temperatura en el mismo modo que A es análogo a la entropía. (…)* **La ley cero**. *La gravedad superficial, κ, de un agujero negro estacionario es constante en todo el horizonte de eventos. (…)* **La tercera ley**. *Es imposible, por cualquier procedimiento, reducir κ a cero por una secuencia finita de operaciones"* (Bardeen, Carter, & Hawking, 1973, págs. 7,8,9).

Es importante resaltar que estas leyes son puramente mecánicas deducidas en la relatividad general, aunque asumiendo la ortodoxia que los agujeros negros son abiertos, de ahí su referencia a la termodinámica y ruptura en la simetría del tiempo.

### 2.6.6. La termodinámica de los agujeros negros

La termodinámica de los agujeros negros parte de asumir las existencias de procesos irreversibles como el ingreso de masa-energía a sus interiores, es deseable que ellos tengan una entropía que aumente con el ingreso de materia (para no desaparecer la entropía de los objetos que caen en él), y también que tengan una temperatura no-nula[50], con la correspondiente radiación térmica que consuma su masa-energía, pues sin ella se conduciría a una violación de la segunda ley de la termodinámica (un flujo de energía y partículas del cuerpo frio al cuerpo caliente):

> *"Sin tal emisión, la segunda ley generalizada sería violada por, por ejemplo, un agujero negro inmerso en una radiación de cuerpo negro a una temperatura inferior que el agujero negro"* (Hawking, 1975, pág. 5).

Esta radiación termal son partículas que se crean en el horizonte de eventos:

> *"Es visto que el campo gravitacional de un agujero negro creará partículas virtuales y las emitirá al infinito tal como uno espectaría si el agujero negro fuera un cuerpo ordinario a temperatura κ/2π, donde κ es la gravedad superficial del agujero negro"* (Hawking, 1975, pág. 3)

---

[50] Si tiene entropía entonces debe tener también una temperatura no-nula, pues así $\delta Q = TdS \neq 0$, si $T = 0$ entonces $\delta Q = 0$ y no entraría energía al interior (en el sentido fenomenológico).

El mecanismo de esa emisión ha sido teorizado por Stephen Hawking[51] y llamada *radiación de Hawking*, que no es detallado aquí; esta radiación es muy débil[52] y "evapora" al agujero negro[53], hasta su eventual desaparición[54].

La entropía y temperatura de un agujero negro son ($c = G = \hbar = k_B = 1$):

$$S_{AN} = \frac{A}{4} \qquad\qquad T_{AN} = \kappa/2\pi \qquad [2.69]$$

A este nivel, se considera la entropía generalizada para todo el universo $S_U$:

$$S_U = S + \frac{1}{4}\sum_i A_i \qquad [2.70]$$

Donde $S$ es la entropía convencional fuera de todos los agujeros negros y $\sum_i A_i/4$ es la entropía de todos los agujeros negros ($A_i$ es el área del horizonte de eventos del i-ésimo agujero negro, ecuación [2.67]); así, si el área disminuye $-\Delta a$ por evaporación, entonces $S$ aumenta al menos en $\Delta a/4$, y viceversa (si disminuye la entropía convencional entonces aumenta aún más el correspondiente área[55]), tanto el ingreso de materia al agujero negro como la evaporación son procesos irreversibles y deben aumentar la entropía de todo el universo $\delta S_U > 0$.

La Tabla N° 2.1 compara las 4 leyes mecánicas de los agujeros negros y de la termodinámica no-relativista; sin embargo, aunque ya se tiene el significado físico para la temperatura del agujero negro (la radiación de Hawking), no hay el significado físico para la entropía de un agujero negro.

---

[51] La idea fue sugerida a Hawking por los físicos rusos Yákov Zeldóvich y Alekséi Starobinski.
[52] Apreciable sólo en agujeros negros muy pequeños, o para escalas cósmicas de tiempo.
[53] El agujero negro seguiría absorbiendo e interactuando con la radiación termal del fondo cósmico, sólo cuando el universo entero se enfríe lo suficiente para que los agujeros negros supermasivos que queden sean más calientes, conseguirán evaporarse de manera neta.
[54] Esto conduce a la paradoja de la pérdida de la información en los agujeros negros, ver Anexo A.2
[55] Un objeto con masa $m$, carga $q$ y momento angular $l$ porta una entropía menor o igual a si fuera un agujero negro con esos parámetros, con área $a$; luego el aumento del área será mayor que $a$.

TABLA N° 2.1

COMPARACIÓN ENTRE LAS LEYES DE LOS AGUJEROS NEGROS Y
LOS DE LA TERMODINÁMICA

| Ley | Agujeros Negros | Termodinámica |
|---|---|---|
| Principio cero | En el equilibrio los agujeros negros tienen una gravedad superficial $k$ constante. | En el equilibrio termodinámico los cuerpos tienen la misma temperatura. |
| Primer principio | Conservación de la energía: $c^2 dm = \dfrac{c^2}{8\pi G} k dA + \Omega dL + V dQ$ | Conservación de la energía: $dU = \delta Q - \delta W$ |
| Segundo principio | En todo proceso irreversible el área del agujero negro siempre aumenta: $\dot{A} > 0$ | En todo proceso irreversible la entropía de un sistema aislado aumenta: $\dot{S} > 0$ |
| Tercer principio | Es imposible obtener $k = 0$ por un proceso físico. | Es imposible obtener $T = 0$ por un proceso físico finito. |

Fuente: Elaboración propia.

# CAPÍTULO III

# VARIABLES E HIPÓTESIS

### 3.1. Variables de la investigación

En esta tesis se considera la emergencia del mundo clásico a partir del mundo cuántico, de esta manera las variables independientes son las propiedades y leyes de los sistemas microscópicos tanto en su evolución (ya sea unitaria o mediante el colapso) como en sus interacciones (absorción y emisión de fotones), mientras que las variables dependientes son las propiedades y leyes emergentes de los sistemas clásicos-macroscópicos, ya sean las ecuaciones reversibles de la mecánica clásica, como la irreversibilidad termodinámica.

#### 3.1.1. Variables independientes

Mecánica cuántica: De la evolución unitaria, colapso del estado cuántico, entrelazamiento cuántico, perturbación dependiente del tiempo, emisión y absorción estimulada.

#### 3.1.2. Variables dependientes

Mecánica clásica y el principio del incremento de la entropía.

### 3.2. Operacionalización de variables

La operacionalización de las variables es presentada en la Tabla N° 3.1:

TABLA N° 3.1

OPERACIONALIZACIÓN DE VARIABLES

| VARIABLE | DIMENSIONES | INDICADORES |
|---|---|---|
| VARIABLE INDEPENDIENTE: MECÁNICA CUÁNTICA | Evolución temporal de la mecánica cuántica | Proceso unitario |
| | | Colapso o proceso R |
| | Interacciones mecano-cuánticas | Entrelazamiento cuántico |
| | | Perturbaciones dependientes del tiempo |
| | | Emisión y absorción estimulada |
| VARIABLE DEPENDIENTE: MECÁNICA CLÁSICA Y EL PRINCIPIO DEL INCREMENTO DE LA ENTROPÍA | Evolución temporal de la mecánica clásica | Ecuaciones canónicas de Hamilton |
| | | Determinismo y reversibilidad |
| | | Teorema de Liouville |
| | Evolución temporal de la termodinámica | Principio del incremento de la entropía |
| | | Irreversibilidad y flecha termodinámica del tiempo |

Fuente: Elaboración propia.

### 3.3. Formulación de la hipótesis

#### 3.3.1. Hipótesis general

**Los problemas de la medida y la clasicalización se resuelven y explican en un programa de colapso objetivo donde no se conserva la información.**

#### 3.3.2. Hipótesis específicas

1. Cada teoría (cuántica y clásica) funciona bien en su propio dominio, de manera que no se extiende una al dominio de la otra, la clasicalización y la observación rigen en la intercepción entre ellas.

2. El universo como un sistema aislado no evoluciona según la ecuación de Schrödinger porque es irreversible termodinámicamente, su irreversibilidad está cuantizada o discretizada, y son los procesos R en sus microsistemas.

3. Los objetos del mundo clásico no se encuentran en superposición de estados porque son observados por su entorno, y actúan como observadores de los demás sistemas; es decir, el mundo clásico es un mundo de observadores.

4. El mecanismo de observación ocurre porque el observador (objeto clásico) se enreda con el microsistema y le extiende sus colapsos.

5. La irreversibilidad termodinámica se origina al nivel fundamental en los colapsos de estados (de sus microsistemas) por ser procesos irreversibles.

6. La clasicalización y el problema de la medida no se pueden resolver en el marco de la decoherencia (u otro de colapso no objetivo) porque no se puede explicar la irreversibilidad del nivel clásico en base a la evolución unitaria (que es reversible); la decoherencia explica la pérdida de la coherencia, pero no la reducción del estado cuántico (la superposición de estados).

7. Los fenómenos cuánticos, tales como el entrelazamiento, no violan la localidad relativista porque en el colapso no hay transmisión de información superlumínica.

8. Los colapsos objetivos son responsables de la termodinámica de los agujeros negros, como extensión de la clasicalización a los agujeros negros.

# CAPÍTULO IV
# METODOLOGÍA

### 4.1. Tipo de investigación

Esta investigación fue de tipo básica al buscar el conocimiento científico que permite una mejor compresión de los sistemas microscópicos (especialmente, el rol del colapso), de cómo se transita y emerge al mundo clásico, y de cómo un microsistema interactúa con los objetos del mundo clásico mediante la observación; esta investigación también fue teórica, pues no hay recolección de datos empíricos; así, el desarrollo de este trabajo va desde mejorar la comprensión de la dinámica de los microsistemas, hasta tener aplicación directa en problemas como *la paradoja de la información*, de gravedad cuántica.

### 4.2. Diseño de la investigación

La hipótesis (general) a demostrar consta de la siguiente estructura:

$$A \text{ y } B \text{ se resuelven en } (X, a)$$

Donde:

$A$: Problema de clasicalización

$B$: Problema de la medida

$X$: Programa que resulta de reinterpretar la mecánica cuántica, con colapso objetivo

$a$: No conservación de la información

Primero se construyó un programa $(X, a)$ de la mecánica cuántica donde el colapso es un proceso objetivo en el que no se conserva la información, y la hipótesis se demostró al resolver $A$ y $B$ (el problema de la medida y la clasicalización) usando $(X, a)$; esta demostración de la hipótesis general no fue trivial porque los problemas $A$ y $B$ no tienen una respuesta física adecuada y son tratados con generalidades y muy vagamente; en esta investigación se superó las contradicciones teóricas descritas en la sección [2.5] que impedían resolver $A$ y $B$; para ser preciso, se demostró las ecuaciones canónicas de Hamilton y el segundo principio de la termodinámica a partir de la ecuación de Schrödinger y la regla de Born, para esto se hizo uso del teorema de Ehrenfest (ver sección [2.2.2]) y una teoría de la información que he desarrollado para este fin (Anexo B); además, he obtenido el *teorema de la clasicalización*, que generaliza y resume este programa.

En el programa $(X, a)$ se emplea una *evolución alternada de procesos U y* R, que resulta de una asunción que se introdujo (y cuya demostración excedería los límites de esta investigación), y trata sobre la repentina activación de los procesos R en un grupo de muchos microsistemas, favorecida por ciertas condiciones energéticas; otro principio que se introdujo, el cual si es deducible desde la *teoría de la información*, es que la entropía se conserva en una evolución determinista; el programa $(X, a)$ pudo explicar el problema de la flecha del tiempo termodinámica (una ruptura de la simetría temporal) como parte del problema de la clasicalización; la gravitación se abordó en esta investigación: he sustentado que la entropía de un agujero negro debe conservarse en una teoría determinista y reversible, como la relatividad general o el proceso U, esto es contradictorio con la termodinámica de los agujeros negros, así la extensión del programa $(X, a)$ a los agujeros negros planteó solución también a este problema incluso en ausencia de una teoría de gravedad cuántica[56], así también se dio respuesta a la paradoja de la pérdida de información en los agujeros negros.

### 4.3. Población y muestra
No aplica.
### 4.4. Técnicas e instrumentos de recolección de datos
No aplica.
### 4.5. Procedimientos de recolección de datos
No aplica.
### 4.6. Procesamiento estadístico y análisis de datos
No aplica.

---

[56] Como se carece de una teoría de gravedad cuántica, la solución de este problema fue en forma de conjetura física, no obstante, fue concluyente (como un teorema) desde la teoría de la información.

# CAPÍTULO V

# RESULTADOS

## 5.1. Los límites de validez de la interpretación de Copenhague

La interpretación oficial de la mecánica cuántica, la de Copenhague, defiende la unitaridad en la evolución de cualquier sistema físico aislado, incluso de sistemas clásicos (con observadores dentro), en consecuencia la ecuación de Schrödinger sigue siendo válido al nivel clásico; esto es rechazado en esta investigación, sustentando que al dominio clásico la unitaridad de la mecánica cuántica deja de ser válida; en esta investigación se ha introducido una nueva interpretación de colapso objetivo para construir el *programa de la clasicalización*; en la Tabla N° 5.1 se presenta una comparación entre el mundo cuántico y clásico:

TABLA N° 5.1

COMPARACIÓN ENTRE EL MUNDO CUÁNTICO Y CLÁSICO

|  | Mundo cuántico | Mundo clásico |
|---|---|---|
| Estado Físico | Es una superposición coherente de estados (vector en un espacio de Hilbert). | Es un estado único y bien definido, no hay superposición de estados (vector en el espacio de fases). |
| Evolución en el tiempo | Mientras no se realice la medida, la evolución es unitaria, determinista, reversible, y simétrica en el tiempo. | Evolución determinista y reversible en la mecánica clásica, es asimétrica e irreversible en la termodinámica: existe la flecha del tiempo. |
| Rol del observador | Provoca el colapso, del estado del microsistema, en un autoestado del observable. | Ninguno |

FUENTE: Elaboración propia.

La paradoja del gato de Schrödinger[57] es una crítica a la interpretación oficial que expone sus límites en la interacción de los sistemas microscópicos con los objetos del mundo clásico; el empleo de este experimento mental es simbólico y representativo, existen versiones "más realistas", en la Figura N° 5.1 a) se muestra que a un objeto clásico en reposo se le asociaría un paquete de ondas con valor medio del momento lineal nulo ( $\langle \hat{P} \rangle = 0$ ), de manera que en un tiempo clásico (sin observar) el paquete de ondas se expandiría en todas direcciones (describiendo un objeto clásico en superposiciones hasta que se haga una medida sobre él), en b) un fotón y un espejo que puede reflejarlo en dos sentidos (y retroceder en sentido opuesto) se enredan cuánticamente, en ambas situaciones un objeto clásico se encuentra en superposición de estados mientras no es observado, que es una contradicción con la física clásica; el hecho que la constante de Planck sea muy pequeña a escala clásica no debe impedir que estos procesos ocurran.

FIGURA N° 5.1

VERSIONES REALISTAS DEL GATO DE SCHRÖDINGER

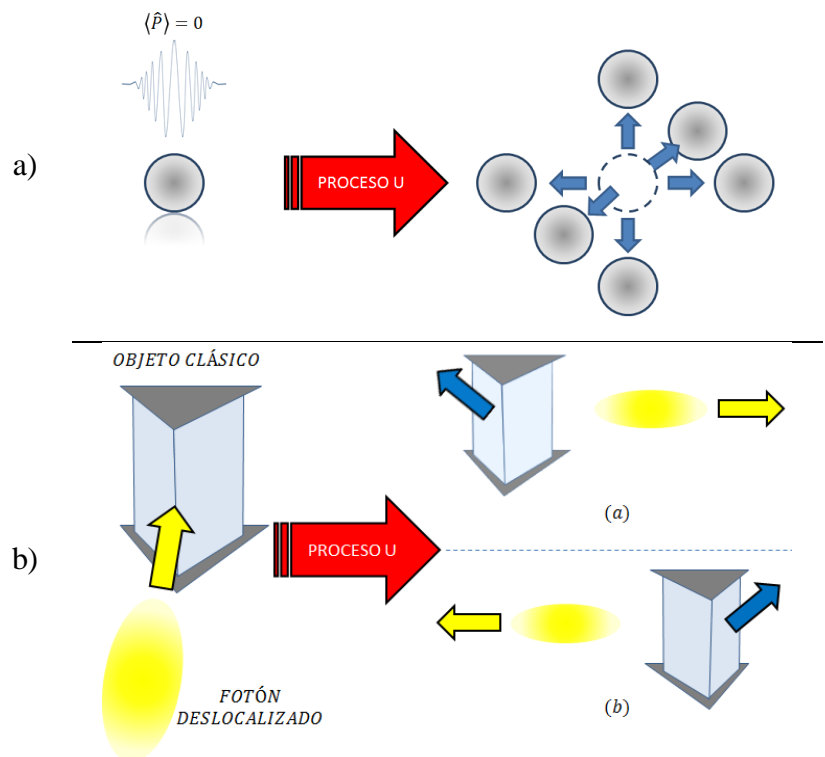

Fuente: Elaboración propia.

---

[57] Ver Anexo A.1 donde se presenta y describe la paradoja del gato de Schrödinger.

### 5.1.1. Superposición de sistemas clásicos

Cuando un microsistema se enreda con el sistema clásico ambos quedan en superposición de estados, esto es permitido para el microsistema, pero no lo es para el sistema clásico, este es el primer límite de la interpretación oficial: la mecánica cuántica deja de ser válida cuando coloca en superposición de estados a sistemas clásicos (típicamente son objetos macroscópicos), la interpretación oficial no tiene restricciones explícitas que impida esto.

### 5.1.2. Proceso U de un sistema clásico

La mecánica cuántica tiene una evolución reversible y simétrica en el tiempo, la coherencia cuántica permite oscilaciones e interferencias entre distintos estados energéticos (microsistemas "congelados"), en cambio los sistemas clásicos son irreversibles (envejecen, se desordenan, aumentan su entropía), una evolución unitaria de ambos sistemas enredados no es posible; la interpretación oficial deja de ser válida cuando pretende aplicar el proceso U a procesos irreversibles.

### 5.1.3. Proceso U en el acto de observación

En la interpretación oficial, si un aparato de medida y un microsistema se colocan en una caja aislada, el estado interno sería el enredo entre ambos sin avanzar irreversiblemente hacia el resultado de la medida, hasta que un segundo observador realice una medición (una versión similar al gato de Schrödinger[58]), así que se puede "apagar" el rol de un observador con respecto a otro, de manera que la realidad física puede ser una de una superposición de varias historias clásicas, todo esto es una contradicción con la física clásica[59]; la interpretación oficial deja de ser válida cuando se pretende aplicar el proceso U a los observadores (y a su mecanismo de observación).

---

[58] En la interpretación oficial se dice que un observador externo a la caja provoca el colapso del sistema enredado *átomo-gato*, el gato como sistema clásico es también un observador.
[59] Si lo que se pretende es tener una descripción unificada que permita la transición entre ambas.

## 5.2. Unificación de los problemas de la clasicalización y la observación, desplazamiento al problema del colapso

### 5.2.1. Límite clásico del enredo cuántico

En las secciones [5.1.1], [5.1.2] y [5.1.3] he mostrado que el entrelazamiento de un microsistema con un sistema clásico (como un observador) no es válida, pero no existe un límite o restricción a esta natural interacción (en el marco de la interpretación de Copenhague); en consecuencia, he introducido un *límite clásico al enredo cuántico* de un microsistema con un conjunto de muchos microsistemas que evolucionan unitariamente, de manera que se produce espontáneamente un colapso como una discontinuidad del proceso U, la introducción de este nuevo límite clásico resuelve las inconsistencias de la interpretación oficial aplicado a objetos del mundo clásico, como en la paradoja del gato de Schrödinger; este nuevo límite clásico se distingue de los precedentes en que no ocurre por un cambio de escala[60], sino que emerge al reunirse un conjunto de condiciones físicas para un ensemble dado, en consecuencia, todos los microsistemas que participan del enredo colapsan como un conjunto simultáneamente (el colapso es espontáneo, irreversible e indeterminista); este límite clásico es también un límite a la validez de la interpretación oficial, pues la unitaridad deja de ser válida a partir de este nivel, desde que el colapso es objetivo.

Es conveniente precisar que en esta investigación no he explicado el origen de los colapsos, sino que lo he postulado como un límite clásico (al entrelazamiento por el proceso U, establecido en la sección [2.1.3]), la aparición espontánea de este colapso no es determinada por las condiciones físicas, sino que estas condiciones favorecerían la probabilidad de su ocurrencia, y son presentadas en la sección [5.3.3]; los criterios para postular este límite clásico son, por un lado la participación de muchos microsistemas, de modo que cada microsistema individual o aislado puede conservar su evolución unitaria mientras no es observado (consistencia con la teoría cuántica), y por otro lado, la presencia de entrelazamiento entre ellos (debida al proceso U) exige la interacción de sus constituyentes.

---

[60] La acción de Planck, la indeterminación en las variables canónicas o la discontinuidad se hace despreciables a dimensiones clásicas

### 5.2.2. Evolución alternada de los procesos U y R

El límite clásico al enredo conduce a una dinámica microscópica que alternan los procesos U y R de forma aleatoria en el tiempo[61] (siguiendo una distribución de Poisson[62]), como se muestra en la Figura N° 5.2, y que al nivel macroscópico emerge una dinámica regular (suave y continua) que lleva al mundo clásico.

FIGURA N° 5.2

EVOLUCIÓN ALTERNADA ENTRE EL PROCESO U Y EL PROCESO R

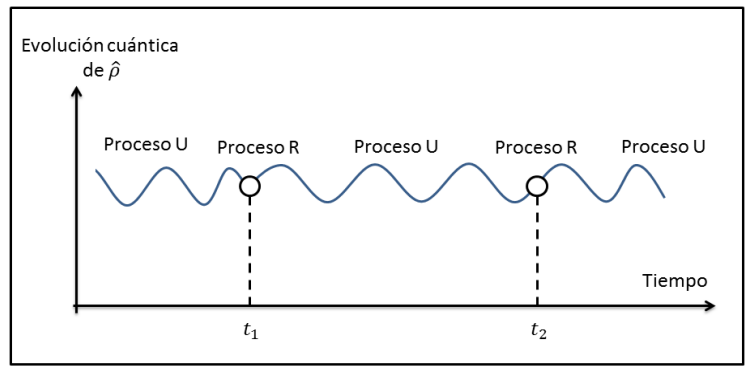

Fuente: Elaboración propia.

En la Figura N° 5.2, la curva ondulada representa el proceso U (ecuación de Schrödinger o de Liouville para $\hat{\rho}$) mientras que las discontinuidades en las curvas representan al proceso R que son saltos cuánticos a los tiempos $t_1$ y $t_2$, (regla de Born); los microsistemas pueden interactuar con los sistemas clásicos en el proceso U (interactuando directamente con sus microsistemas constituyentes), y en el proceso R se da la observación y la introducción de irreversibilidad en el tiempo al mundo clásico, la predominancia del proceso U o del proceso R determina al nivel macroscópico el tipo de clasicalización que se obtiene (mediante el *caos profundo*, sección [5.3.1]).

El mínimo elemento del mundo clásico, *la partícula clásica*, es el mínimo ensemble de microsistemas que pueden activar los procesos R y establecer la dinámica alternada (de forma regular y estable) el cual debe tener todos los atributos de los sistemas clásicos (mecánicos y térmicos), si las partículas clásicas están inmersas en una red de microsistemas (como en un sólido) entonces cada una es el grupo de microsistemas que activa su colapso (simultáneo entre sus microsistemas) independiente de los demás microsistemas, de modo que el límite espacial de las partículas clásicas es aleatorio y no sincronizado en el tiempo, colapsando desordenadamente en el tiempo y el espacio (distribución de Poisson). El tiempo entre dos colapsos consecutivos (para cada partícula clásica) debe seguir una distribución de caída exponencial[63] con un tiempo medio $\tau$ mucho menor que las escalas de tiempo macroscópicas, luciendo suaves y continuos a este nivel; los sistemas clásicos no pueden ser definidos por una función de onda y son mejor aproximados usando la mecánica clásica y la termodinámica.

---

[61] Las condiciones que favorecen la activación de los colapsos pueden ser fijas o estacionarias, como los procesos R se activan de forma aleatoria, da tiempo para el proceso U entre colapsos.

[62] La distribución de Poisson es dada en el anexo C.2; esta idea fue originalmente introducida por la teoría GRW (sección [2.2.3]), y es deseable que esta investigación sea consistente en esto con la teoría GRW; en la sección [5.3.3] es abordado.

### 5.2.3. Unificación en el problema del colapso

La clasicalización y el mecanismo de la observación son dos problemas que pueden ser tratados en un solo marco unificado[64], el cual es reducido a un enigma por resolver: el problema del origen del colapso, que no ha sido abordado en esta investigación; la clasicalización y la observación son dos manifestaciones de un mismo fenómeno (el colapso o proceso R), y pueden ser expresado uno en términos del otro de forma equivalente: la clasicalización puede ser entendida como resultado del "mundo de los observadores cuánticos", donde en principio todos los sistemas clásicos son observadores, de manera que el mundo clásico es así porque sus constituyentes se observan a sí mismos en todo el tiempo[65], encontrándose observables clásicos compatibles con los resultados de una medida; o de forma equivalente, la observación es resultado de que los sistemas clásicos extienden su clasicalización a los microsistemas llevándolos a un autoestado de un observable[66], que es un "estado clasicalizado" con respecto al observable, en ambos casos uno de los problemas se presenta como el fundamento (y la solución) del otro, lo que exhibe que ambos se encuentran en un mismo nivel de complejidad que se ha construido a partir del colapso.

---

[63] Esto es consecuencia de que los colapsos deban seguir una distribución de Poisson, anexo C.2.
[64] Esto es un postulado de esta interpretación de colapso objetivo; es posible que para otras interpretaciones de la mecánica cuántica esto no sea así.
[65] Lo que se experimenta en el mundo clásico son los resultados de esas observaciones, así que en todo el tiempo y a través de todo el espacio están ocurriendo los (muchos) colapsos.
[66] Como si el microsistema a observar fuera parte del aparato de medida que por ser clásico se está "definiendo" así mismo.

Este marco unificado es construido en base a la evolución alternada de los procesos U y R que se presentó en la sección [5.2.2], y se le ha dado el nombre de *programa de la clasicalización*[67] donde el conjunto de muchos microsistemas se vuelven objetos clásicos (donde los microsistemas están colapsando todo el tiempo), y la observación es un caso particular donde el microsistema a observar interactúa tan brevemente con una partícula clásica (en su proceso U) del aparato de medida que sólo colapsa una vez.

Como refuerzo, sea el estado de un microsistema $|\psi\rangle$, en la base $\{|a_n\rangle\}$ donde $\hat{A}|a_n\rangle = a_n|a_n\rangle$ y $\lambda_n = \langle a_n|\psi\rangle$, al realizar la medición y encontrarlo en el autoestado $|a_u\rangle$ se puede decir que el estado inicial ha sido "clasicalizado" con respecto al observable $\hat{A}$, no obstante, $|a_u\rangle$ es a su vez un nuevo estado cuántico en superposición de autoestados de algún otro observable $\hat{B}$ que no conmuta con $\hat{A}$ (entonces $|a_u\rangle$ no está clasicalizado para $\hat{B}$); la expresión "clasicalizado" adquiere mejor significado en un contexto estadístico: si a un ensemble de estados puros $\hat{\rho} = |\psi\rangle\langle\psi|$ se realiza la medición a todos sus microsistemas, se obtiene $\hat{\rho}'$, $p_n = |\lambda_n|^2$:

$$\hat{\rho}' = \sum_n p_n |a_n\rangle\langle a_n| \qquad [5.1]$$

Luego las observaciones han clasicalizado el ensemble original $\hat{\rho}$ (con respecto al observable $\hat{A}$); por ejemplo, si $\hat{A}$ es el observable de la posición (con una indeterminación compatible con la resolución del aparto de medida), entonces los vectores $|a_n\rangle$ son puntos más o menos localizados en el espacio mientras que los kets $|\psi\rangle$ pueden ser estados deslocalizados (como los orbitales electrónicos en un átomo), un ensemble descrito por $\hat{\rho}'$ puede ser considerado localizado mientras que $\hat{\rho}$ no; esta es la forma más elemental de clasicalización, al nivel macroscópico se observan todos sus observables y la clasicalización es completa (sus indeterminaciones y no conmutación son despreciables al nivel clásico).

---

[67] Y no algo así como "programa de la observación", porque se ha considerado que el enfoque de la clasicalización es más enriquecido.

### 5.3. La clasicalización

#### 5.3.1. Mecánica clásica y el *caos profundo*

La mecánica clásica es obtenida a partir de la predominancia del proceso U sobre el proceso R en la evolución alternada entre procesos U y R (sección [5.2.2]): el proceso U introduce el determinismo, la reversibilidad y la simetría en el tiempo al nivel clásico, mientras que el proceso R sólo evita que las indeterminaciones de las funciones de onda, sobre todo espaciales, sean apreciables e interfieran a escalas clásicas o macroscópicas; además, el proceso R canjea la información cuántica por la información clásica, pues el colapso destruye la superposición de estados, y con ello la información de sus fases relativas quedando sólo un estado clásico; no obstante, la mecánica clásica es siempre aproximada y falla cuando el régimen caótico, ver sección [2.3.3], se hace tan sensible a las condiciones iniciales que es afectado por los saltos cuánticos del proceso R que se vuelve predominante sobre el proceso U, a este régimen se le ha llamado *caos profundo*[68], aquí las trayectorias del espacio de fases divergen tanto (por un gran exponente de Lyapunov positivo, sección [2.3.3]) que los procesos R de los microsistemas se amplifican y expresan a escala clásica.

En esta sección se presenta la obtención de las ecuaciones de Hamilton (que sintetizan la mecánica clásica) a partir de la evolución alternada para una partícula clásica (que es el mínimo ente físico al que se le puede aplicar la mecánica clásica): sea un conjunto de partículas clásicas (definida en la sección [5.2.2]) que interaccionan entre sí y en el seno de un campo externo, todas estas interacciones son descritas por el hamiltoniano cuántico $\hat{H} = \hat{H}(\hat{q}_j, \hat{p}_j, t) = \hat{T}(\hat{p}_j) + \hat{V}(\hat{q}_j, t)$ donde $\hat{q}_j$ y $\hat{p}_j$ son los operadores asociados a las variables de las partículas clásicas (cada partícula clásica es tomada como un solo ente cuántico, mientras no hay colapso), en el proceso U el teorema de Ehrenfest (sección [2.2.2]) permite obtener la relación entre valores medios, sean los operadores de coordenadas y momentos generalizados en la representación de momentos y coordenadas respectivamente:

$$\hat{q}_j = i\hbar \frac{\partial}{\partial p_j} \qquad \hat{p}_j = \frac{\hbar}{i}\frac{\partial}{\partial q_j} \qquad [5.2]$$

---

[68] Porque es el régimen caótico donde se desciende a los fundamentos del nivel clásico, tan profundo que se da con los procesos R, que corrompen las ecuaciones clásicas deterministas y reversibles.

Al reemplazar [5.2] en la ecuación [2.34] (y considerando que los operadores $\hat{q}_j$ y $\hat{p}_j$ no dependen explícitamente del tiempo) se obtienen:

$$\frac{d\langle\hat{q}_j\rangle}{dt} = \langle\frac{\partial \hat{H}}{\partial p_j}\rangle \qquad \frac{d\langle\hat{p}_j\rangle}{dt} = -\langle\frac{\partial \hat{H}}{\partial q_j}\rangle \qquad [5.3]$$

Estas ecuaciones son válidas en el proceso U, y como se vio en la sección [2.2.2] estas ecuaciones se pueden corresponder a las ecuaciones canónicas de Hamilton (ecuaciones [2.38]) si se cumple que las dispersiones en sus paquetes de ondas se mantienen confinados a una pequeña región, de manera que luzcan localizados al nivel clásico[69] (y se pueda hacer $\langle\hat{q}_j\rangle \sim q_j{'}$ y $\langle\hat{p}_j\rangle \sim p_j{'}$, donde $q_j{'}$ y $p_j{'}$ son las variables clásicas de la coordenadas y el momento), y además que se cumpla $\langle \vec{\nabla}V(\hat{q}_j,t)\rangle \sim \vec{\nabla}V(\langle\hat{q}_j\rangle,t)$ (sección [2.2.2]), lo cual ocurre si la partícula está muy bien localizada como una delta de Dirac con respecto al potencial $V(q_j',t)$ (o equivalentemente, que $V(q_j',t)$ varíe apreciablemente sólo a escalas clásicas); es decir, que al nivel clásico la función de onda para cada partícula clásica no esté deslocalizada (como una versión realista del gato de Schrödinger) y que las interacciones entre ellas y el campo externo sea descrito por potenciales a escala clásica, el proceso R localiza a las funciones de ondas mencionadas, de manera que el tiempo medio entre colapsos $\tau$ (mucho menor que el tiempo de orden clásico) sea tan breve que no permita evolucionar unitariamente a una deslocalización a escala macroscópica, esto garantiza mantener la dispersión en sus paquetes de ondas confinados a una pequeña región, y que las fuerzas que participen de la dinámica clásica sean las que puedan representarse al nivel macroscópico (electromagnetismo y gravitación); con estas condiciones (ecuaciones [5.4]) se justifica que las ecuaciones [5.3] puedan ser identificadas o al menos representadas con las ecuaciones canónicas de Hamilton [2.38]:

$$\langle \hat{q}_j \rangle \sim q_j' \qquad \langle \hat{p}_j \rangle \sim p_j' \qquad \langle \vec{\nabla} V(\hat{q}_j, t) \rangle \sim \vec{\nabla} V(\langle \hat{q}_j \rangle, t)$$

$$\langle \frac{\partial \hat{H}}{\partial p_j} \rangle \sim \frac{\partial H}{\partial p_j'} \qquad \langle \frac{\partial \hat{H}}{\partial q_j} \rangle \sim \frac{\partial H}{\partial q_j'} \qquad \langle \frac{\partial \hat{V}}{\partial q_j} \rangle \sim \frac{\partial V}{\partial q_j'}$$

---

[69] La distribución espacial de su función de onda debe ser como el de una delta de Dirac.

[5.4]

$\hat{H}$ y $\hat{V}$ son el Hamiltoniano y el potencial cuántico, mientras que $H$ y $V$ son sus correspondientes clásicos; las ecuaciones [5.4] nos dice que son válidas de manera aproximadas, los procesos R para cada partícula clásica crean nuevos estados con otros valores medios de coordenadas y momentos $\langle \hat{q}_j \rangle$ y $\langle \hat{p}_j \rangle$, que evolucionan según [5.3] en el proceso U y saltan cuánticamente a nuevos valores de $\langle \hat{q}_j \rangle$ y $\langle \hat{p}_j \rangle$ en los procesos R, e introducen desviaciones en la trayectoria como en una *caminata aleatoria* (ver anexo C.1) en las trayectorias del espacio de fases:

FIGURA N° 5.3

EVOLUCIÓN DE UNA PARTÍCULA CLÁSICA

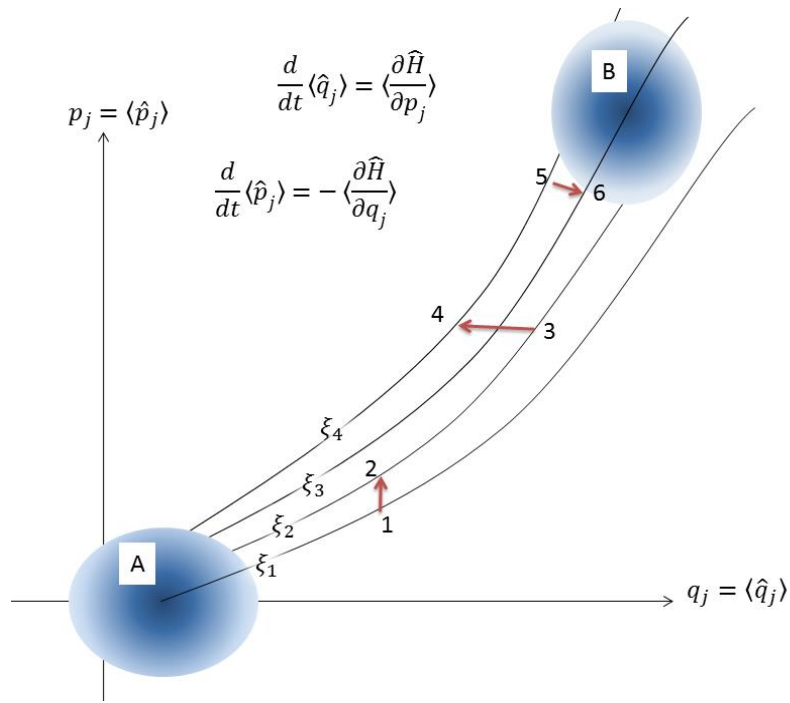

Fuente: Elaboración propia.

En la Figura N° 5.3 se ha exagerado la deslocalización de la función de onda (para la partícula clásica) así como de sus trayectorias $\xi_i$, ($i = 1, 2, 3, 4$) que son soluciones de las ecuaciones [5.3], en el proceso U recorre cada curva $\xi_i$ en la que se encuentre de manera determinista, mientras que en el proceso R saltaría entre ellas aleatoriamente (flechas rojas); si el exponente de Lyapunov para las curvas $\xi_i$ es muy grande, se llega al caos profundo, y deja de cumplirse la mecánica clásica, para dar lugar a la termodinámica.

**5.3.2. Irreversibilidad e indeterminismo en la mecánica estadística**

La predominancia del proceso R sobre el proceso U en la evolución alternada introduce indeterminismo e irreversibilidad al nivel clásico; en esta sección se explica el aumento de entropía debido única y exclusivamente al proceso R, haciendo uso de la *teoría física de la información* que se ha presentado en el anexo B.

*Conservación de la entropía.*

Sea un sistema clásico aislado, de muchas partículas fuera del equilibrio térmico, su evolución mecánica está determinado por las ecuaciones de Hamilton[70] [2.38] y conserva la entropía, como se estableció en la sección [2.5.2] por criterios de simetría y reversibilidad; una demostración directa se obtiene al derivar en el tiempo la entropía estadística, establecida en las ecuaciones [2.54] y [2.53], y considerar el teorema de Liouville (ecuación [2.42]), como se muestra:

$$S(t) = -k_B \int_{\xi(X)} \ln \rho(X,t) \cdot \rho(X,t) d\Gamma \qquad \frac{d}{dt}S(t) = -k_B \int_{\xi(X)} \frac{d\rho}{dt}[1 + \ln \rho] d\Gamma$$

con $d\Gamma = \prod_{k=1}^{2N} dx_k$, y del teorema de Liouville $\frac{d\rho}{dt} = 0$, se anula la integral definida y se obtiene la conservación de la entropía estadística:

$$\frac{d}{dt}S(t) = 0 \qquad [5.55]$$

---

[70] Que no han sido resueltas analíticamente debido al gran número de grados de libertad, pero aun así se puede demostrar que conserva la entropía en el tiempo.

Desde la teoría de la información se sustenta la conservación de la entropía debido al determinismo de la mecánica clásica, las ecuaciones [B.12], Anexo B, establecen que se conserva la información $Y$ cuya medida, según la sección [B.4], es proporcional a la entropía estadística, como se indica:

$$|Y| = -\ln[P] = \ln(N) \qquad\qquad S = k_B|Y| = k_B\ln(N) \qquad [5.56]$$

Donde $N = 1/P$ es el número de microestados independientes que tiene el sistema al inicio del tiempo[71], $|Y|$ y $N$ son independientes del tiempo y conservan $S$:

$$\frac{dS}{dt} = k_B \frac{d}{dt}|Y| = 0 \qquad [5.57]$$

Las ecuaciones [5.55] y [5.57] son equivalentes, y su interpretación es que se conserva el número de microestados desconocidos[72], para comprender mejor esto se hace uso de un concepto introducido: *los microestados prohibidos*.

*Microestados prohibidos*

Si a un tiempo inicial dado, la ignorancia en la información del microestado del sistema se reduce a $N$ posibles microestados independientes, con entropía $k_B\ln(N)$, su evolución mecánica-determinista preserva la información e ignorancia del estado inicial[73]; si el sistema evoluciona al macroestado al que (en el equilibrio) le corresponde $N' \gg N$ microestados, los $N'$ microestados presentan correlación o dependencia entre ellos de manera que sólo tengan $N$ combinaciones independientes, así la conservación de la entropía es debido a la existencia de estas correlaciones que corrige el cálculo de la probabilidad[74] a ser usado en [5.56], habiendo $N' - N$ microestados prohibidos a cada instante del tiempo.

---

[71] La información de un sistema se mide como lo indica [5.56], y la probabilidad se puede interpretar como conocer, a un tiempo dado, el microestado que ocupa de un conjunto de N=1/P microestados posibles, todos ellos independientes, esta medida es equivalente a la entropía física.

[72] En la mecánica estadística se conoce el macroestado y se desconoce el microestado exacto que tiene el sistema, así a cada macroestado le corresponde un conjunto de microestados (a los cuales se les asigna una probabilidad), la entropía viene a ser una medida de la dispersión (en la distribución de la probabilidad) del conjunto de microestados (a los que se les ha asignado una probabilidad).

[73] Ver la conservación de la información, Anexo [B.3.1].

[74] Así en vez de usar $S = k\ln(N')$ se sigue empleando $S = k\ln(N)$, pues las ecuaciones [5.56] sólo se aplican a variables sin correlación entre sí.

FIGURA N° 5.4

MICROESTADOS PROHIBIDOS DEBIDO AL TEOREMA DE LIOUVILLE

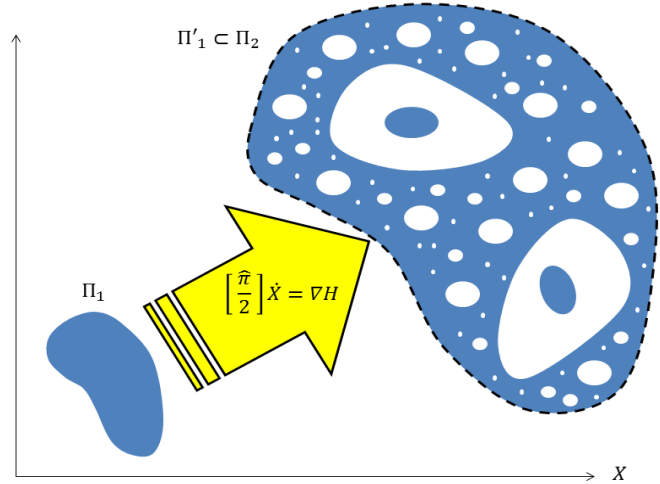

Fuente: Elaboración propia.

En la Figura N° 5.4 $\Pi_1$ y $\Pi_2$ corresponden a los macroestados inicial y final respectivamente, el subconjunto $\Pi'_1 \subset \Pi_2$ (la región en azul) resulta de la evolución determinista de cada punto $X \in \Pi_1$ a $X' \in \Pi'_1$, de manera que $|\Pi_2| \gg |\Pi_1|$ pero $|\Pi'_1| = |\Pi_1|$ (el teorema de Liouville); las áreas blancas que pertenecen a $\Pi_2$ representan los microestados prohibidos para el sistema a un instante dado.

### *Entropía mutua.*

La información sobre las correlaciones (o microestados prohibidos) es tratado en términos de la información mutua (o entropía mutua), sección [B.1.2]; sea el sistema compuesto por dos subsistemas $A$ y $B$, representados por las variables $X_A$ y $X_B$, que interactúan en $t \in \langle t_0; t_1 \rangle$ provocando el pase del microestado del sistema de $\Pi_1$ a $\Pi'_1 \subset \Pi_2$, antes de la interacción sus variables son independientes y no compartían ninguna información en común, así el balance de entropía al tiempo $t_0$:

$$S_{A,B}(t_0) = S_A(t_0) + S_B(t_0) \qquad [5.58]$$

Tras la interacción, al tiempo $t_1$ las variables $X_A$ y $X_B$ quedan correlacionadas y no todas sus combinaciones son posibles (así la variable del sistema $X$ no se encuentra en los microestados prohibidos al tiempo $t_1$), y es medida con la información o entropía mutua $S^M_{A,B}(t_1) > 0$ como se muestra en la Figura N° 5.5, así la entropía del sistema general viene expresado por:

$$S_{A,B}(t_1) = S_A(t_1) + S_B(t_1) - S^M_{A,B}(t_1) \qquad [5.58]$$

FIGURA N° 5.5

ENTROPÍA O INFORMACIÓN MUTUA

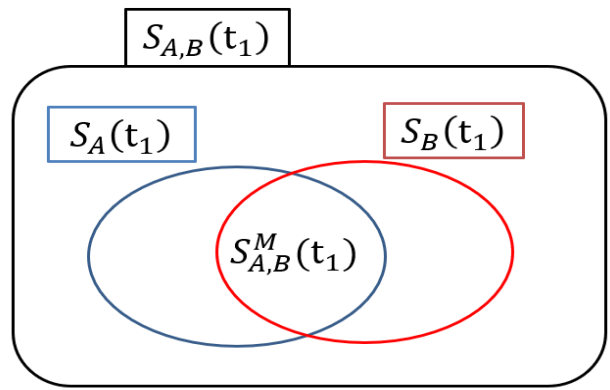

Fuente: Elaboración propia.

Como la entropía $S_{A,B}$ del sistema general se conserva (ecuación [5.57]):

$$S_{A,B}(t_1) = S_{A,B}(t_0)$$
$$\qquad [5.59]$$
$$S_A(t_1) + S_B(t_1) = S_A(t_0) + S_B(t_0) + S^M_{A,B}(t_1)$$

Desde que $S^M_{A,B}(t_1) > 0$ la suma de las entropías de $A$ y $B$ por separado aumentan ($S_A(t_1) + S_B(t_1) > S_A(t_0) + S_B(t_0)$), y la conservación de la entropía se explica con la entropía mutua[75], este fenómeno no puede conducir a una flecha del tiempo (el aumento de la entropía mutua con el paso del tiempo) porque es posible elegir otras variables $Y_A$, $Y_B$ al tiempo $t_1$ que sean independientes entre sí, entonces estarían correlacionados[76] al tiempo $t_0$, y se observaría un decremento de la entropía mutua con el paso del tiempo (en la misma cantidad), lo que es consistente con la reversibilidad de la mecánica clásica.

---

[75] Esto significa que en una evolución determinista de un sistema de muchas partículas sería tan complejo que podría parecer que se desordena el sistema total (y con ello, el aumento de su entropía), pero esto es compensado por la creación de correlaciones entre sus partes.

[76] La correlación aparece porque interaccionan "hacia atrás en el tiempo" con la misma mecánica.

*Información en el proceso R*

Al ocurrir el colapso se destruye el estado previo $|\psi\rangle$ y se crea un nuevo estado $|\varphi\rangle$ de manera indeterminista, desde la perspectiva de la información ocurre la pérdida y creación de información simultáneamente, la parte de la información del estado anterior $I(\psi)$ que sobrevive al colapso y se encuentra presente en la información del nuevo estado $I(\varphi)$ es la información mutua $I_{\psi,\varphi}^{M} = I(\psi) \cap I(\varphi)$:

FIGURA N° 5.6

INFORMACIÓN MUTUA EN EL COLAPSO

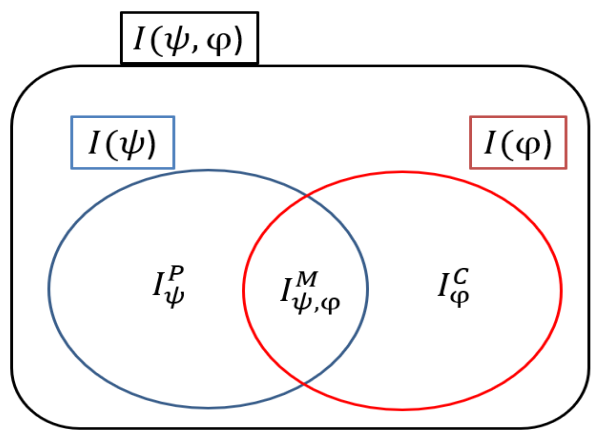

Fuente: Elaboración propia.

La información perdida y creada, $I_{\psi}^{P}$ y $I_{\varphi}^{C}$ respectivamente, tienen la misma medida[77], la cual sólo depende de su probabilidad $|\langle\varphi|\psi\rangle|^2$, como se muestra:

$$|I_{\varphi}^{C}| = |I_{\psi}^{P}| = -2\ln(|\langle\varphi|\psi\rangle|) \qquad [5.60]$$

La información cambia en el colapso, pero conserva su medida; la irreversibilidad de este proceso rompe la simetría temporal del proceso U[78].

---

[77] Debido a la simetría en la probabilidad del colapso (ver ecuación [2.8]).
[78] Realmente los colapsos no distinguen "futuro de pasado", sólo conectan estados aleatorios en la secuencia de los eventos causales, la flecha del tiempo aparece al poner una condición de frontera en un tiempo pasado $t_0$ (no infinito) de un estado de máximo orden; no obstante, toda la historia del universo puede admitir simetría temporal respecto a $t_0$, ver anexo E.

*Aumento de entropía por violación del teorema de Liouville debido al proceso R*

Cuando el proceso R se hace predominante sobre el proceso U introduce al nivel clásico indeterminismo e irreversibilidad, el aumento de la entropía se debe al proceso de creación y destrucción de información en los procesos R: en la sección [5.3.1] se mostró que las trayectorias de la mecánica clásica corresponden al valor medio de los observables de posición y momento del estado cuántico para toda la partícula clásica (en su evolución unitaria), así cada vez que ocurre el colapso (del estado general) de $|\psi\rangle$ a $|\varphi\rangle$ se salta de la trayectoria $\xi_\psi$ a $\xi_\varphi$ que son soluciones de las ecuaciones [5.3] para los estados $|\psi\rangle$ y $|\varphi\rangle$ respectivamente, la predominancia del proceso R sobre el proceso U se define entonces cuando las curvas $\xi_\psi$ y $\xi_\varphi$ son distinguibles en el espacio de fases al nivel clásico, esto ocurre en el caos profundo, puesto que el estado general $|\psi\rangle$ puede colapsar a uno de un conjunto de estados superpuestos $|\lambda\rangle, \lambda \in \Lambda$, cada uno con su correspondiente trayectoria $\xi_\lambda$ y probabilidad $|\langle\lambda|\psi\rangle|^2$ (están normalizadas sobre el conjunto $\lambda \in \Lambda$), esto crea una bifurcación en las trayectorias del espacio de fases porque cada trayectoria $\xi_\psi$ se ramifica en un conjunto de trayectorias $\xi_\lambda$, $\lambda \in \Lambda$, en el caso del ensemble (descrito al inicio de esta sección) la distribución de probabilidad $\rho(X,t)$ deja de cumplir con el teorema de Liouville: cada microestado al tiempo $t_0$ y con probabilidad $\rho_*$ evoluciona al tiempo $t_1$ a varios microestados bifurcados cuya suma de todas sus probabilidades $|\langle\lambda|\psi\rangle|^2$ es $\rho_*$, así pueden ocupar los microestados prohibidos y conseguir que la distribución de probabilidad $\rho(X,t_0)$ se vuelva más homogénea, difusa o dispersada, y la entropía estadística para esta nueva distribución de probabilidad $\rho'(X,t_1)$ aumenta[79]; la entropía mutua disminuye con el tiempo:

$$\frac{d}{dt}S^M_{A,B}(t) < 0 \qquad [5.61]$$

---

[79] Si la distribución de probabilidad es menos localizada, entonces la entropía estadística aumenta.

El proceso R disminuye la información o entropía mutua al poder ocupar microestados prohibidos, en la ecuación [5.58] para un tiempo dado $t$, si se considera al proceso R solamente en la entropía mutua[80]:

$$S_{A,B}(t) = S_A(t) + S_B(t) - S_{A,B}^M(t)$$

$$\frac{d}{dt}S_{A,B}(t) = \frac{d}{dt}S_A(t) + \frac{d}{dt}S_B(t) - \frac{d}{dt}S_{A,B}^M(t)$$

$$\frac{d}{dt}S_{A,B}(t) = -\frac{d}{dt}S_{A,B}^M(t) > 0 \qquad [5.62]$$

las ecuaciones [5.62] y [2.46] son equivalentes, la entropía generada es debido a la pérdida de la información mutua entre sus partes; en la sección [2.3.3] se estableció que en los sistemas Hamiltonianos no aparecen atractores ni repulsores debido al teorema de Liouville, al dejar de cumplirlo (puede ocupar los microestados prohibidos) se establecen nuevas dinámicas difusivas con un intrínseco sentido privilegiado del tiempo (como la ecuación del calor o las definiciones de las fuerzas de fricción[81]) y de procesos irreversibles.

La mecánica estadística necesita introducir el principio de equipartición de la energía[82] porque la mecánica clásica impide que un sistema llegue al equilibrio térmico (que aumente su entropía) y a la equipartición, en cambio se requiere del proceso R para aumentar la entropía y llegar a la equipartición[83]; es decir, añadir la equipartición a la mecánica clásica (al nivel estadístico) es equivalente a considerar el rol predominante del proceso R sobre el proceso U (caos profundo), y permite construir el desarrollo expuesto en la sección [2.4.2]; un sistema fuera del equilibrio ocupa una región $\Gamma^* \subset \Gamma(E) = \{x \epsilon \xi_X / H(X) = E\}$ del espacio de fases $\xi_X$, cuya evolución hacia el equilibrio aumenta la entropía (mediante la violación del teorema de Liouville) "saltando" sólo a los otros estados $x$ que conserven la energía total; es decir, $x \in \Gamma(E)$, hasta cubrir toda la región $\Gamma(E)$; el proceso R redistribuye la energía interna del sistema hasta llegar a la equipartición, sin redistribución de la energía tampoco hay procesos R activados y se cumple el teorema de Liouville.

---

[80] Esto es, que se conserva la entropía de cada subsistema por separado.
[81] Por ejemplo, la fuerza de fricción proporcional a la velocidad: $\vec{F} = -b\vec{v}$ con $b > 0$, rompe la simetría del tiempo en un sistema masa-resorte, e introduce una dinámica difusa e irreversible.
[82] Este principio establece que en el equilibrio térmico la energía se distribuye con igual probabilidad entre todos sus grados de libertad.
[83] En efecto, la entropía es máxima en el caso equiprobable (esto es, la equipartición).

### 5.3.3. El mundo macroscópico es clásico porque es térmico

La mecánica clásica permite hacer una descripción del mundo macroscópico desligado de los fenómenos térmicos[84], esto es sólo una abstracción, si se llegara a 0 K se exhibirían las propiedades cuánticas, se apagarían los colapsos[85] y desaparecería la clasicalización, el mundo clásico es intrínsecamente térmico[86]; la ecuación de Schrödinger no puede recrear la difusión de la energía interna[87] para un ensemble de microsistemas fuera del equilibrio, como si lo hace la ecuación del calor, la comparación de ambas ecuaciones revela su diferencia matemática:

$$i\hbar \frac{\partial \psi(\vec{r},t)}{\partial t} = \left[-\frac{\hbar^2 \nabla^2}{2m} + V(\vec{r},t)\right]\psi(\vec{r},t) \qquad [5.63]$$

$$\frac{\partial T(\vec{r},t)}{\partial t} = \frac{k}{c_p \rho} \nabla^2 T(\vec{r},t) \qquad [5.64]$$

En la Ecuación de Schrödinger [5.63] se tiene la unidad imaginaria en el término de la derivada parcial temporal, mientras que en la ecuación del calor [5.64] no lo hay, esto hace que las soluciones de la Ecuación de Schrödinger tengan términos de la forma $e^{-i\omega t}$ que le da reversibilidad en el tiempo, mientras que las soluciones de la ecuación del calor son del tipo caída exponencial $e^{-b^2 t}$ que le da un carácter difusivo, asintótico y por consiguiente irreversible con el paso del tiempo, definiendo una flecha del tiempo y el aumento de la entropía[88].

---

[84] Las partículas pueden moverse e interactuar como objetos carentes de temperatura y entropía.
[85] Los procesos R introducen la discontinuidad que no permiten "congelar" los macrosistemas en una evolución unitaria y reversible, sino en el irreversible mundo térmico.
[86] Ya que la redistribución de la energía es la que se correlaciona con la actividad de los procesos R, que son necesarios para la clasicalización, sin redistribución de la energía no hay mundo clásico.
[87] De la evolución de un ensemble aislado (conserva la energía) el operador de densidad se conserva.
[88] Si la Ecuación de Schrödinger no tuviera la unidad imaginaria describiría sistemas de difusión irreversible con el paso del tiempo, tal como se observa en el mundo macroscópico-térmico.

La conexión física entre la irreversibilidad de los colapsos y la del mundo clásico-macroscópico (la flecha termodinámica del tiempo) se encuentra en el intercambio de energía entre los microsistemas: a una temperatura no nula los microsistemas están en distintos estados de energía inmersos en un campo de radiación térmica (gas de fotones y fonones), los microsistemas excitados pueden relajarse emitiendo energía como partículas que se integran al campo de radiación térmica, de igual manera pueden adquirir energía del campo de radicación y excitarse, en la sección [2.1.6] se explica la absorción y emisión que fundamenta esta dinámica, el sólo hecho de intercambiar aleatoriamente la energía conlleva a los procesos irreversibles, y conduce al equilibrio termodinámico; no obstante, si se apagan los colapsos todos los microsistemas se encontrarían gobernados por la ecuación de Schrödinger[89] sin poder intercambiar (aleatoriamente) sus energías con el campo de radiación térmica, porque se requiere del proceso R para ocurrir la emisión y absorción de fotones y fonones, como se mostró en la sección [2.1.6]; así, la interacción unitaria (sin colapso) de un microsistema con un fotón que lo atraviesa modificando el Hamiltoniano del microsistema aislado $\widehat{H}$ añadiéndole un término perturbativo $\widehat{H}^I(t)$ que es simétrico en el tiempo (ecuación [2.27]), induce resonancia de acuerdo a la teoría de perturbaciones dependientes del tiempo (sección [2.1.5]) que lleva del estado de energía $E_n$ a los estados $E_{n+1}$ y $E_{n-1}$, como se muestra en la Figura N° 5.7:

---

[89] No es del todo apropiado recurrir sólo a la ecuación de Schrödinger para describir el sistema macroscópico completo porque se requiere considerar a los fotones y fonones que claramente no conservan el número de partículas; es más apropiado un tratamiento de campos cuánticos (sin colapsos), pero esto no afecta el propósito de la explicación.

FIGURA N° 5.7

HAMILTONIANO DEL MICROSISTEMA

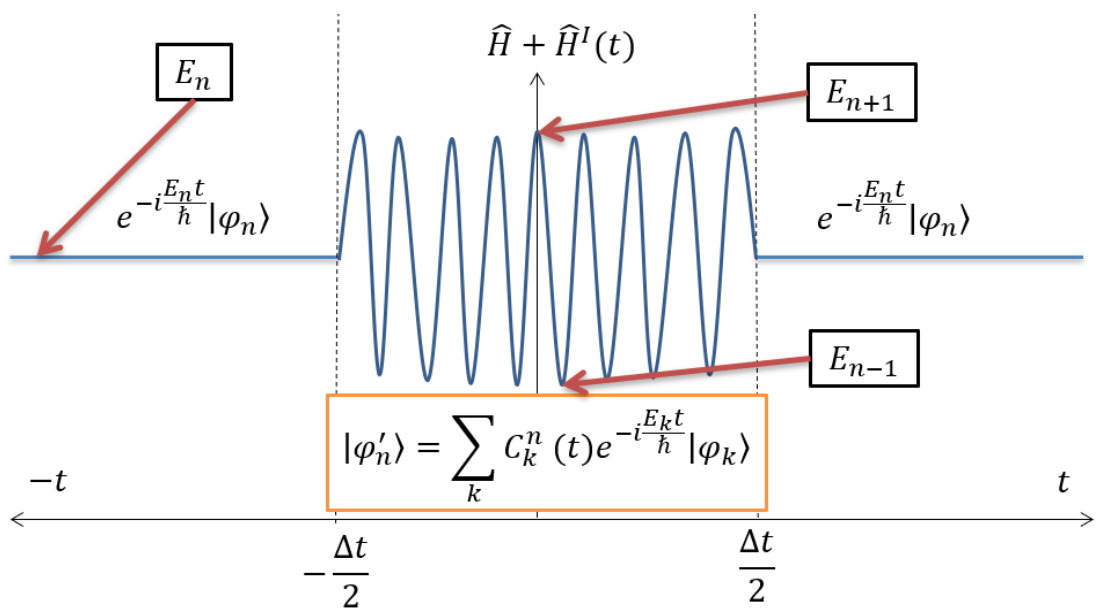

Fuente: Elaboración propia.

En la Figura N° 5.7 el fotón interactúa con el microsistema entre los tiempos $-\Delta t/2$ y $\Delta t/2$ de forma armónica, antes y después de la interacción (sin colapso) el microsistema se encuentra en el auto estado $|\varphi_n\rangle$ de $\hat{H}$ (sin perturbar), durante la interacción[90] su estado viene dado por la ecuación [5.65]:

$$|\varphi'_n(t)\rangle = \sum_k C_k^n(t) e^{-i\frac{E_k t}{\hbar}} |\varphi_k\rangle \qquad [5.65]$$

En consistencia con la ecuación [2.22] de la teoría de perturbaciones dependientes del tiempo; se aprecia la simetría temporal de la interacción debido al proceso U; así un ensemble de microsistemas en interacción unitaria de los fotones o fonones[91] que componen la energía térmica se enredan cuánticamente de muchas maneras[92] evolucionando desde un estado factorizable, ecuación [5.67], a uno entrelazado, ecuación [5.68]: sea $|\varphi_{n_k}(k)\rangle$ el estado del microsistema "$k$" que es autoestado de su Hamiltoniano $\widehat{H}^{(k)}$ como se muestra:

$$\widehat{H}^{(k)}|\varphi_{n_k}(k)\rangle = E_{n_k}^{(k)}|\varphi_{n_k}(k)\rangle \quad [5.66]$$

---

[90] En la interacción el microsistema y el fotón se enredan cuánticamente: el microsistema absorbe el fotón y se excita, o realiza la emisión estimulada, o permanecen independientes, este enredo se mantiene hasta que el fotón atraviesa completamente el microsistema.

[91] La interacción con los fonones es esencialmente la misma: las oscilaciones armónicas en la red cristalina (los fonones propagándose por la red) introducen términos armónicos perturbativos dependientes del tiempo en el Hamiltoniano, que requieren un tratamiento similar.

Donde $n_k$ es el número cuántico de la energía del microsistema "$k$", el estado general de todos los microsistemas viene dado por $|\Psi(t)_{n_1 n_2 \dots n_N}\rangle$, como se muestra:

$$|\Psi(t)_{n_1 n_2 \dots n_N}\rangle = \prod_k e^{-i\frac{E_{n_k} t}{\hbar}}|\varphi_{n_k}(k)\rangle \quad [5.67]$$

En efecto, cada combinación de los números cuánticos $n_1 n_2 \dots n_N$ establece una distribución de energía entre los microsistemas ($|\Psi(t)_{n_1 n_2 \dots n_N}\rangle$ no considera la energía del gas de fotones y fonones); en la interacción unitaria con el gas de fotones y fonones se evoluciona al estado general $|\Psi'(t)_{n_1 n_2 \dots n_N}\rangle$ que es la superposición de los posibles estados $|\Psi(t)_{l_1 l_2 \dots l_N}\rangle$ donde los números cuánticos $l_k$ "*suben o bajan*[93]" con respecto a $n_k$ de acuerdo a absorber o emitir un fotón, respectivamente, en consistencia con la ecuación [2.22]:

$$|\Psi'(t)_{n_1 n_2 \dots n_N}\rangle = \sum_{l_1 l_2 \dots l_N} C(t)_{n_1 n_2 \dots n_N}^{l_1 l_2 \dots l_N} |\Psi(t)_{l_1 l_2 \dots l_N}\rangle \quad [5.68]$$

Donde $C(t)_{n_1 n_2 \dots n_N}^{l_1 l_2 \dots l_N}$ es determinado según el procedimiento de la sección [2.1.5] donde $\widehat{H}^I(t)$ es la interacción con los fotones y fonones; las ecuaciones [5.68] y [5.67] establecen un bucle si se pone en acción el proceso R en todo el estado cuántico del ensemble de microsistemas, como una *partícula clásica*:

---

[92] Cada fotón deslocalizado sobre muchos microsistemas tiene una probabilidad $P_j$ de ser absorbida por un microsistema "$j$" y no por el resto, así el estado general considera la superposición de cada posible absorción o emisión estimulada entre cada fotón incidente (del campo de radiación) permutado entre los microsistemas.

[93] Formalmente se emplean los "*operadores escalera*" para hacer esto.

FIGURA N° 5.8

BUCLE EN LA DINÁMICA DE LOS ESTADOS DE LOS MICROSISTEMAS

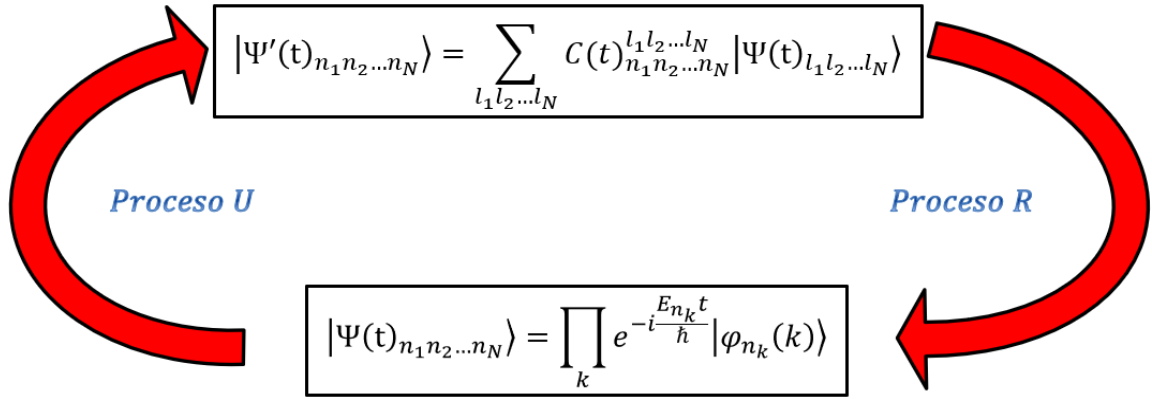

Fuente: Elaboración propia.

En la Figura N° 5.8: el estado factorizable $|\Psi(t)_{n_1 n_2 \ldots n_N}\rangle$ evoluciona unitariamente a un estado entrelazado $|\Psi'(t)_{n_1 n_2 \ldots n_N}\rangle$ (entre los microsistemas), al ocurrir el proceso R regresa a un nuevo estado $|\Psi(t)_{n'_1 n'_2 \ldots n'_N}\rangle$ cerrando el bucle y modificando aleatoriamente los números cuánticos $n_1 n_2 \ldots n_N$ en cada ciclo, redistribuyendo la energía interna hasta alcanzar el equilibrio termodinámico o la equipartición (como se dedujo al final de la sección [5.3.2]); en el Anexo D se muestra una simulación computacional donde al intercambiar aleatoriamente las energías entre partículas clásicas (que al inicio tenían una distribución arbitraria) convergen a la distribución clásica de Maxwell-Boltzmann, este bucle permite a un estado arbitrario llegar al equilibrio, y explicar así el aumento de la entropía, cada proceso clásico-térmico elemental (intercambiar aleatoriamente la energía entre partículas clásicas) se encuentra en el proceso R[94].

Como consecuencia directa, todo proceso de intercambio de energía (ya sea en una reacción química, colisión inelástica o fricción) es irreversible[95], y requiere del proceso R, y es también indeterminista; para el flujo de calor $\delta Q$ en un sistema:

---

[94] Entre dos colapsos consecutivos sólo hay proceso U y no se puede definir un proceso térmico.

[95] Es importante aclarar que la presencia del colapso siempre introduce una irreversibilidad intrínseca, incluso si al nivel macroscópico el macroestado evoluciona reversiblemente ($\delta S = 0$), la evolución del microestado es irreversible porque no hay evolución determinista-unitaria que devuelva el estado inicial y su información.

$$\begin{aligned}\delta Q \neq 0 \quad &\rightarrow \quad Proceso\ R\ activado \\ Proceso\ R\ apagado \quad &\rightarrow \quad \delta Q = 0\end{aligned} \qquad [5.69]$$

Es posible que el proceso R esté activo y no haya transferencia neta de calor[96] ($\delta Q = 0$), las expresiones [5.69] dan una exhibición macroscópica de los colapsos; los procesos R se pueden encender o apagar de acuerdo a la posibilidad de redistribuir la energía interna (hay probabilidad no nula de colapsar a un estado distinto del que se originó el enredo), cuando la temperatura se aproxima a 0 K la energía y la densidad de los fotones y fonones se hacen tan débiles que casi no son suficientes para inducir la absorción y emisión, entonces los colapsos se ralentizan al escasear la energía interna, y pueden apagarse incluso antes de llegar a 0 K, entonces la materia es gobernada únicamente por el proceso U y no hay clasicalización (en la sección [5.5.1] se explica la superconductividad debido a que se apagan los procesos R), esta descripción es consistente con el tercer principio de la termodinámica (*principio de Nernst*) donde los procesos clásicos se ralentizan en la proximidad a 0 K; en el límite al enredo cuántico, sección [5.2.1], el proceso R aparece en un conjunto de muchos microsistemas entrelazados (ecuación [5.68]) y no en un conjunto de microsistemas factorizables (ecuación [5.67]), ni mucho menos un microsistema individual; la idea más simple sobre la probabilidad de la ocurrencia de un colapso en el tiempo es que siga una distribución de Poisson[97] con parámetro $\lambda$ o tiempo medio entre colapsos $\tau = 1/\lambda$, Anexo [C.2], entonces el tercer principio de la termodinámica puede ser expresado como sigue:

$$\lim_{U \to 0} \lambda = 0 \qquad [5.70]$$

Donde el tiempo medio $\tau$ se hace infinito: los procesos clásicos se ralentizan porque se demora más tiempo cada colapso; de igual manera, conforme disminuye el número de microsistemas los colapsos se pagarán, de acuerdo a la sección [5.2.1]:

---

[96] Por ejemplo, en el equilibrio térmico no hay flujo neto de calor, pero ocurren los procesos R.
[97] Esta idea no es nueva y fue originalmente presentada en la teoría GRW de localización espontánea, donde los microsistemas colapsan espontáneamente incluso aislados e individualmente, y sin tener que estar excitados, siguiendo una distribución de Poisson en el tiempo.

$$\lim_{N \to 1} \lambda = 0 \qquad [5.71]$$

De manera que un microsistema y un fotón que interactúan confinados y aislados pueden permanecer gobernado por el proceso U.

### 5.3.4. El mecanismo de la observación es la redistribución de la energía

En la sección [5.2.3] se estableció que el mecanismo de la observación es un caso particular de la clasicalización, en el que un microsistema interactúa tan brevemente con el aparato de medida que sólo colapsa una vez; para que ocurra la observación es necesario que se envíe una señal física del microsistema (a observar) a una partícula clásica (en su proceso U) del aparato de medida (este es el protocolo más simple de comunicación), entonces se establece el enredo cuántico entre ambos:

$$|A\psi\rangle = \sum_k \lambda_k |A_k\rangle |\psi_k\rangle \qquad [5.72]$$

Donde $|A\psi\rangle$ es el estado cuántico entrelazado, $|A_k\rangle|\psi_k\rangle$ es el estado de la partícula clásica (del aparato de medida) y del microsistema tras el colapso (y la obtención de la medida), la señal física porta energía, además de información, que interactúa con los microsistemas de la partícula clásica en el proceso U y es absorbida en el proceso R por uno de sus microsistemas, el entrelazamiento entre el microsistema a observar y la señal física[98] conduce a que el estado cuántico del microsistema a observar colapse a la vez que la señal física es absorbida (y su energía es redistribuida como parte de la energía interna de la partícula clásica), tras el colapso se rompe el entrelazamiento de la ecuación [5.72] y los demás colapsos en la partícula clásica no afectan al microsistema observado, así que colapsa una sola vez; en contraste, los microsistemas al interior del aparato de medida se están "observando así mismos" de manera regular y permanente al nivel clásico.

El colapso en el microsistema y la absorción de la señal física es un proceso simultáneo, esto es abordado en la relatividad especial en la sección [5.7.1].

---

[98] Para cada estado $|\psi_k\rangle$ del microsistema la señal física porta un estado cuántico.

## 5.4. Interpretación de colapso objetivo como el programa de la clasicalización

### 5.4.1. Programa de la clasicalización

Esta interpretación de colapso objetivo que se ha propuesto permite desarrollar un programa para investigar la clasicalización; en las teorías de colapso objetivo precedentes (sección [2.2.3]) se resolvió el problema de la superposición de estados al nivel clásico mediante la localización espontánea (siendo la posición el observable privilegiado en el que ocurren los colapsos objetivos):

FIGURA N° 5.9

LOCALIZACIÓN ESPONTÁNEA

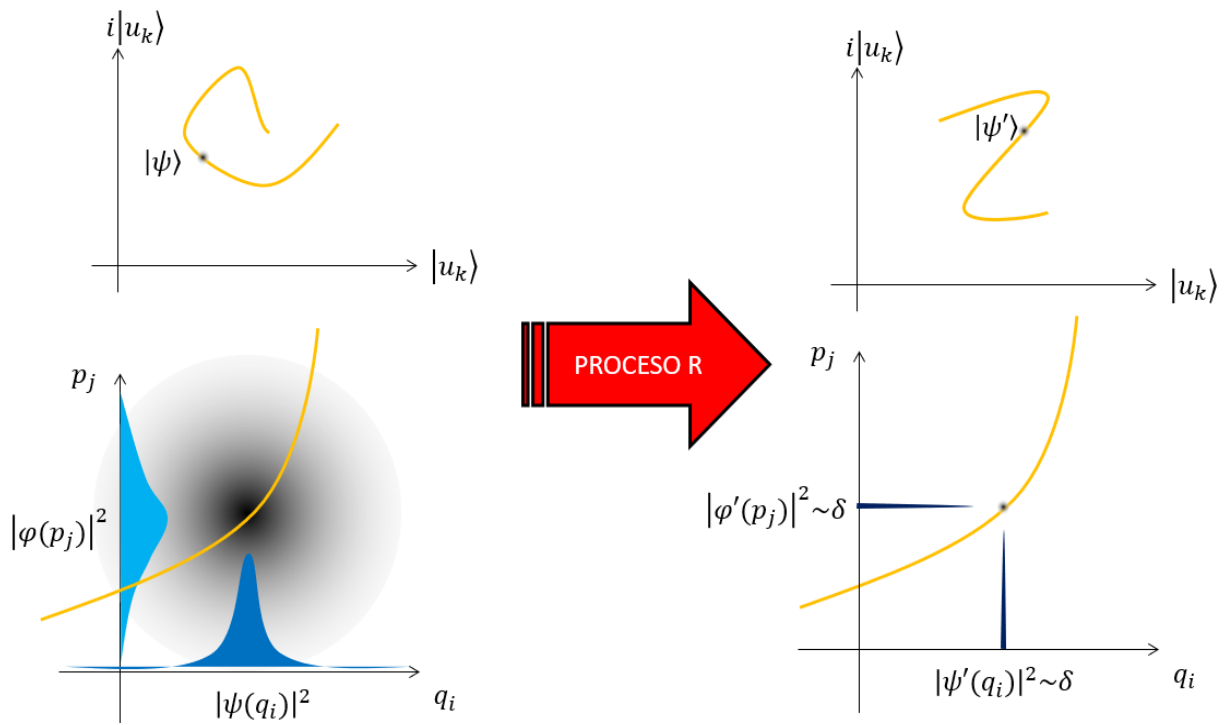

Fuente: Elaboración propia

En la Figura N° 5.9 el estado cuántico del microsistema $|\psi\rangle$ es un punto en el espacio de Hilbert, la evolución en el tiempo es una trayectoria descrita por la ecuación de Schrödinger, mientras que en el espacio de fases el estado clásico correspondiente está indeterminado, sólo sus valores medios evolucionan según las ecuaciones [5.3], el proceso R transforma el estado $|\psi\rangle$ a $|\psi'\rangle$ más localizado.

En la Figura N° 5.10 se representa la relación entre escala clásica y cuántica del tiempo: al nivel cuántico-microscópico se tiene la evolución unitaria y el colapso objetivo por separado, al nivel clásico-macroscópico, en cada elemento de tiempo es indistinguible el colapso de la unitaridad (están "uniformemente mezclados"), las variables clásicas-macroscópicas que se definen varían con suavidad y continuidad (son diferenciables temporalmente):

FIGURA N° 5.10

ESCALAS DE TIEMPO CLÁSICO VS. CUÁNTICO

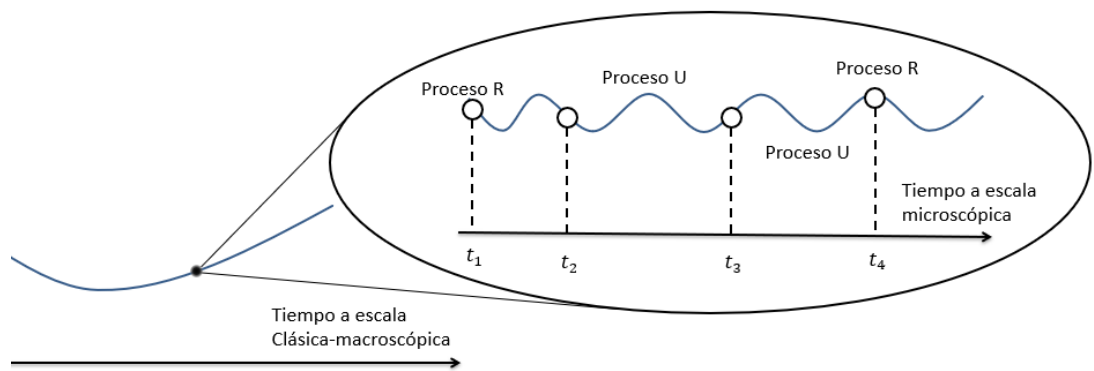

Fuente: Elaboración propia.

La clasicalización, sección [5.3], ha sido resumida en la Figura N° 5.11, mostrando que hay dos formas de clasicalización, con caos profundo o sin caos profundo:

FIGURA N° 5.11

LA CLASICALIZACIÓN

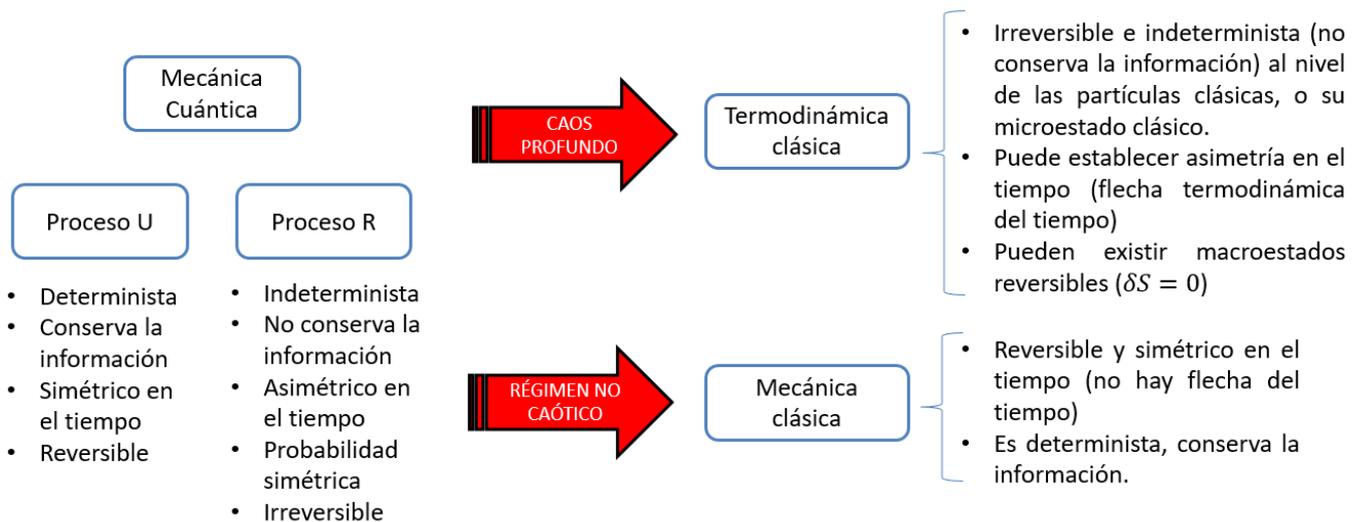

Fuente: Elaboración propia

En principio, todo el mundo clásico es irreversible, indeterminista y no conserva la información cuántica (debido al proceso R), sólo como una buena aproximación puede converger a evoluciones clásicas reversibles y deterministas que preservan una información clásica, como la mecánica clásica o los procesos termodinámicos reversibles; el mundo clásico va desde el mundo microscópico-clásico en el que habitan las partículas clásicas (las primeras partículas no-cuánticas) hasta el macroscópico (a colación, todos los entes clásicos son observadores); el microestado clásico es el que se define (aproximadamente) como un punto en el espacio de fases de un sistema de muchas partículas, se puede descender (y salir del mundo clásico) al mundo microscópico donde se encuentran los microsistemas (átomos, moléculas, etc), los cuales tienen la información (cuántica) completa del microestado del sistema, se debe considerar dos microestados para cada sistema: el microestado clásico y el microestado (cuántico); así por ejemplo, las ecuaciones de Hamilton se pueden aplicar al microestado clásico (dejan de ser válidas en el caos profundo) mientras que la ecuación de Schrödinger se aplicaría al microestado (en el brevísimo proceso U).

Es importante resaltar que la termodinámica al nivel macroscópico puede ser reversible o irreversible, de acuerdo a cómo evolucione el macroestado, pero al nivel del microestado clásico es siempre irreversible e indeterminista (y no conserva la información clásica completa): un conjunto de partículas clásicas fuera del régimen de caos profundo pueden evolucionar mecánicamente hacia un nuevo microestado clásico, y retornar al microestado clásico inicial si se invierten sus velocidades, en cambio la evolución mecánica de las partículas clásicas en régimen de caos profundo que componen un sistema termodinámico ya no cumplen las ecuaciones canónicas de Hamilton, y no retornarían al microestado clásico inicial al invertir sus velocidades[99],

---

[99] Una imagen muy común de la mecánica estadística es que si en un fenómeno térmico e irreversible (como una copa que se rompe al caer) se invirtieran todas las velocidades de las partículas instantáneamente, se observaría la evolución hacia atrás en el tiempo exactamente de los eventos ocurridos en orden inverso (los trozos recomponiendo la copa, la cual se eleva del piso); en este programa se afirma que eso no ocurriría, sino que seguirían evolucionando inexorablemente hacia el desorden (la mayoría de los trozos se aglomerarían pero no recompondrían la copa, la cual implosionaría y rebotaría en otra explosión) porque la evolución no es determinista ni reversible; así el problema de cómo un sistema de muchas partículas evoluciona irreversiblemente hacia el desorden, cuando supuestamente sus partículas evolucionan reversiblemente, se responde también

el microestado clásico no evoluciona reversiblemente incluso si el macroestado evoluciona reversiblemente ($\delta S = 0$); es decir, existen casos donde la irreversibilidad de las partículas clásicas no se expresan al nivel macro, permitiendo fenómenos termodinámicos reversibles, pero lo contrario no es posible: que la evolución reversible del microestado clásico conduzca a una evolución irreversible del macroestado (como se suele afirmar en mecánica estadística), para explicar esto se definen las variables $x_{mc}(t)$ y $X_M(t)$ que representan el microestado clásico y el macroestado del sistema respectivamente, $x_{mc}(t)$ tiene la información completa del sistema, mientras que $X_M(t)$ sólo una pequeña parte de la información total que es deducida a partir de $x_{mc}(t)$, entonces sea la transformación $\hat{T}$ como se muestra:

$$\hat{T}: x_{mc}(t) \to X_M(t) \qquad [5.73]$$

$\hat{T}$ usa la información completa del sistema y entrega sólo una pequeña parte, la del macroestado; $\hat{T}$ es una transformación irreversible e indeterminista ya que no conserva la información[100], ni tiene inverso $\hat{T}^{-1}$ (no puede devolver la información completa del sistema $x_{mc}(t)$ a partir de conocer el macroestado del sistema $X_M(t)$); si $x_{mc}(t)$ es reversible en el tiempo entonces no contiene información que distinga pasado de futuro[101], y cualquier información que se extraiga de $x_{mc}(t)$ tampoco distinguirá pasado de futuro, luego $X_M(t)$ no tiene información que distinga pasado de futuro y es reversible o simétrico, sólo si $x_{mc}(t)$ es irreversible en el tiempo se puede extraer mediante $\hat{T}$ la información que distingue pasado de futuro y obtener un $X_M(t)$ irreversible o asimétrico en el tiempo; es decir, si $x_{mc}(t)$ evoluciona reversiblemente entonces $X_M(t)$ también evoluciona reversiblemente, sólo si $x_{mc}(t)$ evoluciona irreversiblemente, entonces $X_M(t)$ puede ser reversible o no.

---

porque en una breve fracción de tiempo (proceso U) sí evoluciona reversiblemente (y conserva la entropía) pero a la escala clásica del tiempo todo está evolucionando irreversiblemente, y que si toma ese microestado clásico y hace ir hacia atrás en el tiempo (en una línea paralela de tiempo que no contenga su pasado) entonces aumenta la entropía, y sigue aumentando si va al futuro de nuevo.

[100] Por definición $\hat{T}$ sólo pierde información, pero no crea ni añade información en la transformación.
[101] Debido a que las ecuaciones de Hamilton no tienen algún parámetro que distinga pasado de futuro como si lo hace la ecuación del calor.

En la Tabla N° 5.2 se resalta el rol de los procesos U y R en la simetría, reversibilidad y determinismo al nivel clásico (mecánica y termodinámica):

TABLA N° 5.2

EXPRESIÓN DE LOS PROCESOS U, R AL NIVEL CLÁSICO

|  | **Mecánica clásica** | **Irreversibilidad termodinámica** |
|---|---|---|
| **Proceso U** | Se expresa conservando la información clásica, el determinismo, la reversibilidad y la simetría del tiempo. | No se expresa |
| **Proceso R** | Se expresa manteniendo un estado localizado en el espacio de fases (suprimiendo las superposiciones de estados). | • Se expresa suprimiendo la superposición de estados<br>• Se expresa en caos profundo introduciendo el indeterminismo, la irreversibilidad y asimetría en el tiempo |

Fuente: Elaboración propia

El mecanismo de la clasicalización parte de establecer la evolución alternada como un bucle: en el proceso R ocurre la redistribución de la energía interna (quedando en un autoestado del Hamiltoniano general), en el proceso U se evoluciona a una superposición de autoestados del Hamiltoniano general, de manera que la predominancia entre los procesos U y R definen dos tipos de clasicalización: el de la mecánica clásica y el de la termodinámica, además debe explica el mecanismo de la observación.

### 5.4.2. Teorema de la clasicalización

Este teorema es una generalización del programa de clasicalización, y expresa el "espíritu de la investigación":

> *En toda teoría fundamental "X" que sea determinista, simétrica o reversible en el tiempo:*
>
> 1. *No se puede explicar el aumento de la entropía o la flecha termodinámica del tiempo (fenómenos irreversibles), sino que debe conservar la entropía en el tiempo.*
> 2. *El aumento de la entropía, como ruptura de la simetría del tiempo, se origina, a un nivel fundamental, en una serie de eventos "e" donde no se conservan la información ($e \notin X$).*

La demostración de este teorema recae en la clasicalización desarrollada en la sección [5.3], también es sustentado mediante la teoría física de la información presentada en el Anexo B; lo primero a demostrar es P1 que establece equivalencia entre determinismo, simetría y reversibilidad:

*P1: "Una teoría es determinista si es simétrica o reversible en el tiempo"*

Si es reversible en el tiempo, significa que todos los procesos que ocurren como una función del tiempo $f(t)$ tienen también un proceso inverso $g(t)$ que actúa como yendo hacia atrás en el tiempo: $g(t) = f(t_0 - t)$, entonces hay el mismo número de funciones $f(t)$ y $g(t)$, ambas soluciones de la teoría en cuestión; lo que es, que sea simétrico en el tiempo[102]; así se prueba que el ser reversible o simétrico son equivalentes; estas funciones pueden expresarse como las ecuaciones [B.12], en las que conservan la información física, luego esto es el determinismo.

*P2: "Si X es determinista, X no puede explicar el aumento de la entropía o la flecha termodinámica del tiempo (fenómenos irreversibles)".*

*P3: "Si X es determinista, la entropía debe conservarse en el tiempo".*

---

[102] Porque describe tanto procesos hacia "adelante en el tiempo" como hacia "atrás en el tiempo".

Para demostrar P2: al ser *X* determinista es también reversible y simétrica, así de existir fenómenos (para sistemas con muchos grados de libertad) descritos por *X* con aumento de entropía en el tiempo, deben existir también otros fenómenos en los que la entropía disminuya en el tiempo exactamente de la misma forma, exactamente con la misma probabilidad, luego al promediar sobre un amplio rango de microestados las variaciones de la entropía deberían compensarse y conservar la simetría al nivel estadístico: al no existir una dirección privilegiada del tiempo, no se puede establecer una flecha termodinámica ni la existencia de fenómenos irreversibles, luego *P2* es cierta. La conservación de la entropía ha sido establecida en la sección [5.3.2] (ecuaciones [5.55] y [5.57]) para la mecánica clásica, esto se puede generalizar a cualquier teoría determinista, pues el teorema de Liouville se basa en la no-bifurcación de las trayectorias, lo cual es válido para cualquier teoría determinista, luego *P3* es cierta; así es demostrado el primer punto del teorema.

Para el segundo punto del teorema, es evidente que el aumento de la entropía es una ruptura en la simetría del tiempo, y que de ocurrir (inexplicable para *"X"*) requiere de elementos *"e"* externos a *"X"*, que no conservan la información, o equivalentemente que *"e"* es indeterminista; esto se demuestra por *reductio ad absurdum*: si *"e"* es determinista o conservativo (aunque ajeno a *"X"*), entonces según el primer punto *"e"* debería preservar la simetría del tiempo, no poder explicar la flecha del tiempo y conservar la entropía, esto contradice su definición, luego *"e"* es indeterminista (no conserva la información).

El uso de este teorema en la clasicalización permite identificar a *"X"* y *"e"* con los procesos U y R de la mecánica cuántica, respectivamente, esto demuestra la interpretación de colapso objetivo: el proceso R debe ser objetivo porque no puede ser reducido o explicado por la unitaridad de la mecánica cuántica[103], pues $e \notin X$.

---

[103] Si el colapso no es objetivo, entonces es aparente y reducido al proceso unitario, como en el programa de la decoherencia, por ejemplo, en el acto de medición la unitaridad del sistema general aislado microsistema-observador se debe preservar.

## 5.5. Interpretación de colapso objetivo en algunos fenómenos físicos

### 5.5.1. Resistencia eléctrica y la superconductividad

La resistencia eléctrica es un fenómeno clásico por ser termodinámico, la ley de joule establece un incremento de entropía por unidad de tiempo, como se expresa:

$$\frac{dS}{dt} = \frac{\delta Q}{T \delta t} = \frac{VR}{T} > 0 \qquad [5.74]$$

Donde $VR$ es la potencia eléctrica disipada en forma de calor, y $T$ es la temperatura, los fenómenos irreversibles como el efecto Joule son resultados de la clasicalización (a escala macroscópica) o de los colapsos (al nivel microscópico); en la superconductividad la corriente eléctrica no tiene ninguna resistencia porque los colapsos están apagados y los electrones evolucionan unitariamente, sin redistribuir la energía interna (la energía eléctrica de la corriente y la energía térmica de la red cristalina), así los electrones pueden conservar su energía cinética. En la Figura N° 5.12 se muestra la actividad del proceso R con la resistencia eléctrica:

FIGURA N° 5.12

RESISTIVIDAD Y PROCESO R EN UN SUPERCONDUCTOR

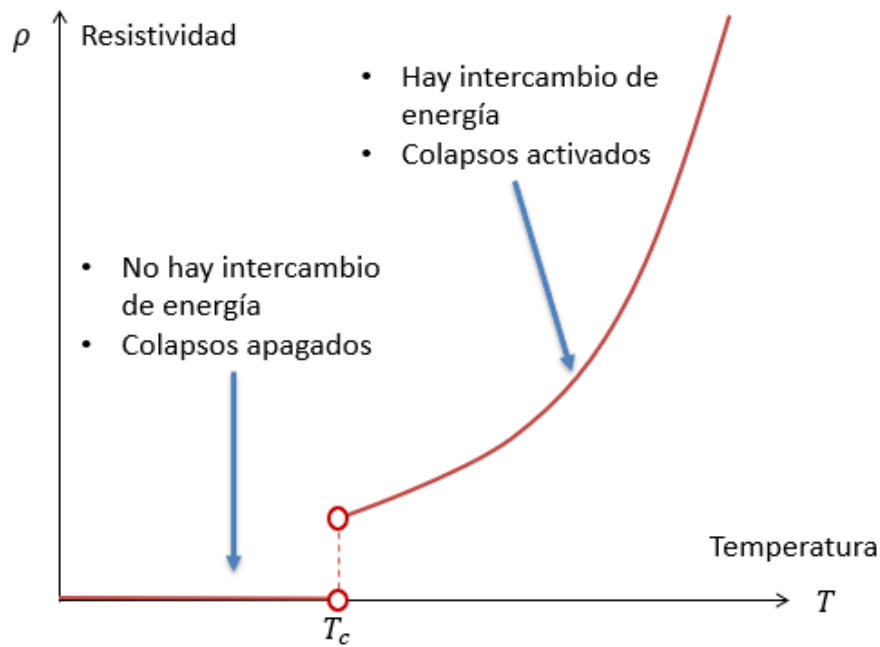

Fuente: Elaboración propia.

Es importante notar que en la superconductividad los colapsos se apagan antes de llegar al cero absoluto (a la temperatura crítica $T_c$) porque los colapsos se activan por la posibilidad de redistribuir la energía: en el modelo de Drude la resistencia eléctrica es producto de la fricción entre la corriente eléctrica y la red cristalina, en el modelo BCS los electrones si bien son fermiones, al descender a la temperatura crítica $T_c$ pueden formar los pares de Cooper y ser estados singletes que se comportan como bosones formando un superfluido de manera que la inexistencia de fricción en su movimiento a través de la red cristalina es la superconductividad; el emparejamiento da lugar a un GAP superconductor[104], el superfluido (la corriente eléctrica) no acepta energía menor al ancho del GAP proveniente de la red cristalina (su energía termal) ni puede descender a un estado de menor energía por ser un estado fundamental, esta imposibilidad de intercambiar energía mantiene apagado a los colapsos a una temperatura no-nula; no obstante, no todos los colapsos se han apagado en el superconductor, pues al seguir teniendo una temperatura no nula sigue teniendo fenómenos térmicos y atributos clásicos.

### 5.5.2. Fotón incidiendo en una pantalla

Una pantalla (placa fotográfica o pastilla CCD[105]) es un observador de cualquier fotón incidente, deslocalizado a escala macroscópica hasta justo antes de tocar la pantalla (con $N$ microsistemas), entonces interactúa con todos los microsistemas de la superficie, pero sólo puede ser absorbido por uno a la vez, el enredo fotón-pantalla (enredo microsistema-observador) es expresado como sigue:

$$|\gamma, P\rangle = \sum_k c_k |\gamma, k\rangle \qquad [5.75]$$

Donde $|\gamma, P\rangle$ representa el estado del fotón y la pantalla entrelazados, $|\gamma, k\rangle$ representa la absorción del fotón por el microsistema $k$, tras la interacción el fotón es

---

[104] GAP superconductor es una banda de energía prohibida, que separa el estado fundamental (condensado de Bose-Einstein) del primer estado excitado para la cuasipartícula (par de Cooper) que corresponde con su disociación (electrones no emparejados).
[105] La pastilla CCD es una colección de sensores fotovoltaicos empleados en las cámaras digitales.

localizado en la pantalla en el microsistema $k$ como un punto, como se aprecia en la Figura N° 5.13, la pantalla actúa como un observador de la mecánica cuántica:

FIGURA N° 5.13

FOTÓN INCIDIENDO EN UNA PANTALLA

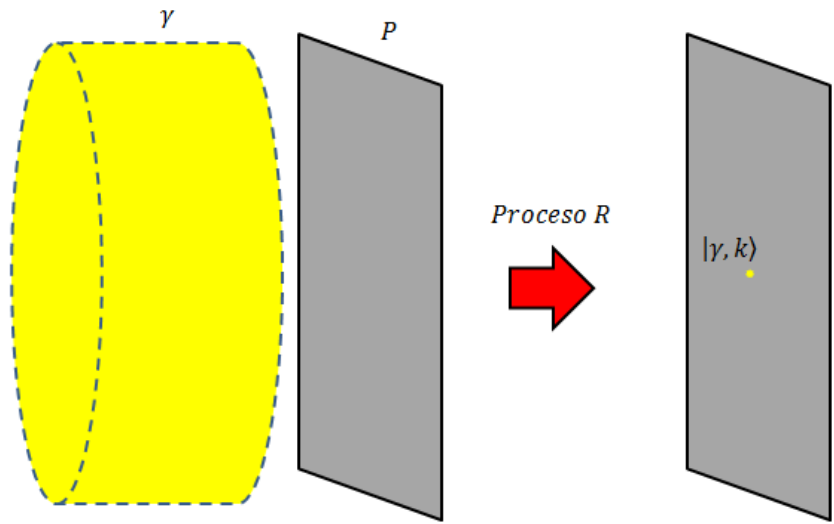

Fuente: Elaboración propia.

Todos los microsistemas de la pantalla no colapsan simultáneamente, sus partículas clásicas van decidiendo si absorben o rechazan[106] al fotón en cada colapso, así en realidad el fotón está interactuando con muchos observadores (las partículas clásicas), sea $\zeta_1, \zeta_2, \zeta_3, \ldots$ la secuencia de partículas clásicas según la ocurrencia del colapso, al primer colapso $\zeta_1$ se "decide" si el microsistema "$i$" absorbe al fotón (con probabilidad $P_i = |c_i|^2$) o si no es absorbido (con probabilidad $P_{\sim i} = 1 - |c_i|^2$), cada probabilidad de ser absorbido es muy pequeña ($P_i \ll 1$), así que probablemente no sea absorbido por el primer proceso R, si ningún microsistema de $\zeta_1$ absorbe al fotón, entonces el estado del fotón colapsa a no-localizado en la región de la pantalla que ocupa la partícula clásica $\zeta_1$; es decir, aparece un "hueco" en la función de onda del fotón que excluye la región mencionada, ahora las probabilidades son un poco mayores por la normalización, al siguiente colapso $\zeta_2$ decide si absorbe al fotón o no, este proceso se repite creando más huecos, disminuyendo la región disponible, y aumentando la probabilidad de ser absorbido hasta que finalmente el fotón es

---

[106] El rechazo consiste en que la función de onda incidente $\psi$ colapsa a otra función de onda $\varphi$ que tiene un "agujero" donde se encuentra el microsistema; esto es, que se anula en esa región.

absorbido por un microsistema, esto es mostrado en la Figura N° 5.14: en (1) el fotón incide sobre la pantalla, en (2) aparece el primer colapso "de rechazo", en (3) ocurre el segundo colapso también de rechazo, en (4) y (5) ya han ocurrido varios colapsos que reducen la región disponible para colapsar, finalmente en (6) colapsa a un solo punto de la pantalla, cada punto negro corresponde a una partícula clásica:

FIGURA N° 5.14

SUCESIÓN DE COLAPSOS EN UN FOTÓN INCIDENTE

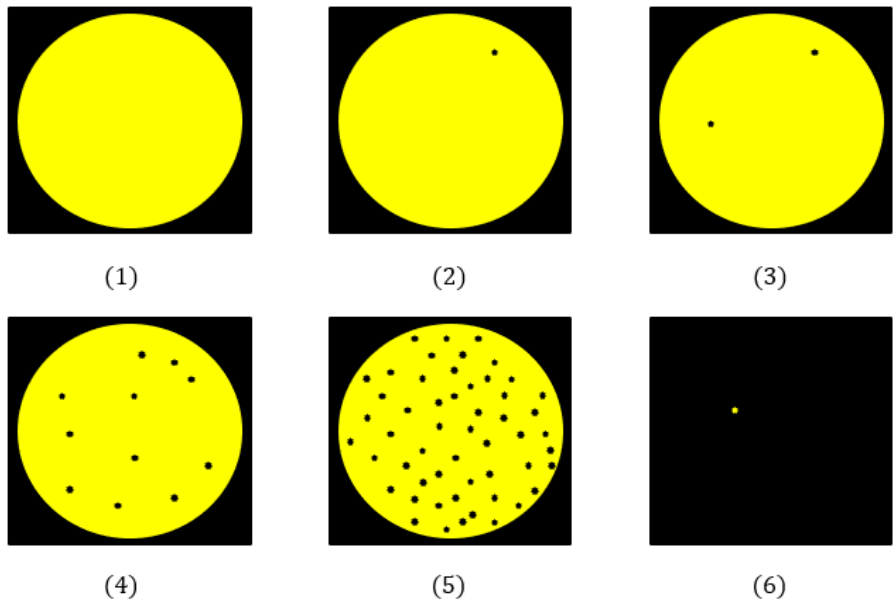

Fuente: Elaboración propia.

Este proceso es consistente con un único colapso de (1) a (6), la probabilidad de llegar a la situación última (6) es invariante de si el fotón colapsa directamente al microsistema $k$, o si lo hace en el $n$-ésimo colapso, pues si una secuencia de colapsos "de rechazo" tiene la probabilidad $q_\sim$ entonces crea un nuevo estado para la ecuación [5.75] que excluyen a los microsistemas que colapsaron, y es normalizado por un factor $1/\sqrt{q_\sim}$, de manera que si en el siguiente colapso el fotón es absorbido por el microsistema $k$, con probabilidad $P'_k = \left|\frac{1}{\sqrt{q_\sim}} c_k\right|^2$, entonces la probabilidad total es $q_\sim P'_k = |c_k|^2$; es decir, la probabilidad de obtener $|\gamma, k\rangle$ es invariante a la secuencia de colapsos; este mecanismo es particularmente importante para explicar cuando el fotón pasa a través del borde de la pantalla, como se muestra:

# FIGURA N° 5.15
## FOTÓN COLAPSANDO SOBRE EL BORDE DE UNA PANTALLA

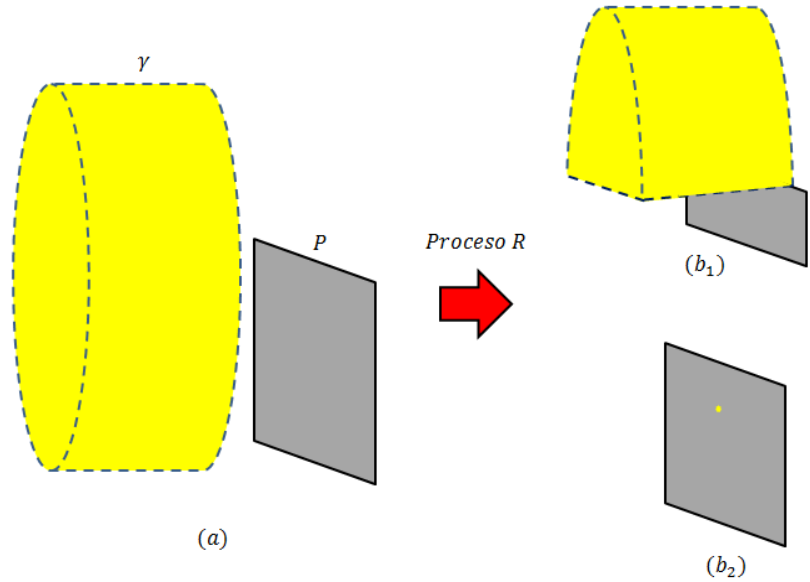

Fuente: Elaboración propia.

En la Figura N° 5.15 un fotón deslocalizado incide sobre el borde de una pantalla, y una parte de su función de onda espacial no incide sobre ningún microsistema, se puede hacer variar el índice $k = 0, 1, 2, 3, \ldots N$ en la ecuación [5.75] donde $k = 0$ corresponde al vacío (y no provocará su colapso), la probabilidad $|c_0|^2$ es (de acuerdo a la imagen) del orden de 0.5 y mucho mayor que la del resto $|c_0|^2 \gg |c_{i \neq 0}|^2$, al ocurrir los colapsos se van excluyendo al conjunto de microsistemas que no absorben el fotón, y se presentan dos casos: uno de ellos absorbe al fotón y cesa el proceso R para la interacción fotón-pantalla $(b_2)$, y que ninguno de los microsistemas absorbe el fotón, quedando intacta la región fuera de la pantalla y libre de moverse $(b_1)$; evidentemente el fotón no se puede partir (se absorbe todo o nada) así que el colapso $|\gamma, P\rangle \to |\gamma, 0\rangle$ es la destrucción del paquete de ondas incidente y la creación de un nuevo paquete de ondas, la indeterminación en la posición se reduce en el plano vertical y la indeterminación en el movimiento aumenta introduciendo ondas planas vertical y perpendicular al fotón incidente que deformará el frente de ondas como "bordeando la pantalla", es justamente esto lo que crea la difracción de la luz al pasar sobre el borde de una pantalla opaca.

### 5.5.3. Experimento de la doble rendija

La interacción fotón-pantalla explica satisfactoriamente el experimento de la doble rendija, que es paradigmático[107] y muy conocido: un fotón (emitido por un microsistema) avanza una distancia macroscópica (adquiriendo un tamaño macroscópico) y atraviesa una pantalla (*pantalla 1*) con dos rendijas (*A* y *B*) antes de impactar en otra pantalla final (*pantalla 2*), como se muestra en la Figura N° 5.16:

FIGURA N° 5.16

EXPERIMENTO DE LA DOBLE RENDIJA

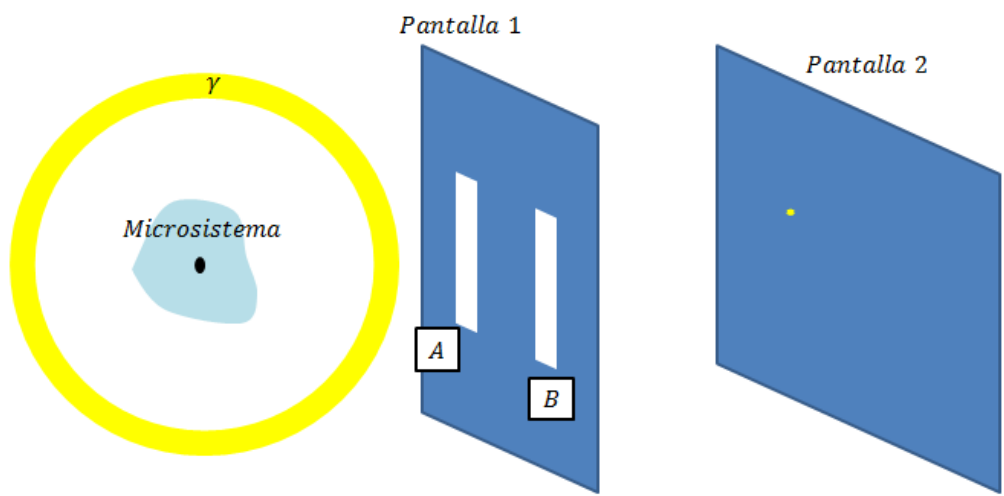

Fuente: Elaboración propia.

Primero se mantiene abierto sólo una rendija en todo el experimento, los fotones enviados uno a uno (deslocalizados al nivel clásico) que atraviesan la *pantalla 1* colapsan a una función de onda con la forma de la rendija abierta y viajan hasta dar con la *pantalla 2* en las que se localizan como un punto, esto se muestra en a) de la Figura N° 5.17; luego se realiza el experimento con ambas rendijas abiertas, entonces la función de onda del fotón colapsa a un estado con la forma de ambas rendijas, que incluso separadas y desconectadas son un solo fotón, la difracción (en la forma del fotón) al atravesar cada rendija permite la interferencia del fotón consigo

---

[107] En opinión de Richard Feynman este experimento encierra casi toda la esencia de la mecánica cuántica, en su libro de mecánica cuántica (Feynman, R. & Leighton, R. & Sands, M. (1963). *The Feynman Lectures on Physics, Vol 3, Quantum Mechanics*. Massachusetts: ADDISON-WESLEY.**)**

mismo, creando un patrón de interferencias cuando los fotones son localizados en la pantalla 2, como se muestra en b):

FIGURA N° 5.17

PATRONES DE INTERFERENCIAS

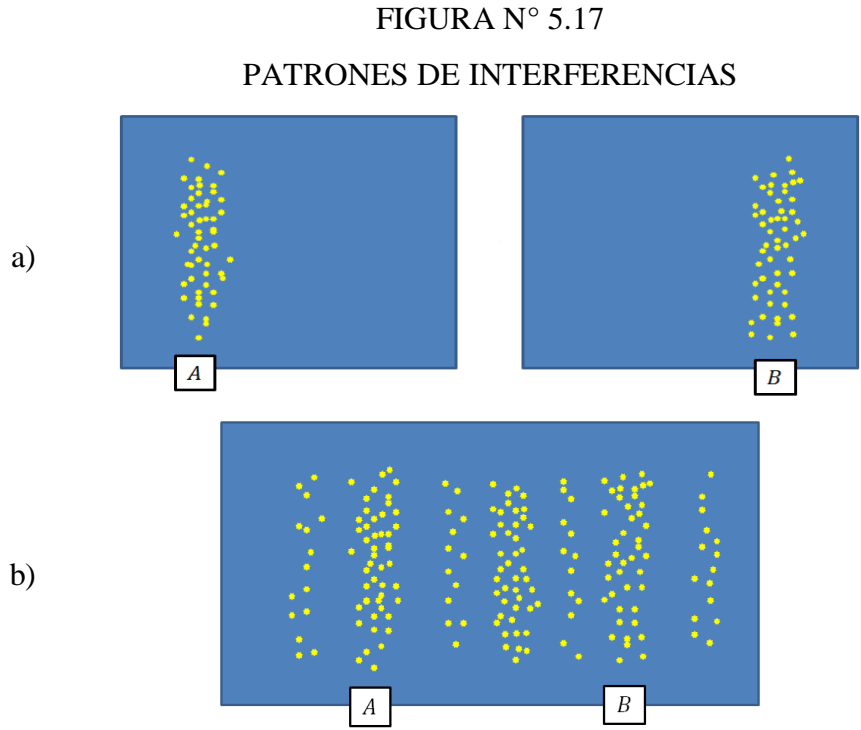

Fuente: Elaboración propia.

Esta es la explicación de esta interpretación de colapso objetivo, que se ha introducido, para el experimento de la doble rendija, y evita el problema de "por cual rendija el fotón ha pasado" porque no interpreta al fotón como un punto del que se desconoce su posición (y solo se conoce su probabilidad), sino que el fotón está distribuido en el espacio y cambia de forma tanto en el proceso U como en el proceso R (la probabilidad no es del desconocimiento de su posición, sino de que se transforme en un estado más localizado).

Finalmente, si se instala un sistema de medición que permita determinar porqué rendija cada fotón está pasando[108], se forzaría al fotón a "localizarse" en una de las rendijas aleatoriamente justo al atravesar la *pantalla 1*, lo que naturalmente sería el caso a), al enviar muchos fotones se tendría la superposición directa de ambos patrones de a), sin la interferencia característica de b).

---

[108] Un buen ejemplo de ese sistema de medición consistiría en un sistema que detecte "empujes" horizontales sobre la pantalla 1, con la fuente colocada simétricamente en medio de ambas rendijas, de manera que si el fotón pasa por una rendija intercambia un cantidad de movimiento casi opuesto a si pasaría por la otra rendija, estos impulsos serían muy pequeños pero en principio mensurables;

### 5.6. Decoherencia, el problema de la medida y la clasicalización

#### 5.6.1. Definición de colapso objetivo

El proceso R, sección [2.1.2], puede ser interpretado como un evento real y objetivo (independiente del observador que realiza la medida) o como un evento aparente y relativo al observador, esta elección realmente va más allá de las creencias o concepciones filosóficas y ha sido presentado como un problema físico; si la evolución del sistema general aislado observador-microsistema sigue siendo unitaria al realizarse la medida, entonces el colapso del estado del microsistema es relativo al observador, porque para él la evolución del microsistema deja de ser unitaria (y experimenta la irrupción del proceso R), mientras que en todo el sistema general aislado no ha ocurrido ninguna irrupción en su evolución unitaria, así deben existir mecanismos unitarios que en la interacción observador-microsistema los lleve al estado posterior a la medida (en esto se ha referido intensamente a la decoherencia cuántica); el colapso objetivo es definido cuando ocurre a todo nivel: en el acto de medición el estado cuántico general observador-microsistema también colapsa, deja de ser unitario, y no puede ser explicado en base a mecanismos unitarios (que preservan la unitaridad) ni como relativo al observador, además es indeterminista[109].

Aunque para muchos investigadores es deseable conservar la unitaridad en el sistema general aislado observador-microsistema[110], su preservación tanto en el

---

este ejemplo se encuentra en el libro *Quantum Mechanics* de Cohen-Tannoudji y Bernard Diu (capítulo 1, complemento D1).

[109] A diferencia de la física clásica, que interpreta al acto de medición como un proceso determinista, de manera que si bien la observación perturba al sistema al medir (por la interacción), todo el sistema aislado evoluciona deterministamente, conservando la información física; esta situación es análoga a la conservación de la energía: al interactuar se puede cambiar la energía del sistema en estudio, pero la energía del sistema total aislado se conserva, este tipo de razonamiento aplicado a preservar la unitaridad cuántica en el acto de medición (y la clasicalización) es justamente un colapso aparente o relativo (la no-conservación de la energía al interactuar con el sistema bajo estudio es aparente o relativo al observador, pues la energía total se conserva).

[110] De ahí que los intentos en resolver la paradoja de la pérdida de la información en los agujeros negros, Anexo [A.2], sea predominante la postura de preservar la unitaridad.

problema de la medida como en la clasicalización es contradictoria: el mundo clásico debería evolucionar de forma reversible[111], con mediciones "hacia atrás en el tiempo", y esto nos devuelve a las contradicciones en las secciones [2.5.2] y [2.5.3].

Sea el acto de medición al tiempo $t_0$, donde $|\psi(t)\rangle$ y $|A(t)\rangle$ son los estados del microsistema y del aparato de medida[112] antes de la medición, $|a\rangle$ y $|A_a\rangle$ los estados del microsistema y del aparato de medida justo después de la medición, si el colapso no fuera objetivo el sistema general aislado $|\varphi(t)\rangle$ evolucionaría unitariamente entre los tiempos $t_1$ y $t_2$ tal que: $t_0 \in [t_1; t_2]$, como se muestra:

$$\widehat{U}(t_2, t_1)|\varphi(t_1)\rangle = |\varphi(t_2)\rangle \qquad [5.76]$$

El colapso objetivo se define como la imposibilidad de que exista algún operador $\widehat{U}(t_2, t_1)$ que cumpla [5.76], como el microsistema y el observador antes y después de la medida son independientes, entonces $|\varphi(t)\rangle$ es factorizable:

$$\lim_{t_2 \to t_0^+} \widehat{U}(t_2, t_1) \cdot |\psi(t_1)\rangle|A(t_1)\rangle = |a\rangle|A_a\rangle \qquad [5.77]$$

Donde el límite lleva $t_2$ a un tiempo inmediatamente posterior al colapso; el determinismo del hipotético proceso U implica que la reducción al autoestado $|a\rangle$ no es un proceso aleatorio, aunque lo es en apariencia al observador con probabilidad $|\langle a|\psi(t_0)\rangle|^2$ como un proceso además irreversible[113], el aparato de medida introduciría la información sobre a qué autoestado reducirse[114]; al ser objetivo el colapso, el proceso R actúa también sobre $|\varphi(t)\rangle$.

---

[111] El universo, como sistema aislado, debería evolucionar reversiblemente.
[112] En este caso se asumirá que puede representarse por un ket y evolucionar unitariamente.
[113] La evolución en todo el sistema general es reversible, mientras que en sus componentes (microsistema y aparato de medida) es irreversible, esta es una contradicción similar al de la mecánica clásica y la termodinámica, sección [2.5.2]; si se acepta que la irreversibilidad termodinámica pueda aparecer de la reversibilidad clásica, entonces es posible aceptar esto.
[114] Ya que esta información no existe en el microsistema (por definición), y por conservación de la información en ambos; esta información se encontraría en el observador como "ruido blanco".

### 5.6.2. La decoherencia no explica el colapso

Sea $|\psi(t)\rangle = \sum_i \lambda_i(t)|i\rangle$ y $|\epsilon\rangle$ el estado de un microsistema y de su entorno[115] respectivamente, al interactuar entre sí va del estado separable al entrelazado (asumiendo una evolución unitaria) donde $|\epsilon_i\rangle$ es el estado que el entorno adquiere si el microsistema se reduce a $|i\rangle$, similar al acto de observación, el estado general es:

$$|\psi,\epsilon(t)\rangle = \sum_i \lambda_i(t)|i\rangle|\epsilon_i\rangle \qquad [5.78]$$

La decoherencia aparece cuando el entorno introduce "perturbaciones" aleatorias en las fases de cada amplitud, pasando de $\lambda_j(t)$ a $e^{i\phi_j(t)}\lambda_j(t)$ donde $\phi_j(t)$ es una variable aleatoria, desfasándolas, provocando interferencias destructivas entre sí, y anulando los términos de interferencia $\lambda_i(t)^*\lambda_j(t)$, $(i \neq j)$ al promediarse en el tiempo[116] en el estado puro $|\psi,\epsilon(t)\rangle\langle\psi,\epsilon(t)|$, entonces se obtiene el estado $\hat{\rho}$:

$$\hat{\rho} = \sum_j |\lambda_j(t)|^2 |j\rangle\langle j| \otimes |\epsilon_j\rangle\langle\epsilon_j| \qquad [5.79]$$

La interpretación de $\hat{\rho}$ no debe ser entendido como una mezcla (estadística), pues se trata de un microsistema individual, debe ser interpretado como una superposición incoherente[117], y no constituye una explicación del colapso porque no destruye la superposición de estados, como se consigne en la ecuación [5.77] en la cual es explícito el resultado de la medida, mientras que en $\hat{\rho}$ se sigue teniendo la misma ignorancia sobre el resultado de la medida; no se debe usar la decoherencia para explicar el colapso, sobre todo si se pretende preservar la unitaridad[118]; así

---

[115] El entorno es clásico, pero en la decoherencia se asume que se lo puede representar por un ket.
[116] El tiempo clásico $\Delta t$ es infinito a escala microscópica:

$$\langle e^{i\phi_j} e^{i\phi_k}\rangle = \lim_{\Delta t \to \infty} \frac{1}{\Delta t} \int_t^{t+\Delta t} dt' \cdot e^{i(\phi_j(t')-\phi_k(t'))} = \delta_{jk}$$

[117] Así, por ejemplo, en el experimento de la doble rendija, sección [5.5.3], un fotón que atraviesa ambas rendijas en superposición coherente describiría el patrón de la Figura N° 5.17 b), pero si pierde la coherencia entonces sería la superposición de ambos casos de a).
[118] Si la decoherencia ocurre preservando la unitaridad de todo el sistema aislado, entonces debería ocurrir el proceso inverso por reversibilidad: superposiciones (incoherentes) que ganan coherencia,

mismo, la decoherencia no explica la clasicalización porque entra en una argumentación circular: la explicación de la pérdida de coherencia en el estado puro $|\psi, \epsilon(t)\rangle\langle\psi, \epsilon(t)|$ parte de reconocer un entorno clásico preexistente que es justamente lo que se desea explicar, y vicia la argumentación sobre la clasicalización, porque al tener un mundo clásico previo es posible decir que su explicación de la clasicalización (mediante la irreversible y difusiva "filtración" de la coherencia del estado puro al entorno) recae en el entorno clásico ya existente, y no en el mecanismo mismo de la decoherencia.

### 5.7. La clasicalización en los agujeros negros

#### 5.7.1. La simultaneidad del proceso R en la relatividad especial

Se ha descrito al colapso como un proceso instantáneo en toda la función de onda, incluso deslocalizado a escala macroscópicas, que se reduce a un nuevo estado como "localizado en un punto" (ver sección [5.5.2]), como un caso particular, un estado enredado colapsa instantánea y simultáneamente incluso si los microsistemas están separados una distancia macroscópica; al asumir el colapso como objetivo (como se usó en la sección [5.5.3]) debe ser explicado en el marco de la relatividad, donde la simultaneidad es relativa.

Sea la función de onda $\psi(\vec{r}, t)$ (en azul) que colapsa al tiempo $t_0$ al estado más localizado $\varphi(\vec{r}, t)$ (en naranja), para un observador inercial $O$, como se muestra en a) de la Figura N° 5.18, para otro observador inercial $O'$ que se mueve hacia la derecha (ejes de coordenadas en rojo) el colapso no es un acto instantáneo y simultáneo sino que es un proceso (de ahí denominarlo "proceso R") que ocurre desde la parte más distante a la derecha en el eje $x'$ (evento A) hasta cerca del observador $O'$ (evento B) entre los tiempos $t'_1$ y $t'_2$, como se muestra en b):

---

así la explicación de que la coherencia se "filtra" al entorno como un proceso difusivo tendría su inverso: coherencia que "emana" del entorno y se concentra en algún microsistema particular.

FIGURA N° 5.18

COLAPSO DE LA FUNCIÓN DE ONDA

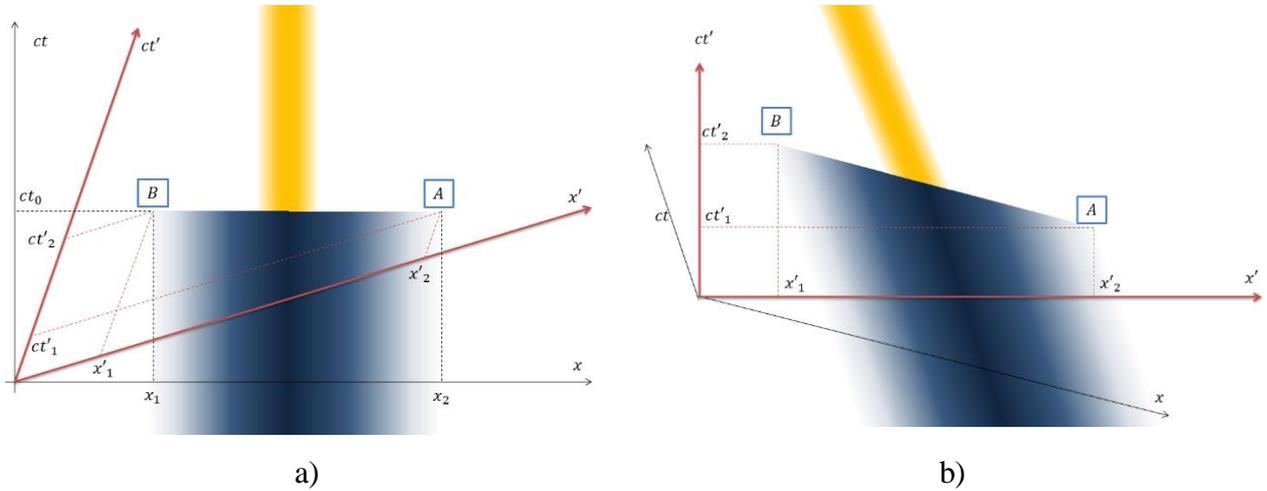

a)                                      b)

Fuente: Elaboración propia.

Para el observador $O'$ la creación del nuevo estado, con la nueva función de onda $\varphi'(\vec{r}', t')$ más localizada, en la proximidad al evento B puede parecer paradójica: existe un "remanente fantasmagórico" de la función de onda anterior $\psi'(\vec{r}', t')$ (que deja de existir) coexistiendo en simultaneidad con la nueva función de onda creada, para esta interpretación no hay inconveniente, ni supone una contradicción de principio, la existencia de tal "remanente fantasmal" próxima al evento B, para esto basta con interpretar al principio de complementariedad[119] de Bohr: los aspectos de onda $\psi'$ y de partícula $\varphi'$ pueden ser simultáneos para el observador $O'$, pero no pueden ser observados simultáneamente por $O'$ (ni por algún otro observador)[120], porque no hay ninguna señal proveniente de $\psi'$ (cuando ya existe $\varphi'$) y otra proveniente de $\varphi'$ que converjan a algún observador antes que se destruya $\psi'$ (y con ello su señal, porque anula la probabilidad de dicha emisión e interacción y envío de su información cuántica).

Sean dos microsistemas enredados, uno de ellos es acelerado hasta alcanzar velocidades relativistas, entonces establece su propia simultaneidad como se muestra

---

[119] La complementariedad nos dice que los atributos de ondas o de partículas son dos naturalezas complementarias de los sistemas físicos, que depende del tipo de experimento que se realiza, pero que no son observados simultáneamente.

[120] En general, no se puede observar simultáneamente los estados previos y posteriores al colapso.

en la Figura N° 5.19 (líneas rojas), si un observador mide localmente al microsistema en reposo al tiempo $t_0$, el otro colapsa en la intercepción de la recta $t = t_0$ y su trayectoria (curva azul), si el observador se mueve con el otro microsistema y realiza una medida local en el tiempo propio $\tau_0$, entonces el microsistema estacionario colapsa en la intersección entre su trayectoria (eje $ct$) y la prolongación de su recta roja de simultaneidad:

FIGURA N° 5.19

SIMULTANEIDAD DEL MICROSISTEMA VIAJERO

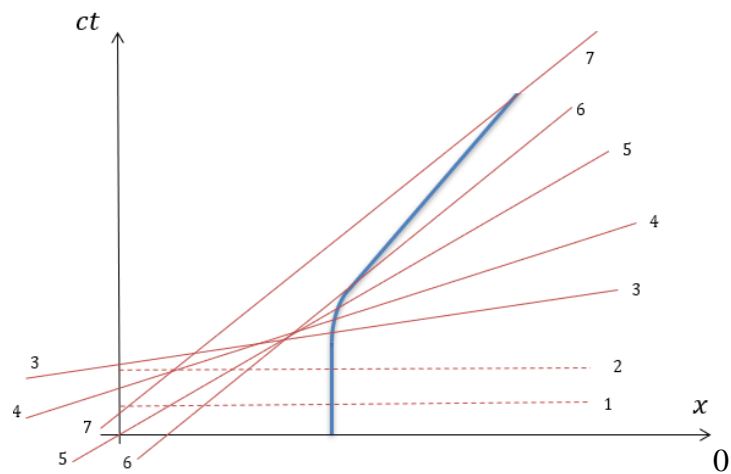

Fuente: Elaboración propia.

El colapso no implica una transmisión superlumínica de información porque no hay ninguna información transmitida entre los microsistemas enredados[121].

### 5.7.2. Contradicciones teóricas

En la sección [2.6.1] se estableció la reversibilidad y simetría de la relatividad general en el tiempo, de igual manera en [2.6.2] el diagrama de Penrose para la métrica de Schwarzschild exhibe tal simetría entre agujeros negros, agujeros blancos y la inversión temporal $t \to -t$; por otro lado, en la sección [2.6.5] y [2.6.6] la segunda ley mecánica de los agujeros negros o, equivalentemente, el segundo principio de la termodinámica en los agujeros negros establece una ruptura de la simetría de la relatividad general (en el tiempo) que no se puede reducir o explicar en el marco de una teoría reversible y simétrica[122]; así, la existencia de procesos irreversibles constituye una contradicción con la relatividad general.

---

[121] El colapso de un estado entrelazado ocurre simultáneamente como un todo deslocalizado, pero sin transmisión de señal.

Los procesos irreversibles, como son descritos en la sección [2.6.3], requiere del aumento de la masa irreducible $M_{irr}$ en un sentido privilegiado del tiempo; el ingreso de materia o energía al agujero negro es un proceso irreversible, pues implica un incremento en $M_{irr}$, y en consistencia el teorema del agujero negro (sección [2.6.4]) establece que estos son cerrados: no entra (ni sale) materia o energía al interior, conservando $M_{irr}$ y desarrollando sólo procesos reversibles y simétricos en el tiempo[123], esto conduce a una eterna caída (o eterno escape, en un agujero blanco); así, la contradicción entre la termodinámica de los agujeros negros y la relatividad general es análogo a la contradicción entre la mecánica clásica y la termodinámica o física estadística vista en la sección [2.5.2]. para la relatividad general los agujeros negros son cerrados[124] mientras que en su termodinámica ellos son abiertos[125].

Al considerar la mecánica cuántica se llega a dos contradicciones más (debido a la radiación de Hawking) al asumir la interpretación de Copenhague donde se preserva la unitaridad en todo sistema (incluso de un sistema clásico como un agujero negro). Por un lado, se tiene la contradicción de que la mecánica cuántica exige la superposición de estados mientras que la relatividad general no, similar a lo expuesto en la sección [2.5.1], y por otro lado la contradicción entre la termodinámica y la mecánica cuántica en la evolución de un estado puro a un estado mezcla (para un ensemble de microsistemas que caen al interior de un agujero negro) y que ha dado lugar a la paradoja de la información (Ver Anexo A.2), el caso no relativista se discutió en la sección [2.5.3]; estas tres contradicciones teóricas son mostrados en la Figura N° 5.20, en clara analogía a la Figura N° 2.5:

---

[122] La relatividad general al ser intrínsecamente reversible en el tiempo, no puede explicar el origen de la flecha del tiempo termodinámica.
[123] El que los agujeros negros sean cerrados, innatos y eternos, garantiza que todos los procesos elementales sean reversibles (pues se conserva la masa irreducible) y en consecuencia simétricos.
[124] Esta afirmación es sostenida tanto en la reversibilidad y simetría (sección [2.6.1] y [2.6.3]) como en el teorema de los agujeros negros, sección [2.6.4].
[125] Desde antes de la termodinámica se argumentaba que los agujeros negros sean abiertos por cuestiones astronómicas (que consuman materia estelar), pero al nivel teórico son mejor sustentados por la termodinámica, la cual incluye también su evaporación.

FIGURA N° 5.20

CONTRADICCIONES TEÓRICAS SOBRE LOS AGUJEROS NEGROS

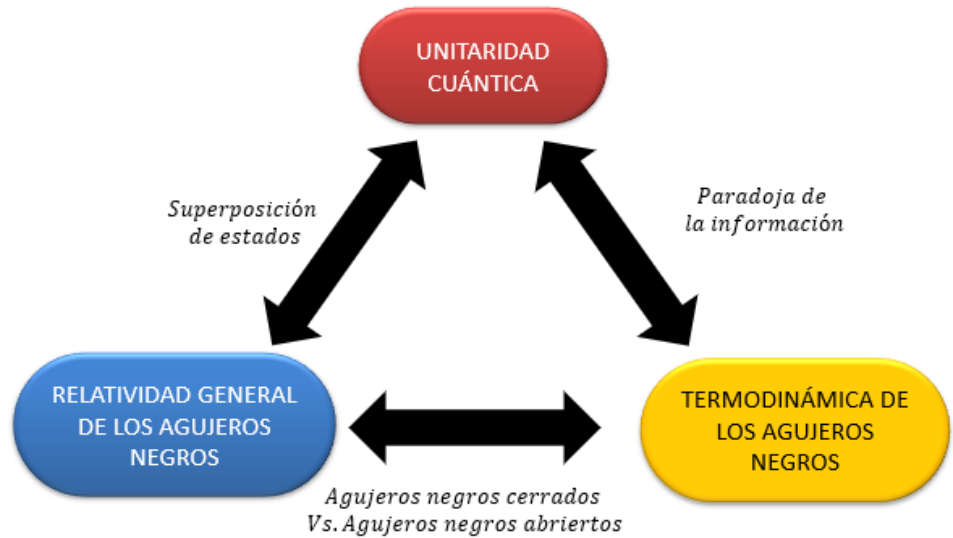

Fuente: Elaboración propia

**5.7.3. La relatividad general y la termodinámica de los agujeros negros como la clasicalización de una (desconocida) teoría de gravedad cuántica**

Esta sección es una extensión natural del programa de la clasicalización (sección [5.4]) a los agujeros negros. En la actualidad no se dispone de una teoría completa de gravedad cuántica, sin embargo, las contradicciones teóricas (sección [5.7.2]) pueden ser abordados por este programa, del teorema de la clasicalización (sección [5.4.2]) aplicado a los agujeros negros se deduce que *"X"* y *"e"* constituyen una teoría completa de gravedad cuántica donde *"X"* es una teoría de gravedad cuántica puramente unitaria y *"e"* son los colapsos (en el marco de la relatividad), lo cual resuelve las contradicciones teóricas dadas en la sección [5.7.2].

En la Figura N° 5.21, en una teoría de gravedad cuántica la evolución unitaria de los agujeros negros son cerrados ($\delta Q = 0$) e innatos (no hay colapso gravitacional) donde la materia está cayendo o escapando eternamente (sin cruzar el horizonte de eventos), con evolución reversible: $dS = \delta Q/T = 0$ (la evolución unitaria del agujero negro conserva su entropía); para la termodinámica de los agujeros negros se requiere $\delta Q \neq 0$ (ingreso de energía-momento o evaporación) con un $dS > 0$ para todo el universo, y sin conservar la información.

## FIGURA N° 5.21
### ACTIVACIÓN DEL PROCESO R EN EL AGUJERO NEGRO

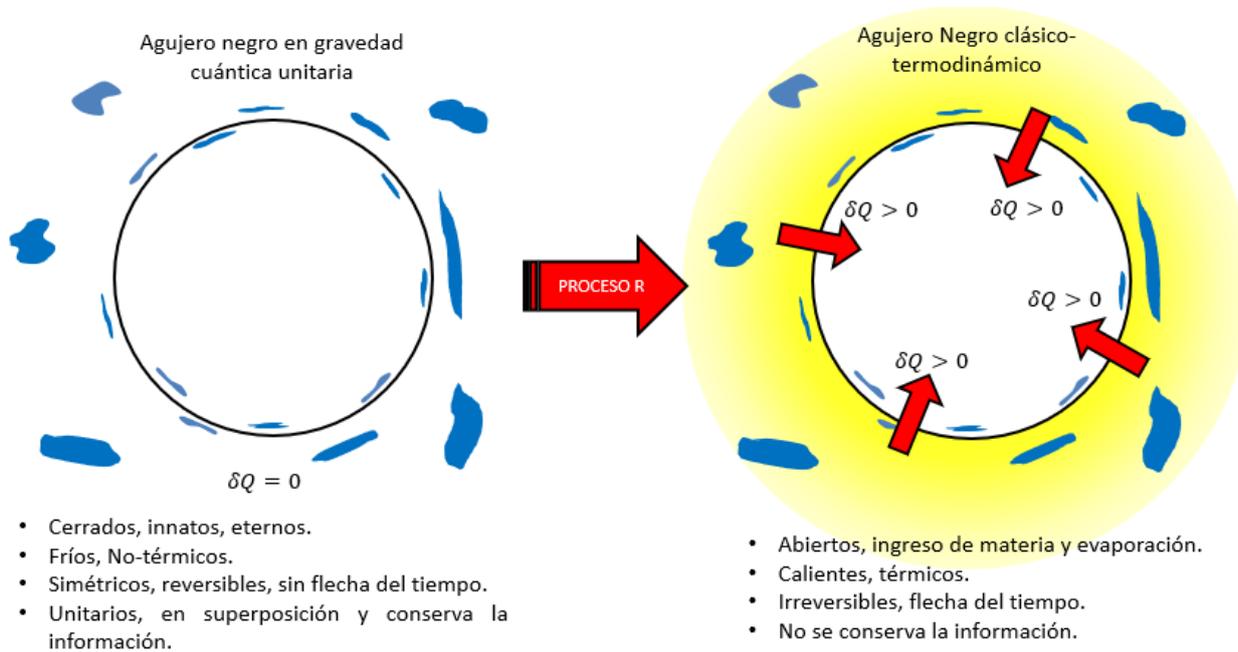

Fuente: Elaboración propia.

Una teoría de gravedad cuántica debe separar explícitamente el proceso U del R, la predominancia de un proceso sobre otro deviene en dos formas de clasicalización, como se muestra en la Figura N° 5.22:

## FIGURA N° 5.22
### CLASICALIZACIÓN DE UNA TEORÍA DE GRAVEDAD CUÁNTICA

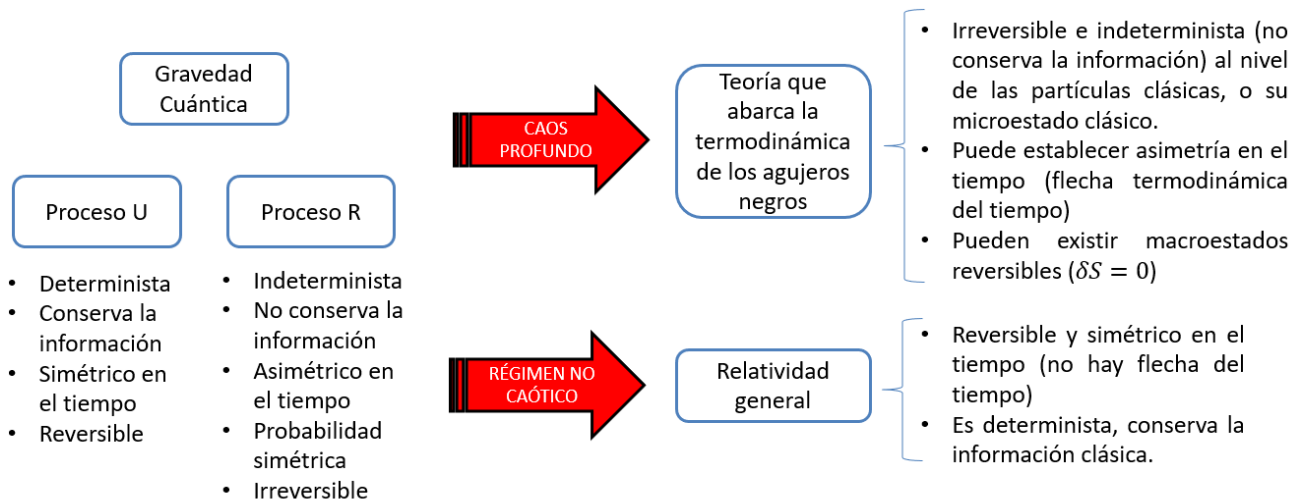

Fuente: Elaboración propia.

El rol de los procesos U y R son presentados en la Tabla N° 5.3:

TABLA N° 5.3

EXPRESIÓN DE LOS PROCESOS U, R AL NIVEL CLÁSICO

|  | **Relatividad general** | **Irreversibilidad termodinámica de los agujeros negros** |
|---|---|---|
| **Proceso U** | Se expresa conservando la información clásica, el determinismo, la reversibilidad y la simetría del tiempo. | No se expresa. |
| **Proceso R** | Se expresa rompiendo la superposición de estados. | • Se expresa suprimiendo la superposición de estados<br>• Se expresa en caos profundo introduciendo el indeterminismo, la irreversibilidad y asimetría en el tiempo. |

Fuente: Elaboración propia

FIGURA N° 5.23

ECUACIONES DESCONOCIDAS DE GRAVEDAD CUÁNTICA

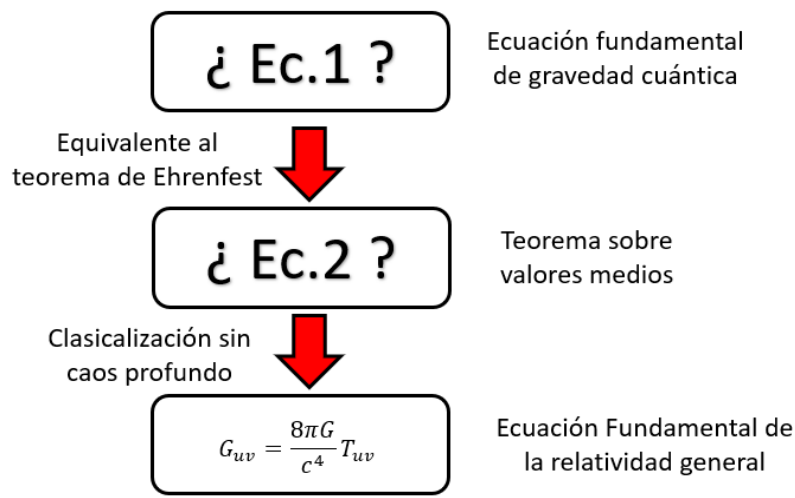

Fuente: Elaboración propia.

En la Figura N° 5.23, el programa de clasicalización sugiere que la ecuación del campo de Einstein [2.56] debe resultar de la clasicalización (sin caos profundo) de una ecuación de valores medios Ec.2, que a su vez resulta de un procedimiento análogo al teorema de Ehrenfest (sección [2.2.2]) a partir de una ecuación fundamental de gravedad cuántica Ec.1 (Ec.1 y Ec.2 hacen referencia a dos ecuaciones desconocidas).

Se puede interpretar la Ec.1 como una asignación de amplitudes de probabilidad a los vectores propios $|\boldsymbol{g}_k\rangle$ y $|\boldsymbol{T}_{k'}\rangle$[126] ($\boldsymbol{g}$ y $\boldsymbol{T}$ es el tensor métrico y de energía-momento) de los observables $\hat{\boldsymbol{g}}$ y $\hat{\boldsymbol{T}}$ (que devuelven los tensores métricos $g_k$ y de energía-momento $T_{k'}$), de manera que el estado cuántico general es de la forma:

$$|\Psi\rangle = \sum_k \psi_k |\boldsymbol{g}_k\rangle \qquad\qquad |\Psi\rangle = \sum_k \varphi_k |\boldsymbol{T}_k\rangle \qquad [5.80]$$

En esta ecuación [5.80] hay información de la interferencia cuántica entre los autoestados, las parejas $(g_k, T_{k'})$ no están limitadas por las ecuaciones de la relatividad general[127]; la solución de la Ec.2 pierde la superposición (evitando así versiones relativistas del gato de Schrödinger[128]) y la información cuántica[129], obteniéndose sólo una ecuación de valores medios (probablemente de la forma de la ecuación del campo de Einstein $\langle\hat{\boldsymbol{G}}\rangle = \frac{8\pi G}{c^4}\langle\hat{\boldsymbol{T}}\rangle$ donde $\boldsymbol{G}$ es el tensor de Einstein[130]), así al nivel clásico se usa la relatividad general como una buena aproximación, pero se está saltando entre los valores medios $\langle\hat{\boldsymbol{G}}\rangle$ y $\langle\hat{\boldsymbol{T}}\rangle$ introduciendo desviaciones (de orden muy inferior a la escala clásica) como en una caminata aleatoria (como se describió en la sección [5.3.1] para el caso no-relativista de los valores medios $\langle\hat{q}\rangle$ y

---

[126] Probablemente no serían una base común para ambos observables (no conmutarían), tampoco serían de índice discreto sino en general de índices continuos.

[127] Como una analogía a la electrodinámica cuántica donde todas las trayectorias son posibles, y no están limitadas por las ecuaciones de la mecánica clásica.

[128] Donde la deslocalización podría llegar a tener un agujero negro deslocalizado en dos regiones macroscópicamente separadas del espacio-tiempo; versiones más realistas y no relativistas han sido comentadas en la sección [5.1].

[129] Al perder la superposición de estados se destruye la información cuántica y se crea la información clásica, la cual por el determinismo de la relatividad general se conserva (fuera de la termodinámica).

[130] También podría ser: $\boldsymbol{G}(\langle\hat{g}\rangle) = \frac{8\pi G}{c^4}\langle\hat{\boldsymbol{T}}\rangle$, donde $\boldsymbol{G}(\langle\hat{g}\rangle)$ es el tensor de Einstein construido a partir del tensor métrico $\langle\hat{g}\rangle$, que es el valor esperado de $\hat{g}$ en $|\Psi\rangle$; o podría adquirir otras formas.

⟨$\hat{p}$⟩ ), el caos profundo aparecería cuando el espacio-tiempo se estira tanto que las desviaciones ya no pueden ser despreciables y la ecuación del campo de Einstein deja de cumplirse estrictamente, esto ocurriría en la proximidad al horizonte de eventos, este caos profundo (como la predominancia de los procesos R sobre los procesos U) conduce a la aparición de la termodinámica en los agujeros negros[131], el determinismo, la reversibilidad y simetría de la relatividad general dejan de ser válido, y la información no es conservada[132], ocurriendo procesos prohibidos por la relatividad general como saltar de una geodésica fuera del horizonte de eventos, a una geodésica al interior del horizonte de eventos, cuando ambas geodésicas se han aproximado los suficiente entre sí (una cantidad finita y no-nula [133]); esto permite el ingreso de materia y energía al interior y con ello la termodinámica de los agujeros negros, este es el rol del proceso R en los agujeros negros.

Las ecuaciones [5.80] deben resultar de una formulación análoga a los caminos de Feynman, donde las métricas pueden adoptar todas las formas geométricamente posibles incluso clásicamente prohibidas o absurdas (como los caminos de Feynman en el espacio-tiempo) con amplitudes cuyas fases varían de acuerdo a la acción de Einstein-Hilbert[134] $S(\boldsymbol{g})$, al superponer sus amplitudes interfieren dando la Ec.1, así también cuando $S(\boldsymbol{g})/\hbar$ varía lentamente las interferencias son constructivas (porque las fases se mantiene casi constantes) y cuando $S(\boldsymbol{g})/\hbar$ varía rápidamente las interferencias son destructivas (porque las fases varían como aleatoriamente), así al nivel clásico la amplitud se hace apreciable para la métrica que resulta de minimizar la acción $\delta S(\boldsymbol{g}) = 0$, y prácticamente se anula en las demás métricas; además, se tienen amplitudes no-nulas para variaciones muy pequeñas de la métrica (que conducen a variaciones en la acción de Einstein-Hilbert del orden de la acción de Planck), lo que justifica las ecuaciones [5.80].

---

[131] Intuitivamente uno puede pensar en atributos termodinámicos intrínsecos, proporcionales al área del horizonte de eventos, al ser regiones donde el espacio y el tiempo se estira infinitamente.

[132] Lo que resuelve así la paradoja de la pérdida de información en los agujeros negros, Anexo [A.2].

[133] Como se vio en las secciones [2.6.2] y [2.6.4] estas geodésicas no tienen continuidad entre sí, aunque se prolonguen hasta el infinito sólo se aproximan asintóticamente.

[134] $S(\boldsymbol{g}) = \int \left[\frac{c^4}{16\pi G} R - \mathcal{L}\right]\sqrt{-g}\, d^4x$ donde $R$ es el escalar de Ricci, $g$ es el determinante del tensor métrico $\boldsymbol{g}$, $\mathcal{L}$ es la lagrangiana de la que se obtiene el tensor de energía-momento: $\boldsymbol{T} = -2\frac{\delta \mathcal{L}}{\delta g} + \boldsymbol{g}\mathcal{L}$, y $\delta S(\boldsymbol{g}) = 0$ conduce a las ecuaciones del campo de Einstein [2.56].

### 5.7.4. Conjetura del reflejo en los agujeros negros

En las secciones [2.1.5], [2.1.6] y [5.3.3] se explicó que cuando un fotón incide en un microsistema (átomo o molécula), que se encuentra en un autoestado de su hamiltoniano, evoluciona unitariamente a una superposición de autoestados de su hamiltoniano sin perturbar (que corresponden a estados en los que se absorbe o emite un fotón), y regresa a su estado inicial cuando el fotón lo atraviesa completamente y se aleja de él, esta transición es reversible y simétrica en el tiempo[135], así la ocurrencia del proceso R durante la interacción permite la absorción y emisión estimulada, que no ocurriría unitariamente[136]; este programa de la clasicalización sugiere que este proceso debe tener su análogo en los agujeros negros (en una teoría completa de gravedad cuántica, aún desconocida) donde los autoestados del hamiltoniano (del microsistema sin perturbar) son análogos a los distintos niveles en la masa irreducible $M_{irr}$ de un agujero negro[137], y el fotón incidente es análogo a partículas que caen (o se alejan) del horizonte de eventos[138]; la interacción unitaria microsistema-fotón es análogo a partículas que pasan cerca del horizonte de eventos, rodeándolo y volviendo a escapar (la asistencia gravitatoria es un proceso reversible), la conjetura que introduzco consiste en que cuando las partículas se aproximan mucho al horizonte de eventos debe aparecer una amplitud de probabilidad no-nula al interior del agujero negro (muy próxima al horizonte de eventos), entonces tiene una probabilidad de ingresar al interior al ocurrir el proceso R; esta idea la planteo en términos generales, pero es necesaria desde este programa de clasicalización, la llamo *conjetura del reflejo* porque la función de onda de las partículas fuera del horizonte de eventos se "reflejan" en el interior (como si de verdad estuvieran atravesando el horizonte de eventos, ignorando la discontinuidad en él)[139], la probabilidad a ambos lados del horizonte de eventos esta normalizada, y la amplitud al interior tiende a anularse cuando la partícula se aleja del horizonte de eventos.

---

[135] En la interacción se tiene un estado entrelazado entre el microsistema y el fotón incidente, antes y después se tienen estados completamente factorizables, de acuerdo a la sección [2.1.3].
[136] Un estado excitado sin perturbar se preserva eternamente.
[137] De manera que el ingreso de materia al agujero negro corresponde con la transición y colapso a un autoestado de mayor masa irreducible.
[138] No siempre van a estar en eterna caída, si el fotón cae con una componente de momento angular no nula puede "rodear" al agujero negro y alejarse de él, como sucede en la lente gravitacional.
[139] Lo mismo para un agujero blanco: se refleja una amplitud en el exterior del agujero negro.

En la Figura N° 2.9 se explicó el ingreso de un cascarón esférico a un agujero negro, en la Figura N° 5.24 se representa la amplitud de probabilidad reflejada al interior del agujero negro conforme el cascarón se aproxima al horizonte de eventos en $r = r_2$, la ocurrencia del colapso permite a los microsistemas del cascarón aparecer en una geodésica del interior como producto de un *salto cuántico* al tiempo $t_1$, finito y posterior a $t_0$, de manera que no se requiere de la eterna caída, ni que las geodésicas sean continuas en el horizonte de eventos:

FIGURA N° 5.24

INGRESO AL INTERIOR DE UN AGUJERO NEGRO

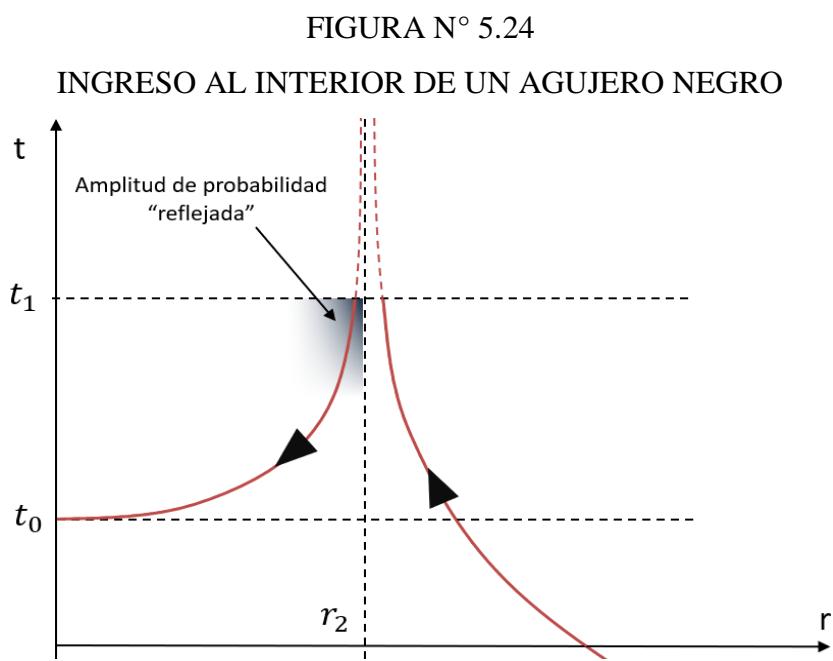

Fuente: Elaboración propia.

Al nivel clásico los colapsos ocurren permanentemente y la materia que cae en el horizonte de eventos se "desgarra" para aparecer al interior de manera aleatoria, ya que cada microsistema tiene una probabilidad de quedarse afuera o ingresar: atravesar el horizonte de eventos es un proceso destructivo que maximiza la entropía.

# CAPÍTULO VI

# DISCUSIÓN DE RESULTADOS

## 6.1. Contrastación de hipótesis con los resultados

### 6.1.1. Contrastación de la hipótesis general

En este trabajo se ha conseguido explicar la clasicalización (sección [5.3]) y el mecanismo de la observación unificados en un marco teórico común donde se alterna el proceso R y el proceso U (sección [5.2]), el colapso del estado cuántico es interpretado como un proceso objetivo (sección [5.6]) en el que la información no es conservada, alcanzando así las metas generales de esta investigación, además he presentado el *programa de la* clasicalización como un marco de trabajo (por ejemplo, sobre gravedad cuántica) y en particular el teorema de la clasicalización que resume este programa.

### 6.1.2. Contrastación de las hipótesis específicas

Las hipótesis específicas son contrastadas de acuerdo al orden establecido en la sección [3.3.2] del CAPÍTULO III:

1.- Aunque se ha dicho que la mecánica cuántica se aplica también al nivel clásico, he demostrado que el mundo clásico es el resultado de una evolución alternada entre los procesos U y R (este último como objetivo, una irrupción en el proceso U), de manera que no se puede aplicar directamente la evolución unitaria de la mecánica cuántica en los sistemas macroscópicos; además, he demostrado que la mecánica clásica es una buena aproximación, y cuando entra al caos profundo deja de ser válida para dar lugar a la termodinámica (la mecánica estadística); así, cada descripción teórica funciona bien en su propio dominio.

2.- El universo no puede evolucionar aisladamente de manera unitaria (por ejemplo, según la ecuación de Schrödinger), la irreversibilidad de los colapsos al nivel microscópico es equivalentes con la irreversibilidad termodinámica al nivel macroscópico (de manera que la mínima irreversibilidad del mundo macroscópico es cada irreversibilidad microscópica del proceso R), así que el hecho que el universo sea térmico-clásico implica su no evolución unitaria.

3.- En la sección [5.2.3] se mostró la equivalencia de enfocar a los sistemas clásicos como compuestos de muchos observadores, o a los observadores como sistemas clasicalizados que "propagan" la clasicalización; en consecuencia, rompen la superposición de estados y están siempre en un autoestado clásico.

4.- En la sección [5.3.4] se explicó el mecanismo de la observación como una interacción irreversible entre un sistema macroscópico y un microsistema, al redistribuir la energía del observador y de la señal física que media la observación.

5.- En la sección [5.2.1] se describió el límite clásico al entrelazamiento cuántico, debido a la activación espontánea del proceso R; en la sección [5.3.2] se demostró que el aumento de la entropía se origina debido al colapso en los estados cuánticos de los microsistemas (lo que rompe el teorema de Liouville), en la sección [5.3.3] se explicó que el rol de los colapsos en los macrosistemas es redistribuir aleatoriamente la energía interna, que es fundamento de la irreversibilidad térmica.

6.- En la sección [5.6] se consideró a la decoherencia como preservando la evolución unitaria al nivel clásico (consistente con la interpretación de Copenhague), en consecuencia, no puede explicar los atributos del mundo clásico, tales como su irreversibilidad termodinámica y la no superposición de autoestados de un observable, ni el mecanismo de la observación; la decoherencia debe ser además reversible en el tiempo si preserva la unitaridad, y no puede explicar el proceso R.

7.- En la sección [5.7.1] se explicó que la relatividad permite un "remanente fantasmal" de la función de onda tras el colapso, sin entrar en contradicción con el principio de complementariedad de Bohr; sin embargo, no se explicó la no-transmisión de información (al colapsar un estado entrelazado) a partir de conceptos más fundamentales, sólo se mostró la consistencia del colapso como un proceso simultáneo en el sistema de referencia del observador (que realiza la medida), y que la no-transmisión de información en el colapso preserva la localidad relativista.

8.- En las secciones [5.7.3] y [5.7.4] se realizó una extensión natural (no forzada) del programa de la clasicalización a los agujeros negros, y se dedujo que la termodinámica de los agujeros negros requiere de los colapsos, con su irreversibilidad y no conservación de la información (indeterminismo).

**6.2. Contrastación de resultados con otros estudios similares**

En las secciones [2.2.3] y [2.2.4] se presentó los modelos de colapso objetivo precedentes, las teorías GRW, CSL y la interpretación de Roger Penrose, las cuales son las primeras teorías en abordar el problema de un colapso objetivo, y en la sección [2.5.4] se presentó *el teorema del agujero negro* cuyos autores sostienen que los agujeros negros son objetos cerrados e innatos.

Las teorías GRW y CSL han servido de inspiración para esta investigación, y por eso se ha considerado los colapsos objetivos como espontáneos y emergiendo en un conjunto de muchos microsistemas, en este trabajo las discusiones se iniciaron al problematizar sobre el rol de la información física en los colapsos objetivos, lo que condujo a su no-conservación en virtud del indeterminismo; posteriormente la investigación fue orientada al problema de la irreversibilidad termodinámica frente a la reversibilidad del proceso unitario y de la mecánica clásica; luego se abordó la *localización espontánea* (postulado fundamental en la teoría GRW, debido al colapso objetivo) como consecuencia de una dinámica más fundamental: la redistribución de la energía (recién aquí es postulado el colapso objetivo como fundamental), este programa de clasicalización dista de la teoría GRW en que esta última no "apaga" (al menos, explícitamente) los colapsos en la proximidad al cero absoluto.

El *teorema del agujero negro*, sección [2.5.4], establece que los agujeros negros son cerrados e innatos, esta interpretación de las ecuaciones de la relatividad general es consistente con el programa de la clasicalización (aplicado a los agujeros negros), sección [5.7.3]; los autores del referido teorema señalan que es demostrado considerando solamente la relatividad general, sin usar la mecánica cuántica; en este programa de clasicalización el proceso R de la mecánica cuántica permite "abrir" los agujeros negros para hacerlos consistentes con la termodinámica.

# CAPÍTULO VII

# CONCLUSIONES

**7.1. Conclusión general**

1.- Al hacer un cambio de interpretación (con respecto a la oficial) sobre la naturaleza del colapso (o, dicho de otra forma, si la unitaridad se preserva en todo el sistema aislado) se puede explicar la clasicalización y el mecanismo de la observación, y como consecuencia la información no se conserva.

**7.2. Conclusiones específicas**

1.- La mecánica cuántica y la física clásica (mecánica y termodinámica) son teorías que funcionan muy bien en sus respectivos dominios (los microsistemas y los macrosistemas, respectivamente), en particular la física clásica es una buena aproximación, y no es requerido extender una teoría al dominio de la otra.

2.- El universo como un sistema aislado contiene una dinámica intrínsecamente indeterminista e irreversible, se crea mucha información con el paso del tiempo y no evoluciona unitariamente, en particular, no evoluciona según la ecuación de Schrödinger.

3.- El mundo clásico es un mundo de observadores cuánticos, todos los entes clásicos, desde la partícula clásica, son observadores y se localizan así mismos.

4.- El mecanismo de observación ocurre porque el observador redistribuye la energía de la señal (que proviene de interactuar con el microsistema entrelazado) con su energía interna, y al ocurrir el colapso objetivo como parte de su clasicalización, colapsa también el microsistema por estar entrelazado, lo que lo localiza en un autoestado del observable en cuestión.

5.- La irreversibilidad termodinámica, el aumento de la entropía, requiere siempre de la redistribución de las diversas formas de energía almacenada en los macrosistemas, lo cual requiere al nivel microscópico que los microsistemas colapsen sus funciones de ondas para intercambiar sus "paquetes de energía" (fotones y fonones) entre ellos, y con su campo de radiación termal; así la irreversibilidad del colapso objetivo se expresa al nivel clásico-macroscópico en la irreversibilidad termodinámica.

6.- La decoherencia (preservando la unitaridad en todo el sistema aislado) no explica el colapso ni la clasicalización porque sólo explica la pérdida de la coherencia, pero no de la superposición, además parte de un entorno ya clasicalizado.

7.- El colapso objetivo en el marco relativista no viola la causalidad ni la localidad relativista porque no hay envío de señal superlumínica, el hecho que la simultaneidad sea relativa permite una "función de onda fantasmal" sin mayor relevancia física.

8.- El colapso objetivo permite a los agujeros negros tener su termodinámica, y en especial los fenómenos térmicos irreversibles como el ingreso de materia y energía a su interior, así como su radiación térmica; en una teoría de gravedad cuántica donde los colapsos están apagados, la evolución unitaria de los agujeros negros los mantiene reversibles, cerrados y sin termodinámica.

# CAPÍTULO VIII

# RECOMENDACIONES

1.- Investigar el colapso del estado cuántico, en el marco de este programa de clasicalización, sobre sus causas o condiciones físicas que provocan o facilitan su activación en un conjunto de microsistemas.

2.- Usar la teoría de la información física que se presentó (Anexo B), para investigar sobre la conservación de la información, el indeterminismo y la entropía.

3.- Evitar usar la decoherencia cuántica para explicar el colapso o la clasicalización, siempre que preserve la unitaridad de todo el sistema general aislado.

4.- Al investigar sobre gravedad cuántica considerar el programa de la clasicalización, sección [5.7.3], en vez del "límite clásico" (esto es, hacer $\hbar \to 0$, $\delta S = 0$ y $c \to \infty$ para el caso no-relativista), y que hay dos maneras de clasicalizar una teoría de gravedad cuántica, donde no se conserva la información cuántica: como relatividad general, y como termodinámica (en particular, de los agujeros negros).

# CAPÍTULO IX

# REFERENCIAS BIBLIOGRÁFICAS

## 9.1.- Bibliografía

## 9.2.- Referencias electrónicas

# ANEXO 1: MATRIZ DE CONSISTENCIA

Como en esta investigación no se emplea población ni muestra no es considerado en esta matriz de consistencia, por razones de espacio.

| PROBLEMA | OBJETIVOS | HIPÓTESIS | VARIABLES | METODOLOGÍA |
|---|---|---|---|---|
| **PROBLEMA GENERAL** ¿Cómo solucionar los problemas de la medida y la clasicalización en una interpretación de colapso objetivo donde no se conserva la información? **PROBLEMAS ESPECÍFICOS** ¿Cuál es el mecanismo de observación? ¿Cómo es la transición del mundo cuántico al mundo clásico? ¿Cómo se relacionan las irreversibilidades del colapso y la termodinámica? ¿Cómo se interpreta el indeterminismo y la irreversibilidad? ¿Puede la decoherencia resolver el problema de la medida y la clasicalización? ¿Cuál es el rol de los colapsos en la termodinámica de los agujeros negros? | **OBJETIVO GENERAL** Resolver y explicar los problemas de la medida y la clasicalización a través de una interpretación de colapso objetivo donde no se conserva la información. **OBJETIVOS ESPECÍFICOS** Reconocer y unificar el problema de la medida y la clasicalización, Explicar el rol de la termodinámica en los dos problemas unificados. Describir esta nueva interpretación, de tipo colapso objetivo. Aplicar la interpretación a ciertas situaciones físicas. Replicar sobre la decoherencia en el problema de la medida y la clasicalización. Aplicar la nueva interpretación al dominio relativista, en el sentido de no propagar señales superlumínicas. Explicar el rol del colapso en la termodinámica de los agujeros negros. | **HIPÓTESIS GENERAL** Los problemas de la medida y la clasicalización se resuelven y explican en un modelo de colapso objetivo donde no se conserva la información. **HIPÓTESIS ESPECÍFICAS** Cada teoría funciona bien en su propio dominio, la clasicalización y la observación rigen entre ellas. El universo no evoluciona unitariamente porque es termodinámicamente irreversible, debido a los colapsos. Los objetos clásicos no se encuentran en superposición de estados porque viven en un mundo de observadores que los colapsan. La observación ocurre porque el observador se enreda con el microsistema y el estado general colapsa. La irreversibilidad termodinámica se origina al nivel microscópico en los colapsos cuánticos, la asimetría temporal no puede reducirse a la mecánica clásica sino a los colapsos. La decoherencia no puede resolver la clasicalización porque no explica el colapso. Los microsistemas no violan la localidad porque no hay envío de información clásica superlumínica. Los colapsos objetivos son responsables de la termodinámica en los agujeros negros. | **VARIABLE INDEPENDIENTE: MECÁNICA CUÁNTICA** La mecánica cuántica: De la evolución unitaria, colapso del estado cuántico, entrelazamiento cuántico, perturbación dependiente del tiempo, emisión y absorción de fotones. **VARIABLE DEPENDIENTE: MECÁNICA CLÁSICA Y EL PRINCIPIO DEL INCREMENTO DE LA ENTROPÍA** 1. La reversibilidad de la mecánica clásica 2. La irreversibilidad de la termodinámica en el incremento de la entropía | **Tipo:** La presente es una investigación básica. **Diseño de la investigación:** La hipótesis (general) a demostrar consta de la siguiente estructura: $A\ y\ B\ se\ resuelven\ en\ (X, a)$ Donde: $A$: Problema de clasicalización $B$: Problema de la medida $X$: Modelo que resulta de reinterpretar la mecánica cuántica, con colapso objetivo $a$: No conservación de la información De manera que primero se construye un $(X, a)$, y la hipótesis se demuestra al resolver $A\ y\ B$ usando $(X, a)$; es decir, construyendo un modelo que resulta de aplicar las reglas de la mecánica cuántica donde el colapso es un proceso objetivo en el que no se conserva la información, y con él resolver el problema de la medida y la clasicalización; esta demostración no es trivial porque los problemas $A\ y\ B$ no tienen una respuesta física adecuada y son tratados con generalidades. |

# ANEXO A: Paradojas físicas

## A.1.- Paradoja del gato de Schrödinger

En una caja cerrada un sistema microscópico (un átomo radiactivo excitado) es enredado con un sistema clásico, térmico y macroscópico (un gato) de manera que si el átomo decae y emite una partícula desencadena una serie de eventos que mata al gato, por el contario si el átomo no decae y la partícula no es emitida el gato permanece vivo; según la interpretación oficial se tendría una superposición coherente de todo el sistema gato-átomo radioactivo en una caja, y que sólo al observar se encontraría al gato o vivo o muerto puesto que la medida provoca el colapso del estado general; en la interpretación oficial se reconoce al observador como un objeto clásico, lo que conduce a dos suposiciones:

1. El Gato actúa como un observador (como suposición de que todo sistema clásico es un observador), y al enredarse con el microsistema provoca un colapso relativo al gato, pero todo el sistema aislado sigue evolucionando según la ecuación de Schrödinger hasta que un segundo observador (el que abre la caja para medir) provoca el colapso de la función del onda del átomo y del gato enredados, pero el sistema total aislado (átomo, gato y observador que abre la caja) siguen evolucionando unitariamente.
2. El gato no actúa como observador (no todo sistema clásico sería un observador) pero si se puede enredar con el microsistema, luego otro objeto clásico "exótico" (el observador) provoca el colapso de la función de onda, el sistema general aislado (incluyendo al observador) sigue evolucionando unitariamente.

La primera hipótesis es más consistente que la segunda, pues no requiere de elementos exóticos, no obstante, lleva a un objeto clásico (gato) a una superposición de estados (vivo y muerto) que es inconsistente con la física clásica, incluso si los sistemas no son observados.

## A.2.- Paradoja de la pérdida de información en los agujeros negros

En la sección [2.6.6] se estableció la termodinámica de los agujeros negros, y que al tener una temperatura y entropía no-nula emite calor en forma de radiación termal, lo que consume la masa-energía del agujero negro hasta su eventual evaporación (siempre que consuma toda la materia a su alrededor y que la temperatura del fondo cósmico sea menor al del agujero negro), la evolución unitaria de un sistema preserva su reversibilidad y determinismo, y con ello la conservación de la información cuántica, así al caer este sistema al interior del agujero negro se esperaría que su información se conservara "en el agujero negro" si se asume la interpretación de Copenhague donde la unitaridad se aplica a todo nivel; este problema es similar a si la información cuántica se conserva tras los procesos clásicos termodinámicos (como la combustión, por ejemplo), la particularidad de los agujeros negros es que nos fuerza a responder a esta cuestión porque al evaporarse debe emitir una radiación termal independiente de la información del sistema que cayó en él, y sin quedar nada más (entonces la información ha desaparecido), la radiación termal no debe poseer la información cuántica del sistema que ingresó porque esto supondría una violación del teorema de Liouville cuántico (ecuación [2.17]) donde un ensemble (incluso en estado puro) evoluciona unitariamente a un estado mezcla de máxima entropía (la radiación termal), ya que el sistema y el agujero negro al estar aislados deben preservar la unitaridad según la interpretación oficial (que limita la aplicación del proceso U al mundo clásico), o equivalentemente, si la radiación de Hawking emitida portara la información cuántica, entonces la radiación presentaría correlaciones y dejaría de ser termal (con entropía maximizada), la paradoja está en que la información cuántica se pierde (se viola la unitaridad) en la evolución aislada del agujero negro y el sistema que cae en él, juntos.

Se ha pretendido solucionar esta paradoja mediante el *principio holográfico* o la *pared de fuego* (incluso S. Hawking teorizó que sobre la superficie del agujero negro hay "pelos" en los que se almacenara la información), de manera que la unitaridad es preservada en ambos, actualmente son las posturas predominantes en la física teórica sobre la gravedad cuántica.

## ANEXO B: Teoría física de la información

Existe la *teoría matemática de la comunicación*, publicada por C. Shannon en 1948, más conocida como "*Teoría de la Información*", en la física teórica es usado en la equivalencia entre la entropía de información y la entropía estadística. En este anexo se presenta este desarrollo propio (*Teoría física de la información*) en el que se define matemáticamente la información y su conservación (debido al determinismo).

### B.1.- Formalismo matemático

#### B.1.1.- Los conjuntos mensurables de información

Sea $\Omega$ el espacio muestral de una variable aleatoria $X$, tras adquirir una información $I$ se reduce la incertidumbre de $\Omega$ a una región $A \subset \Omega$, este es un proceso indeterminista[140] y ocurre con probabilidad $P[A/\Omega]$, así se define matemáticamente la información $I$ usando conjuntos mensurables, cuya medida $|I|$ depende de la probabilidad, como se muestra:

$$I = \{\Omega \to A\} \qquad |I| = -\ln(P[A/\Omega]) \qquad [B.1]$$

La información $I$ es interpretada como una reducción de la ignorancia, y su medida $|I|$ es expresado en *nat* como unidad natural de información[141], así el evento certero $P[\Omega/\Omega] = 1$ no aumenta la información: $I = \{\Omega \to \Omega\} = \emptyset$, y $|I| = 0$, conforme se aumenta la información la probabilidad se reduce; cuando se tienen dos variables aleatorias $X$ y $Y$, sus informaciones $I(x)$ e $I(y)$ pueden tener elementos en común si las variables aleatorias presentan algún tipo de dependencia o correlación, como se muestra:

FIGURA B.1

CONJUNTOS DE INFORMACIÓN

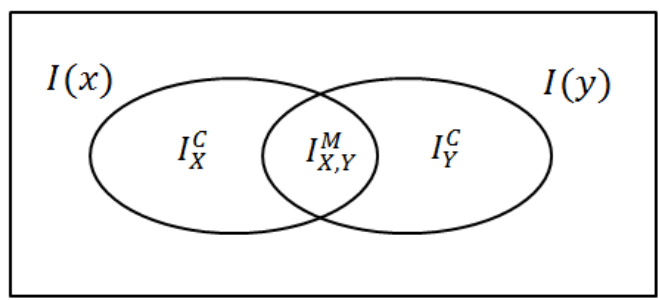

Fuente: Elaboración propia.

---

[140] Porque no se puede conocer a priori la información que se va adquirir.
[141] 1 nat corresponde a 1 bite/$ln(2)$.

La intersección $I_{X,Y}^M = I(x) \cap I(y)$ es la *información mutua*, $I(x,y)$ es la información conjunta o total, $I_X^C$ e $I_Y^C$ son las informaciones contribuidas, entonces se tienen las relaciones:

$$I(x,y) = I(x) \cup I(y) \qquad I_X^C = I(x) - I_{X,Y}^M$$

$$I_{X,Y}^M = I(x) \cap I(y) \qquad I_Y^C = I(y) - I_{X,Y}^M$$

[B.2]

Todas ellas con medidas, según [B.3]:

$$|I(x)| = -\ln[P(x)] \qquad |I_{X,Y}^M| = \ln[\frac{P(x,y)}{P(x)P(y)}]$$

$$|I(y)| = -\ln[P(y)] \qquad |I_X^C| = -\ln[P(x/y)]$$

[B.3]

$$|I(x,y)| = -\ln[P(x,y)] \qquad |I_Y^C| = -\ln[P(y/x)]$$

$P(x,y)$ es la probabilidad conjunta, $P(x)$ y $P(x)$ son las probabilidades marginales, y $P(x/y)$ y $P(y/x)$ son las probabilidades condicionales; las correlaciones entre variables (que da las probabilidades mencionadas) son mostradas en la Figura B.2:

FIGURA B.2

CORRELACIÓN DE VARIABLES X-Y

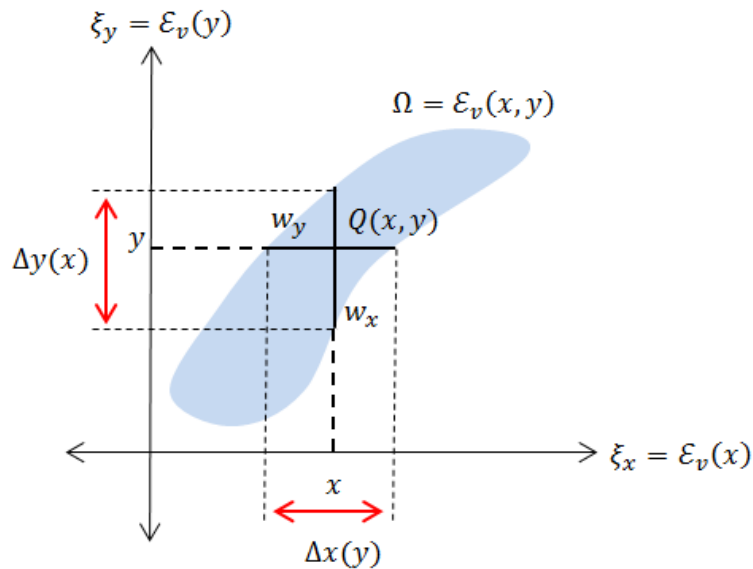

Fuente: Elaboración propia.

Donde $\Omega$ es el espacio muestral de ambas variables en conjunto ($P(x,y) \neq 0$); Las informaciones dadas en [B.2] se representan explícitamente (de acuerdo a la Figura B.2):

$$I(x,y) = \{ \Omega \to Q \} \qquad\qquad I_{X,Y}^M = \{\mathcal{E}_v(x,y) \to W(x,y)\}$$

$$I(x) = \{\mathcal{E}_v(x) \to x\} = \{\mathcal{E}_v(x,y) \to w_x\} \qquad I_X^C = \{\Delta x(y) \to x\} \qquad [\text{B.4}]$$

$$I(y) = \{\mathcal{E}_v(y) \to y\} = \{\mathcal{E}_v(x,y) \to w_y\} \qquad I_y^C = \{\Delta y(x) \to y\}$$

### B.1.2.- Entropía de información

Al nivel estadístico se define la entropía de información $H$ como la media $\langle \ \rangle$ de la cantidad de información $|I|$ multiplicado por el número $N$ de datos, para las cantidades dadas en [B.3] se obtienen las mismas fórmulas de la *teoría matemática de la comunicación*:

$$H(x) = N \langle |I(x)| \rangle = -N \sum_x P(x) \cdot \ln[P(x)]$$

$$H(X;Y) = N\langle |I(x,y)| \rangle = -N \sum_{x,y} P(x,y) \ln[P(x,y)]$$

[B.5]

$$H_{X,Y}^M = N\langle |I_{X,Y}^M| \rangle = N \sum_{x,y} P(x,y) \ln[\frac{P(x,y)}{P(x) \cdot P(y)}]$$

$$H(X/Y) = N\langle |I_X^C| \rangle = -N \sum_{x,y} P(x,y) \ln[P(x/y)]$$

Aquí se suma sobre los valores discretos de $(x,y)$, si es una variable discreta, o se integra sobre una densidad de probabilidad $p(x,y) = dP/dxdy$, si es una variable continua; la entropía mutua $H_{X,Y}^M$ es conocido en la teoría de Shannon como *Información mutua*.

### B.1.3.- Representación y transformaciones

Sean las variables $X$ y $Y$ completamente correlacionadas ($I_X^C = I_Y^C = \emptyset$), de manera que el conocimiento de una permite el conocimiento de la otra; es posible establecer transformaciones reversibles como una función biunívoca entre ellas, mediante el operador $\hat{T}$:

$$\forall \ \hat{T}: I_x \to I_y, \qquad \exists! \ \hat{T}^{-1}: I_y \to I_x \qquad [\text{B.6}]$$

Entonces el operador $\hat{T}$ es reversible y preserva la información (porque permite recuperarla), y en consecuencia conserva su medida: $|I_x| = |I_y|$; las relaciones [B.6] nos indica que existe una información "$Y$" que va de una representación a otra y que no depende del lenguaje o la representación empleada, $Y$ es el invariante de $I_x$ e $I_y$ y se proyecta a cada representación con los operadores del lenguaje $\hat{L}_x$ y $\hat{L}_y$ para cada variable, como se muestra:

$$\hat{L}_x \cdot Y = I_x \quad , \quad \hat{L}_x \cdot Y = I_y \qquad [B.7]$$

Donde: $|Y| = |I_x| = |I_y|$; las ecuaciones [B.6] y [B.7] se pueden generalizar a un conjunto de transformaciones de lenguajes, sean los operadores $\hat{L}_{j,i}$ que cumplen:

$$I_i = \hat{L}_i * Y \qquad I_j = \hat{L}_{j,i} * I_i \qquad \hat{L}_j = \hat{L}_{j,i} * \hat{L}_i \qquad [B.8]$$

Con el elemento inverso: $\hat{L}_{i,j}^{-1} = \hat{L}_{j,i}$; los operadores forman los elementos de un grupo $G$ (la operación del grupo es la composición de los operadores):

$$G = (\{\hat{L}_{i,j}\}, *) \qquad [B.9]$$

**B.2.- Información de un sistema físico**

Se puede tratar la información de un sistema físico, ya sea clásico o cuántico, con el mismo formalismo teórico dado en la sección precedente [B.1]. La información cuántica se distingue de la clásica en que la primera es una reducción de la ignorancia en el espacio de Hilbert, mientras que la otra lo es en el espacio de fases.

**B.2.1.- En mecánica clásica**

La información de un sistema de partículas[142] es la reducción de la ignorancia desde el espacio de fases del sistema X (como el espacio muestral) hasta un punto $(q_i, p_i)_{i=1}^{N}$ que representa su estado (como un resultado aleatorio), así la información física del sistema es:

$$I_s = \{X \to (q_i, p_i)_{i=1}^{N}\} \qquad [B.10]$$

En principio existen infinitos estados, con probabilidades prácticamente nulas, luego la cantidad de información es infinita; al nivel estadístico se calcula la entropía sobre una región $\Gamma \subset X$ en el que se distribuye la probabilidad, según las ecuaciones [B.5].

---

[142] Con coordenadas y momentos $q_i$ y $p_i$, $i = 1, 2, \ldots, N$, se encuentra definido como un punto $(q_i, p_i)_{i=1}^{N}$ en el espacio de fases de $2N$ dimensiones.

**B.2.2.- En mecánica cuántica**

El estado de un sistema microscópico queda definido por su ket $|\psi\rangle$ que es un vector en el espacio de Hilbert[143] $\mathcal{H}$, y contiene toda la información del microsistema; su información es la reducción de la ignorancia desde $\mathcal{H}$ (como espacio muestral) hasta el punto $|\psi\rangle \in \mathcal{H}$ (como evento aleatorio), que es esencialmente la información de sus amplitudes:

$$I(|\psi\rangle) = \{\mathcal{H} \rightarrow |\psi\rangle\} \qquad [B.11]$$

Al nivel estadístico existe una región en el espacio de Hilbert $A \subset \mathcal{H}$, con probabilidades no nulas (sobre el microestado de un ensemble de sistemas microscópicos) que corresponde al estado mezcla, sobre la cual se define la entropía de Von Newman.

**B.3.- Información en la evolución de un sistema físico determinista**

En la mecánica clásica y cuántica la evolución de un sistema físico está dado por las ecuaciones canónicas de Hamilton (ecuación [2.48]) y por la ecuación de Schrödinger (ecuación [2.1]), ambas son deterministas porque dado un punto (condición inicial) se obtiene una trayectoria en su espacio correspondiente, como un punto que evoluciona con el tiempo, el cual es reversible (se puede recuperar el estado inicial tras haber), esto es la conservación de la información debido a transformaciones reversibles, como se muestra:

$$I(t) = \hat{E}(t) * (Y + i_0) \qquad I(t') = \hat{E}(t',t) * I(t) \qquad \hat{E}(t') = \hat{E}(t',t) * \hat{E}(t) \qquad [B.12]$$

[B.12] es en clara analogía a las ecuaciones [B.8], donde $I(t)$ es la información (observable) del sistema físico al tiempo $t$, $\hat{E}(t',t)$ son los operadores que transforman la información del tiempo $t$ al tiempo $t'$, y $\hat{E}(t)$ es el operador de proyección de la información $Y + i_0$ en el tiempo $t$ como "una representación"; $Y + i_0$ al ser invariante al tiempo es conservada y no es observable (por encontrarse "fuera del tiempo"), $i_0$ es la *información de tiempo cero* que fija una proyección $I$ a un tiempo dado $t_0$, la información $Y$ no tiene ninguna referencia al tiempo; esto puede ser resumido en el siguiente enunciado:

"*En un sistema aislado que evoluciona por leyes físicas deterministas se conserva la información*"

---

[143] Realmente, desde todos los puntos $|\psi\rangle \in \mathcal{H}$ donde $|\psi\rangle$ está normalizado.

Es conveniente agregar la ecuación $\hat{E}(t', t) = \hat{E}(t' - t, 0)$ como una propiedad de los operadores de evolución, de que las leyes físicas no dependen del tiempo, son así operadores de avance o retroceso en el tiempo; los operadores $\hat{E}(t', t)$ constituyen un grupo como en [B.9]; estas ecuaciones [B.12] establecen la conservación de la información, y son invariantes a la formulación clásica o cuántica unitaria:

$$\frac{d}{dt}|Y| = 0 \qquad [B.13]$$

**B.4.- Redundancia y conservación de la entropía**

Desde que la información puede intersecarse, entonces es posible presentar redundancia: en la Figura B.1 la información mutua $I_{X,Y}^M$ en una variable es redundante porque ya se encuentra presente en la otra variable; también se puede interpretar que en las transformaciones de lenguajes, sección [B.1.3], la información puede aumentar o disminuir su redundancia pero conserva siempre la misma medida (la información invariante $Y$), así que la información no-redundante es una medida de la mínima información que presenta un sistema; al nivel estadístico la información no redundante ha sido descrito por la entropía de información en la *teoría matemática de la comunicación* de Shannon; así aunque la evolución física de un sistema pueda variar la redundancia al proyectar la información $Y_s$ (de las ecuaciones [B.12]) al tiempo $t$, $Y_s$ debe ser no-redundante a lo largo de toda la evolución temporal puesto que esas redundancias lo introducen los operadores $\hat{E}(t)$ y $\hat{E}(t', t)$, entonces la conservación de $|Y|$ establece la conservación de la información no-redundante, y con ello de la entropía del sistema; esto es, que la entropía se conserva en evoluciones deterministas[144].

---

[144] Es importante imaginar lo descrito: en el caos determinista, al inicio se parte de un estado muy ordenado que evoluciona a un estado más complejo y "desordenado", sin embargo, la información mínima para describirlo en cualquier instante del tiempo ( $Y_s$ ) no cambia, sólo ha aumentado su redundancia.

# ANEXO C: Herramientas estocásticas

## C.1.- Caminata aleatoria

La caminata aleatoria (*Random Walks*) es una técnica de modelación estocástica para caminos que resultan de pasos aleatorios; en este anexo se ha descrito una partícula que avanza o retrocede en una dimensión (la generalización al caso de más dimensiones es inmediata).

Sea una partícula que salta una distancia $d$, con probabilidad $p$ hacia la derecha y con probabilidad $1-p$ hacia la izquierda, así el que tras $n$ saltos $m \leq n$ hayan sido hacia la derecha, viene dado por la distribución binomial $X \sim B(n,p)$, con probabilidad $P(m)$:

$$P(m) = \binom{n}{m} p^m (1-p)^{n-m} \ , \ m = 0, 1, 2, 3, \ldots n \qquad [C.1]$$

Si inicia en $x = 0$, entonces se encuentra en la posición $x = (2m-n)d$, y con desviación estándar $\sigma = \sqrt{np(1-p)}$; sí $p = 0.5$ (el caso simétrico) entonces $X \sim B(n, 1/2)$ está centrada en $n/2$, y $P\left(m = \frac{k+n}{2}\right)$ es la probabilidad para $x = kd$, con $k = -n, \ldots n$, y $\sigma = \sqrt{n}/2$, centrada en $k = 0$; cuando ocurren muchos saltos se toma el límite cuando $n$ tiende a infinito (y $d$ tiende a cero), la distribución binomial $X \sim B(n \to \infty, p = 0.5)$ se aproxima a una distribución normal (teorema del límite central) con función de probabilidad:

$$P(k) = \frac{1}{\sigma\sqrt{2\pi}} e^{-\frac{(k)^2}{2\sigma^2}} \ , \ x = kd, \sigma = \sqrt{n}/2 \qquad [C.2]$$

## C.2.- Distribución de Poisson

La distribución de Poisson modela fenómenos que ocurren aleatoriamente con un valor medio $\langle u \rangle = n/\Delta u$ sobre un intervalo de alguna magnitud[145], sólo tiene un único parámetro $\lambda$ adimensional (definido en el intervalo $\Delta u'$: $\lambda = n\frac{\Delta u\prime}{\Delta u} = \langle u \rangle \Delta u'$, $\lambda$ indica el valor esperado de $n$ en $\Delta u'$, en vez de $\Delta u$)[146]; así, la probabilidad que en ese intervalo $\Delta u'$ se encuentren u ocurran $k$ veces el evento, viene dado por:

$$P(k, \lambda) = \frac{e^{-\lambda}\lambda^k}{k!} \ , \qquad k = 0, 1, 2, \dots \infty \qquad [C.3]$$

---

[145] De $n$ eventos por cada intervalo de magnitud $\Delta u$ (por ejemplo, el número de partículas o eventos $n$ por longitud $\Delta l$, área $\Delta a$, volumen $\Delta v$, o tiempo $\Delta t$).

Donde la media de esta distribución es justamente $\lambda$ (igual su varianza).

Un caso especial es de eventos aleatorios en el tiempo, sea una secuencia aleatoria de pulsos con tiempo medio entre pulsos $\tau$ ( $\langle u \rangle = 1/\tau$ ) que sigue la distribución de Poisson, la probabilidad de contar $k$ pulsos en un intervalo de tiempo $\Delta t$ es $P(k, \Delta t/\tau)$, pues $\lambda = \Delta t/\tau$ en la fórmula [C.3], la densidad de probabilidad en el tiempo de ocurrir un pulso, a un tiempo $t$ de ocurrido un pulso precedente, requiere de variar $\lambda$ con $t$ ($\lambda(t) = t/\tau$), la probabilidad que no ocurra un pulso ($k = 0$) entre $t$ y $t + dt$ es $dP = P(0, \lambda(t)) - P(0, \lambda(t + dt)) > 0$, así la densidad de probabilidad en el tiempo $\rho(t) = dP/dt$ viene a ser (considerando [C.3]) la distribución exponencial con parámetro $\lambda_\tau$:

$$\rho(t) = \lambda_\tau e^{-\lambda_\tau t} \ , \qquad t \geq 0 \qquad [C.4]$$

Donde $\lambda_\tau = \frac{1}{\tau}$ tiene dimensión de frecuencia, $P(t) = \int_0^t \rho(t')dt'$ es la probabilidad que no ocurra ninguna señal hasta el tiempo $t$ (o equivalentemente, que la señal ocurra para un tiempo mayor a $t$ ), y $\rho(t)$ se hace muy pequeño para tiempos mucho mayor a $\tau$.

---

[146] Así por ejemplo si en un campo hay en promedio tres plantas por cada metro y medio cuadrado de área, $n = 3$, $\Delta u = 1.5 \text{ m}^2$ y $\langle u \rangle = 2 \text{ m}^{-2}$, si la distribución de Poisson se evalúa sobre un área de $10\,000 \text{ m}^2$ entonces $\Delta u' = 10\,000 \text{ m}^2$ y $\lambda = 20\,000$ (adimensional) que es el valor esperado en una hectárea.

**ANEXO D: Simulación de la distribución de energía en un gas ideal clásico**

Sea un gas de $N$ partículas, por exceso de idealización con un solo grado de libertad en su energía, que al inicio tienen una energía $E_i$ asignadas aleatoriamente como $E_i = rE_0$, donde $r \in [0; 1]$ es una variable aleatoria uniforme en el intervalo dado, el valor medio de la energía de las partículas es $\langle E_i \rangle = \frac{E_0}{2}$ y la densidad de probabilidad inicial (de la energía) es $\wp_0(E) = 1/E_0$ si $E \in [0; E_0]$, y es $\wp_0(E) = 0$, si $E > E_0$ (una distribución escalón); al correr la simulación computacional, se eligen dos partículas al azar "$i$" y "$j$" y se intercambian sus energías aleatoriamente, obteniéndose $E'_i$ y $E'_j$ con el uso de la variable aleatoria $r \in [0; 1]$:

$$E'_i = r \cdot (E_i + E_j) \qquad\qquad E'_j = E_i + E_j - E'_i \qquad [D.1]$$

Donde la energía total se conserva; luego las partículas "actualizan" su energía y se repite el proceso iterativamente, cada cierto número de iteraciones se toma muestra de la densidad de probabilidad de la energía, contando cuántas partículas tienen energía entre $Q_i$ y $Q_{i+1}$ (donde $Q_i$ es una partición en el espectro de energías de ancho fijo) y dividiendo entre $N$, de manera que $\wp(i) = [n(Q_{i+1}) - n(Q_i)]/N$; así para una simulación de $N = 10^5$, $E_0 = 100$meV, contando y mostrando la distribución $\wp(i)$ cada $n_0$ iteraciones de intercambio de energía, $n$ veces ($n_0$ y $n$ se eligen al ejecutar el programa).

FIGURA D.1

SIMULACIÓN DE LA DENSIDAD DE PROBABILIDAD

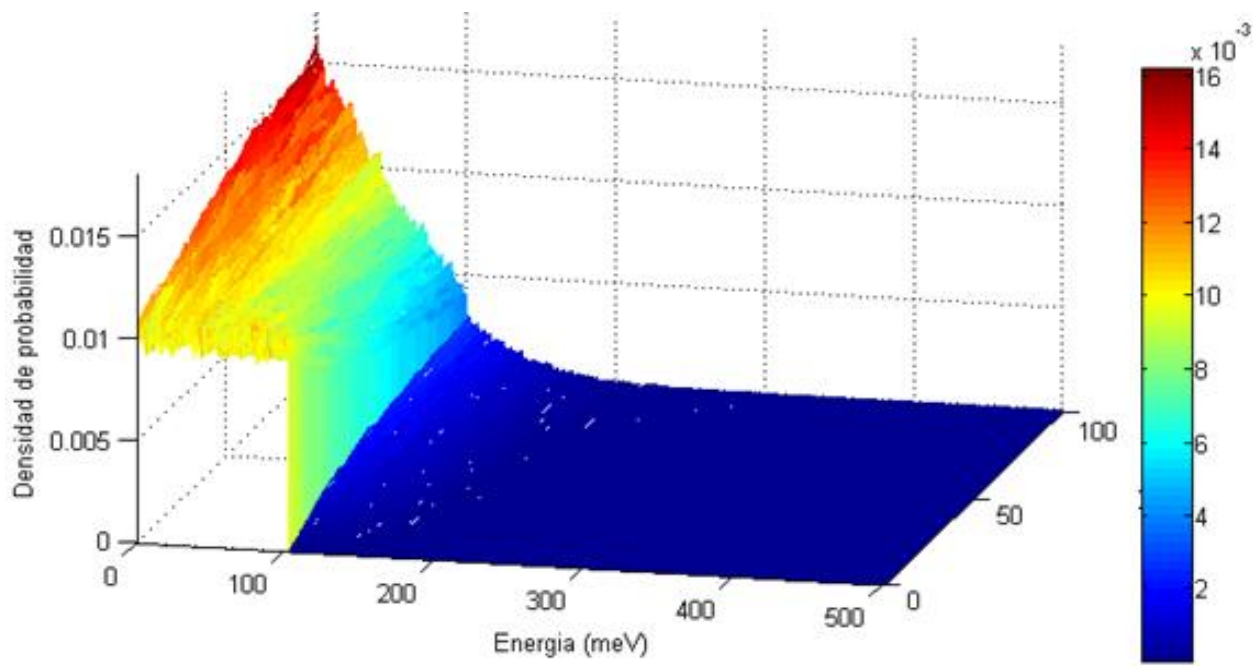

Fuente: Elaboración propia

Para $n_0 = 1000$ y $n = 100$, se grafica en la Figura D.1 donde se muestra cómo la densidad de probabilidad inicial $\wp_0(k)$ que tiene forma de escalón (debido a la asignación aleatoria de la energía para las partículas, según $E_i = rE_0$) se deforma y va adquiriendo una distribución exponencial con el paso de las iteraciones, para $n_0 = 5\text{x}10^4$ y $n = 100$, se grafica en la Figura D.2 donde la densidad de probabilidad es estacionaria (el gas ha llegado al equilibrio térmico):

La distribución de Boltzmann para este gas predice una densidad de probabilidad dada por $\rho(\varepsilon) = \frac{1}{\langle E \rangle} e^{-\frac{\varepsilon}{\langle E \rangle}} = \frac{1}{100\text{meV}} e^{-\frac{\varepsilon}{100\text{meV}}}$, como solución analítica, en la Figura D.2 se compara la distribución inicial $\wp_0(k)$ (en verde), la distribución en el equilibrio (rojo), y la formula analítica de la mecánica estadística (amarillo):

FIGURA D.2

COMPARACIÓN: SIMULACIÓN & SOLUCIÓN ANALÍTICA

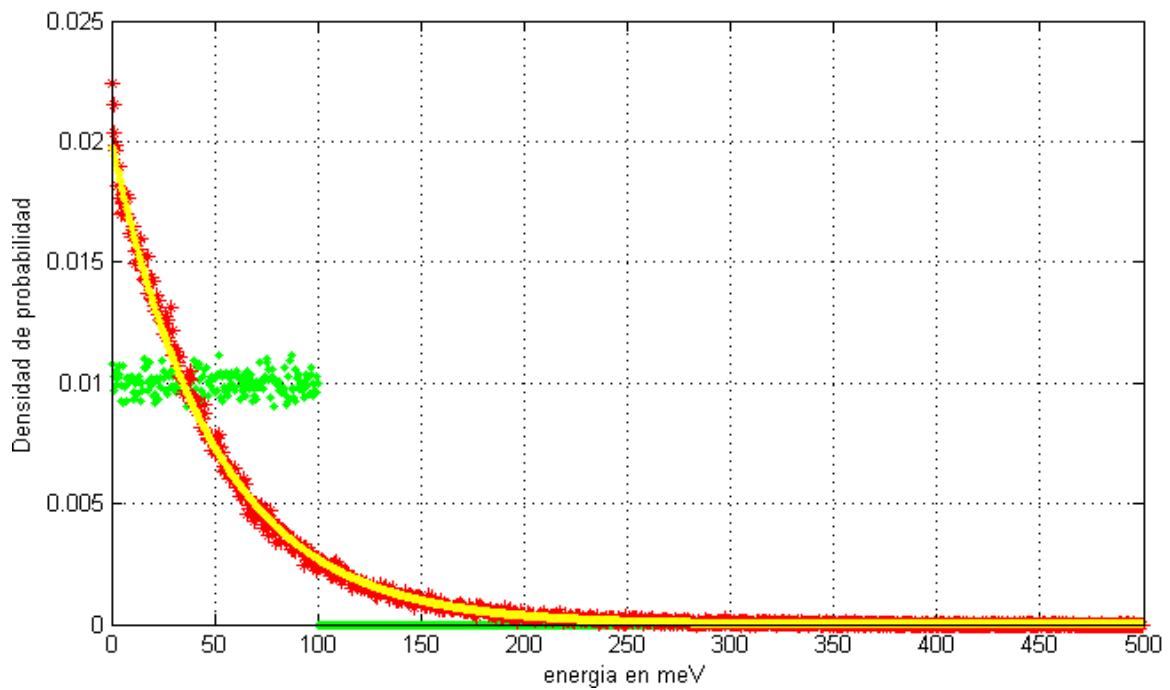

Fuente: Elaboración propia

## ANEXO E: La flecha del tiempo debido a los procesos R

En este anexo se pretende responder la siguiente cuestión: ¿por qué el proceso R establece la irreversibilidad en el tiempo, si de acuerdo a la ecuación [2.8] no depende del sentido del proceso? La respuesta se encuentra en las "condiciones iniciales"; realmente el proceso R no establece la asimetría en el tiempo, sino que permite la introducción de asimetría en el tiempo al fijar un tiempo inicial como el estado más ordenado en todo el tiempo, lo que el proceso U (o cualquier otro que conserve la información) no puede hacer.

Sea el proceso U y R representados por una curva ondulante, y un cirulo respectivamente, la evolución alternada de ellos es una secuencia de curvas interrumpidas por círculos, los trozos de curvas sin círculos conservan la información, son reversibles y en general conservan la entropía, los círculos son realmente invariantes al sentido del tiempo; esto es, que se pueden conectar dos curvas reversibles con distintas entropías a ambos lados del círculo en ambos sentido del tiempo, así que puede describir el aumento de entropía como también su decremento en un sentido del tiempo dado Figura E.1 a); si en el espacio-tiempo se fija un tiempo inicial $t_0$ como el estado más ordenado del universo (esta es la condición inicial), permitiendo que la causalidad conecte círculos con curvas ondulantes, entonces al recorrer el sentido positivo del tiempo se encuentra a las curvas ondulantes conectándose a otras curvas ondulantes (mediante círculos) de manera aleatoria, con probabilidad dada por la ecuación [2.7], aquí es válido el argumento estadístico: es más probable conectarse con curvas de mayor entropía porque son estados más probables[147], de esta manera emerge la flecha termodinámica del tiempo; no obstante, lo mismo es aplicable para la región $t < t_0$, así conforme el tiempo se hace menor la entropía también aumenta, y la flecha del tiempo, de color amarillo en la Figura E.1 b), apunta en ambos sentidos del tiempo con eje de simetría en $t = t_0$, como se muestra, de manera que se retoma la simetría en el tiempo.

---

[147] Los estados desordenados son más abundantes que los estados ordenados, pues el número de microestados correspondientes crece exponencialmente con la entropía, según: $W = e^{S/k_b}$.

FIGURA E.1

PROCESO R Y FLECHA DEL TIEMPO

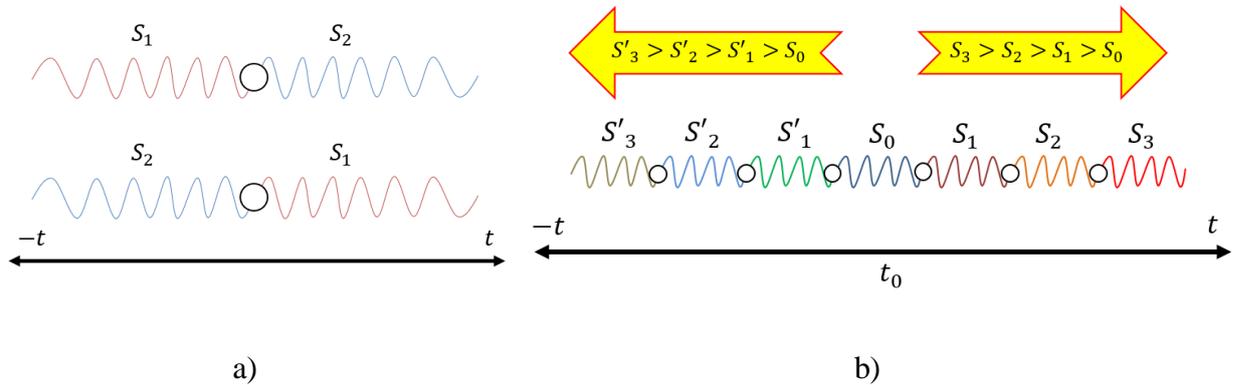

a) b)

Fuente: Elaboración propia.

Este concepto generaliza la segunda ley de la termodinámica pues el aumento de la entropía no es solamente en el sentido positivo del tiempo, sino que puede ser en ambos sentidos (una vez que se define un estado de máximo orden a un tiempo arbitrario $t_0$) de manera que un *Big-Crunch* es posible sin violar la segunda ley de la termodinámica si se define dos tiempos ($t_0$ y $t'_0$) de máximos estados ordenados en el espacio-tiempo (el universo como lo conocemos evolucionaría de uno a otro estado ordenado), y esto podría generalizarse a una secuencia de *Big-Bang* y *Big-Crunch* (modelos cíclicos) tan solo "estableciendo" estados ordenados en ciertos instantes del tiempo.